\documentclass[prd,superscriptaddress,amsfonts,amssymb,amsmath,showpacs,twocolumn,nofootinbib]{revtex4-2}
\usepackage{bm}
\usepackage{amsfonts}
\usepackage{latexsym}
\usepackage{graphicx}
\usepackage{rotating}
\usepackage{amsmath}
\usepackage{palatino}
\usepackage{xcolor} 
\usepackage{mathpazo}
\usepackage{xcolor}
\usepackage{rotating}
\usepackage{adjustbox}
\usepackage{tensor}
\usepackage{textcomp}
\usepackage{float}
\usepackage{booktabs}
\usepackage{dcolumn}
\usepackage[titletoc]{appendix}
\usepackage{booktabs}
\usepackage{multirow}
\usepackage{hyperref}
\hypersetup{colorlinks,citecolor=blue}
\usepackage{amsmath}
\usepackage{xcolor}
\usepackage{orcidlink}
\usepackage{epsfig}
\usepackage{caption}
\usepackage{subcaption}
\usepackage{commath}
\hypersetup{colorlinks,citecolor=blue}
\hypersetup{colorlinks=true,linkcolor=magenta,filecolor=magenta,    urlcolor=blue}

\def\be{\begin{equation}}
\def\ee{\end{equation}}
\def\bea{\begin{eqnarray}}
\def\eea{\end{eqnarray}}

\definecolor{owngreen}{rgb}{0.0, 0.5, 0.0}

\begin{document}

\title{Is Dark Energy Dynamical in the DESI Era? A Critical Review}

\author{Salvatore Capozziello}
\email{capozziello@na.infn.it}
\affiliation{Dipartimento di Fisica ``E. Pancini", Universit\`a di Napoli ``Federico II", Complesso Universitario di Monte Sant’ Angelo, Edificio G, Via Cinthia, I-80126, Napoli, Italy,}
\affiliation{Istituto Nazionale di Fisica Nucleare (INFN), sez. di Napoli, Via Cinthia 9, I-80126 Napoli, Italy,}
\affiliation{Scuola Superiore Meridionale, Largo S. Marcellino, I-80138, Napoli, Italy.}

\author{Himanshu Chaudhary}
\email{himanshu.chaudhary@ubbcluj.ro,\\
himanshuch1729@gmail.com}
\affiliation{Department of Physics, Babeș-Bolyai University, Kogălniceanu Street, Cluj-Napoca, 400084, Romania}
\affiliation{Research Center of Astrophysics and Cosmology, Khazar University, Baku, AZ1096, 41 Mehseti Street, Azerbaijan}

\author{Tiberiu Harko}
\email{tiberiu.harko@aira.astro.ro}
\affiliation{Department of Physics, Babeș-Bolyai University, Kogălniceanu Street, Cluj-Napoca, 400084, Romania}
\affiliation{Astronomical Observatory,
19 Ciresilor Street, Cluj-Napoca 400487, Romania}

\author{G. Mustafa}
\email{gmustafa3828@gmail.com}
\affiliation{Department of Physics,
Zhejiang Normal University, Jinhua 321004, People’s Republic of China}

\begin{abstract}
We investigate whether the recent DESI DR2 measurements provide or  not  evidences for dynamical dark energy by exploring the $\omega_0\omega_a$CDM model and its extensions with free $\sum m_{\nu}$ and $N_{\mathrm{eff}}$. Using a comprehensive MCMC analysis with a wide range of cosmological datasets including DESI~DR2 BAO and Ly$\alpha$ data, CMB compressed likelihoods, BBN, cosmic chronometers, and multiple Type~Ia supernova compilations we assess the statistical preference for departures from $\Lambda$CDM. We find that neither $\Lambda$CDM nor $\omega_0\omega_a$CDM reduces the sound horizon by the $\sim 7\%$ required to alleviate the Hubble tension. DESI~DR2 consistently favors the quadrant $\omega_0 > -1$ and $\omega_a < 0$, indicating a preference for dynamical dark energy of the Quintom-B type at the $\lesssim 3\sigma$ level for most dataset combinations, rising to $\sim 3.8\sigma$ only when the DES-SN5Y supernova sample is included. Allowing $\sum m_{\nu}$ and $N_{\mathrm{eff}}$ to vary does not alter this preference and yields neutrino mass constraints consistent with $\sum m_{\nu} \lesssim 0.1$\,eV, with nonzero masses detected up to the $\sim 1.5\sigma~+$ level. The systematics diagnosis shows that the preference for dynamical dark energy is biased by the low-$z$ ($z<0.1$) DES–SN5Y SNe~Ia sample from the CfA/CSP sample. When these low-$z$ SNe~Ia are excluded, our analysis no longer requires a dynamical dark energy and fully restores the $\Lambda$CDM model. The reconstructed evolution of $\omega(z)$ and $f_{\mathrm{DE}}(z)$ shows a transition from the phantom to the quintessence regime by crossing the phantom divide. Overall, DESI~DR2 provides valuable new insights into dark energy but does not yet challenge the $\Lambda$CDM paradigm. Forthcoming Stage~IV surveys including DESI~DR3, Rubin Observatory, Euclid, Roman Space Telescope, and the Simons Observatory will be crucial for determining whether these hints of dynamical dark energy persist or are due to statistical fluctuations or residual systematics in low-redshift supernova samples.
\end{abstract}

\maketitle
\tableofcontents

\section{Introduction}\label{sec_1}
The modern cosmology was born in the years 1922-1924 through the works by Alexander Friedmann \cite{friedman1922krummung,friedmann1924moglichkeit}, who proposed a drastic change in the existing scientific and philosophical paradigm of his time, namely, that the Universe is static, and the possible dynamical evolution takes place only on small astrophysical scales. Soon after, the dynamic large scale cosmological evolution was confirmed by Hubble's observations  \cite{hubble1929relation}, who also found the mathematical expression of the law of the expansion of the Universe. The theoretical model based on the Friedmann equations also made the cosmological constant $\Lambda$, introduced by Einstein in 1917 in his static model of the Universe \cite{einstein1917kosmologische}, unnecessary from the point of view of the interpretation of the cosmological data, and it was mostly ignored in the discussions of the evolution and structure of the Universe. 

However, the cosmological constant still attracted the attention of theoreticians due to its interesting features \cite{weinberg1989cosmological}, and, in this sense, one can say that it was always present, in some form or another, in the forefront of theoretical investigations. The fundamental question of whether it is a physical or a geometrical quantity is still unsolved. From a theoretical perspective, investigations using a quantum field theoretical approach led to the so-called cosmological constant problem \cite{weinberg1989cosmological}. One of the basic assumptions on the cosmological constant is that it is the quantum vacuum energy density. The cosmological constant problem  consists of the very important disagreement between the small observed value of $\Lambda$, and the large theoretical estimate of the quantum vacuum energy density. The effective cosmological constant, which can be determined from the observed cosmological expansion rate, is expected to be composed of a bare value plus the quantum vacuum energy contribution. However, theoretical quantum field calculations show that the latter is between 50 and 120 orders of magnitude larger than the
value of $\Lambda$ obtained from cosmological observations \cite{bengochea2020can}. Therefore, to explain its observed value, an extreme fine tuning of the bare cosmological constant is required.  Hence, the vacuum energy interpretation of $\Lambda$ remains at least problematic \cite{bengochea2020can,kundu2023does}. For a discussion of the various proposals to solve the cosmological constant problem, see \cite{carroll2001cosmological}.  

In its simplest formulation, the standard (before 1998) cosmological models were based on the Friedmann equations, written down for a flat, homogeneous, and isotropic  Friedmann-Lemaitre-Robertson-Walker metric as
\begin{equation}\label{Fra}
  H^2\equiv\left(\frac{\dot{a}}{a}\right)^2=\frac{8\pi G}{3}\rho, 
 \end{equation}
 \begin{equation}\label{Frb}
 \dot{H}= \frac{\ddot{a}}{a}=-4\pi G\left(\rho+p\right),
\end{equation}
\begin{equation}\label{Frc}
\frac{\ddot{a}}{a}=-\frac{4\pi G}{3}\left(\rho+3p\right),
\end{equation}
where $H$ is the Hubble parameter, $\rho$ and $p$ denote the baryonic matter component in the Universe, satisfying the energy conditions $\rho \geq 0$, $\rho +p\geq 0$, and $\rho+3p\geq 0$. From the Friedmann equations we naturally obtain the law of the conservation of energy
\begin{equation}\label{Frd}
\dot{\rho}+3H(\rho+p)=0.
\end{equation}
From Eq.~(\ref{Frc}) is immediately follows that if the matter content of the Universe  consists only of baryonic matter,  $\ddot{a}<0$, $\forall t>0$, which implies that the Universe is decelerating. This conclusion is also confirmed by the expression of the deceleration parameter $q$,
\begin{equation}
q=\frac{d}{dt}\frac{1}{H}-1=\frac{1}{2}\frac{\rho+3p}{\rho},
\end{equation}
whose positive sign indicates a decelerating evolution. Hence, for a Universe consisting of pressureless dust, the prediction for the present day value of the standard (before 1998) cosmological  model was $q(0)\approx 1/2$. However, the problem of the presence/absence of the cosmological constant in the Friedmann equations continued to be of major interest for the cosmologist.

A theory-independent and direct method to measure the cosmological constant relies on the determination of the functional form of the scale factor as a function of the cosmological time. This is not an easy task,  but if enough precise information about the dependence of a distance measure on redshift can be obtained, then one constrains the value of the cosmological constant. What astronomers determine is the distance in terms of the ``distance modulus $m - M$, where $m$ denotes the apparent magnitude of the source and $M$ is its absolute magnitude. The distance modulus is determined by the luminosity distance $d_L$ through the relation \cite{carroll2001cosmological}
\begin{equation}
m-M=5{\rm log}_{10}\left[d_L\;{\rm (Mpc)}\right]+25. 
\end{equation}
The apparent magnitude can be easily measured,  but it is very difficult to obtain the absolute magnitude of a distant object. However, a very important advance in this field did occur after the possibility of using Type Ia supernovae as standard candles was suggested \cite{goobar1995feasibility}. Type Ia supernovae have an approximately uniform intrinsic luminosity (absolute magnitude $M \sim -19.5$,  comparable to the brightness of the host galaxy in which they explode. They can be detected at high redshifts ($z \sim 1$), and thus can be used for the efficient study of the cosmological effects. The fact that all SNe Ia have similar intrinsic luminosities can be explained  through their explosion mechanism, which takes place when a white dwarf, accreting mass from a companion star, explodes by exceeding  the Chandrasekhar mass limit \cite{liu2023type}. The Chandrasekhar mass limit is usually considered to be nearly-universal quantity, and therefore the supernova explosions are also of nearly-constant luminosity. A scatter of approximately 40\% in the peak brightness still exists in the nearby supernovae, which can be explained by the differences in the composition of the atmospheres of the white dwarfs. 

The use of the observations of the type Ia supernovae led in the 1990's to the amazing discovery of the recent acceleration of the Universe \cite{riess1998observational,perlmutter1999measurements,huterer1999prospects,betoule2014improved,scolnic2018complete,abbott2019first}, representing one of the most important discoveries in the history of cosmology. The simplest possibility to explain the observational data is to reinsert in the gravitational field equations Einstein's ``old" cosmological constant $\Lambda$, and to assume that the Universe consists of matter and a vacuum component \cite{abdalla2022cosmology,navas2024review}. Although the cosmological constant is an excellent fit to the observational data, the corresponding $\Lambda$CDM model constructed through the inclusion of $\lambda$ faces numerous theoretical and observational challenges. From a theoretical perspective, the problem of the nature and interpretation of $\Lambda$ is still an open problem. Is it a fundamental constant of nature that could be considered as such together with the other constants of nature $(c,G,\hbar,e)$, or is it a derived quantity? It is interesting to note that the cosmological constant can be expressed in terms of some fundamental constants as $\Lambda =\hbar ^2G^2m_e^6c^6/e^{12}\approx 1.4\times 10^{-56}\;{\rm cm^{-2}}$ \cite{boehmer2008physics}, where $m_e$ is the electron mass.   

Even though $\Lambda$ is a very good fit to the current data, the observations can also be explained by assuming the presence in the Universe a component called dark energy, which does not cluster on small scales (to avoid its detection through the determinations of the matter density parameter). Moreover, the dark energy has redshift dependence as the Universe expands, and thus it can explain the accelerated expansion of the Universe. 

There are at least three possibilities for introducing an effective dark energy term into the mathematical formalism of the Einstein equations. The first is to assume that dark energy is essentially a geometric effect, and thus the geometric part of the Einstein field equations must be modified by adding a new geometric term, so that $G_{\mu \nu}+G_{\mu \nu}^{(dark)}=8\pi G T_{\mu \nu}$. This approach may be called the modified gravity approach which can consist in extending General Relativity including further curvature terms. See e.g. \cite{Capozziello1, Capozziello2,Capozziello3,Nojiri:2017ncd,Odintsov2}. One can also modify the matter part of the Einstein equations, by adding the dark energy term as a matter term, with the field equations becoming $G_{\mu \nu}=8\pi G T_{\mu \nu}+T_{\mu \nu}^{(dark)}$, in an approach called a modified matter model. Finally, one could also assume the existence of a non-minimal coupling between matter and geometry, which has an effect equivalent to adding a dark energy term in the field equations \cite{harko2018extensions}. 

In the modified matter approach one usually parameterizes the new component $X$ through an equation of state of the form $p_X=\omega _X \rho_X$, where $p_X$ is the effective pressure and $\rho_X $ is the effective energy density of the dark energy. $\omega _X$, the parameter of the dark energy equation of state, can be wary between $\omega _X=-1$, a case corresponding to the cosmological constant, and $\omega _X=0$, corresponding to the presence of baryonic matter only. 

The simplest cosmological model for a dark energy component consists of a single, slowly-rolling scalar field $\phi$, in the presence of a self-interacting potential $V(\phi)$, called a quintessence field. The field satisfies the generalized Klein-Gordon equation \cite{carroll2001cosmological}
\begin{equation}
\ddot{\phi}+3H\dot{\phi}+V'(\phi)=0.
\end{equation}

The energy density of the quintessence field is $\rho_\phi=\dot{\phi}^2/2+V(\phi)$, while its pressure is $p_\phi=\dot{\phi}^2/2-V(\phi)$, giving for the parameter of the equation of state the expression
\begin{equation}
\omega_\phi=\frac{p_\phi}{\rho_\phi}=\frac{\dot{\phi}^2/2-V(\phi)}{\dot{\phi}^2/2+V(\phi)},
\end{equation}
which is generally a function of time. When the field varies slowly, $\dot{\phi}^2<<V(\phi)$, we have $w_\phi \approx -1$, and thus the scalar field can act as an effective cosmological constant. Many types of dark energy models have been proposed in the literature, and for a discussion of their classification see \cite{motta2021taxonomy}.  In particular, recent observational results obtained from the DESI collaboration seem to favor the so-called $\omega _0\omega_a$CDM parameterization \cite{adame2025desi,cortes2024interpreting,giare2024robust,karim2025desi}.   

The theoretical and observational issues related to the acceleration of the Universe are still a topic of debate in the scientific community. In \cite{chung2025strong} it has been pointed out that the post-standardization brightness of SNe Ia is influenced by the properties of their host galaxies, such as mass and star formation rate, both of which are closely related to progenitor age. Thus post-standardization SN Ia luminosity may vary with progenitor age, an effect that is not properly accounted for in SN cosmology. The direct and extensive age measurements of SN host galaxies reveal a significant, 5.5$\sigma$,  correlation between standardized SN magnitude and progenitor age. This correlation introduces a serious systematic bias with redshift in SNe Ia cosmology \cite{chung2025strong}. When three cosmological probes (SNe, BAO, and CMB) are combined, a significantly stronger, greater than $9\sigma$ tension with the $\Lambda$CDM  model is found, suggesting a time-varying dark energy equation of state in a non-accelerating Universe \cite{son2025strong}.

Over its nine-year observational program, successive Wilkinson Microwave Anisotropy Probe (WMAP) data releases tightened the constraints on the dark energy equation of state, while remaining broadly consistent with a cosmological constant. The first-year results of WMAP1~\cite{spergel2003first} predict $\omega = -0.98 \pm 0.12$, consistent with $-1$ within uncertainties. 

The improved polarization measurements in WMAP3~\cite{spergel2007three} refined this estimate to $\omega = -0.967^{+0.073}_{-0.072}$, providing a more precise confirmation of the $\Lambda$CDM predictions. 

A more extensive parameter analysis was performed in WMAP5~\cite{komatsu2009five}, which constrained both the parameterization of $\omega$CDM and $\omega_0\omega_a$, resulting in $\omega = -1^{+0.12}_{-0.14}$ together with $\omega_0 = -1.06 \pm 0.14$ and $\omega_a = 0.36 \pm 0.62$.  

The WMAP7 release~\cite{jarosik2011seven} predicts $\omega = -1.10 \pm 0.14$, and in the $\omega_0\omega_a$CDM model it yields $\omega_0 = -0.93 \pm 0.13$ and $\omega_a = -0.41^{+0.72}_{-0.71}$. Finally, the nine-year mission data in WMAP9~\cite{bennett2013nine} further stabilized the constraints, obtaining $\omega = -1.073^{+0.090}_{-0.089}$, and confirming the absence of any statistically significant deviation from the $\Lambda$CDM model. 

It is still important to note that when cosmological parameter constraints were derived from the WMAP data, they were often combined with complementary observational datasets, including BAO measurements from the 2dFGRS~\cite{percival20012df} and 6dFGS~\cite{beutler20116df} surveys, type~Ia supernova catalog such as SuperNova Legacy Survey (SNLS)~\cite{astier2006supernova} and Union2~\cite{suzuki2012hubble}, as well as small-scale CMB experiments and independent determinations of the Hubble constant $H_0$, all of which tightened the allowed parameter space.

In 2013, the Planck mission released its first-year data~\cite{ade2014planck}. When combined with WMAP9 and the SNLS dataset, it yielded $\omega = -1.13^{+0.13}_{-0.14}$, favoring the phantom regime up to the $2\sigma$ level. The Joint Light-curve Analysis (JLA) dataset released in 2014 \cite{betoule2014improved} improved the spectral calibration of the SNLS data, resulting in a deviation of 1.8$\sigma$ relative to SNLS-3. When Planck13 is combined with the JLA dataset, the resulting constraints shift $\omega$ back toward $-1$, consistent with $\Lambda$CDM. Furthermore, when Planck15 \cite{ade2016planck,ade2016planckmodified} is combined with the JLA data set whose improved supernova brightness calibration removed the apparent phantom behavior ($\omega < -1$) the result yields $\omega = -1.006^{+0.085}_{-0.091}$.

In 2018, the largest combined sample of SNe Ia, consisting of a total of 1048 SNe Ia ranging from $0.01 < z < 2.3$, known as the Pantheon Sample, was released \cite{scolnic2018complete}. When combined with Planck15, it shows $\omega = -1.026 \pm 0.041$, and the combination of Pantheon with Planck15 and BAO yields $\omega = -1.014 \pm 0.040$. In the case of the $\omega_0\omega_a$CDM model, the combination of Pantheon + Planck15 gives $\omega_0 = -1.009 \pm 0.159$ and $\omega_a = -0.129 \pm 0.755$, while Pantheon + Planck15 + BAO yields $\omega_0 = -0.993 \pm 0.087$ and $\omega_a = -0.126 \pm 0.384$. 

In the same year, the release of Planck18 \cite{aghanim2020planck} showed that using only temperature anisotropies (TT+lowE) leads to a preference for phantom-like behavior, with $\omega_0 = -1.54^{+0.59}_{-0.48}$. Including high-$\ell$ polarization (TT, TE, EE+lowE) reduces the uncertainty, but still favors phantom-like behavior, yielding $\omega_0 = -1.52^{+0.56}_{-0.45}$. When Planck lensing is added, the constraints tighten further to $\omega_0 = -1.54^{+0.51}_{-0.41}$. However, once the BAO data are included, the preferred values shift toward the cosmological constant, predicting $\omega_0 = -1.03^{+0.10}_{-0.11}$, thus restoring consistency with the $\Lambda$CDM model. It also shows that Planck18, when used alone, indicates a preference for phantom dark energy up to ($>2\sigma$) \cite{escamilla2024state}.

In addition, new observational data such as DES Y1 \cite{troxel2018dark,abbott2018dark} and DES Y3 \cite{abbott2022dark,abbott2023dark} show a preference for $\omega < -1$. Consequently, the new catalog of SNe Ia, known as Pantheon$^{+}$, consisting of 1701 SNe Ia, presents similar predictions. When combined with CMB and BAO data, it yields $\omega = -0.978^{+0.024}_{-0.031}$, while the Pantheon$^{+}$ dataset alone yields $\omega = -0.90 \pm 0.14$. In the case of the $\omega_0\omega_a$CDM model, the Pantheon$^{+}$ dataset alone yields $\omega_0 = -0.93 \pm 0.15$ and $\omega_a = -0.1^{+0.9}_{-2.0}$. When CMB and BAO measurements are included, the constraints tighten considerably, yielding $\omega_0 = -0.841^{+0.066}_{-0.061}$ and $\omega_a = -0.65^{+0.28}_{-0.32}$. Despite these shifts in the best-fit values, the results remain consistent with $\Lambda$CDM within the $2\sigma$ confidence level. This trend is further supported by the Union3 compilation~\cite{rubin2025union}. In the case of the $\omega$CDM model, Union3 alone yields $\omega = -0.735^{+0.169}_{-0.191}$. When combined with BAO and CMB measurements, yields $\omega = -0.957^{+0.034}_{-0.035}$. For the $\omega_0\omega_a$CDM model, the joint CMB+BAO+Union3 $\omega_0 = -0.744^{+0.100}_{-0.097}$ and $\omega_a = -0.79^{+0.35}_{-0.38}$. 

These results show a preference for dynamical dark energy models over the $\Lambda$CDM scenario at the $1.7$-$2.6\sigma$ level, depending on the choice of dataset combination.

In 2024, the Dark Energy Spectroscopic Instrument (DESI) released its first-year of BAO measurements~\cite{adame2025desi}. When combined with CMB, Pantheon$^{+}$, Union3, and DESY5, these BAO data favor a dynamical dark energy scenario over $\Lambda$CDM at the levels of $2.5\sigma$, $3.5\sigma$, and $3.9\sigma$, respectively. In 2025, DESI published its second BAO dataset~\cite{karim2025desi}, which further strengthened this indication: the DR2 results show a preference for dynamical dark energy at $2.8\sigma$, $3.8\sigma$, and $4.2\sigma$ when combined with CMB, Pantheon$^{+}$, Union3, and DESY5. Consistent support is also found in Ref.~\cite{cuceu2025desi}, where the Best-Ly$\alpha$ sample combined with CMB and Pantheon$^{+}$, Union3, and DESY5 yields preference levels of $1.62\sigma$, $2.50\sigma$, and $2.65\sigma$, respectively. It is worth noting that both DESI DR1 and DESI DR2 favor a dynamical dark energy scenario characterized by $\omega_0 > -1$, $\omega_a < 0$, and $\omega_0 + \omega_a < -1$. Recently, an improved cosmological analysis of the DESY5 dataset~\cite{popovic2025dark} reported $\omega_0 = -0.803 \pm 0.054$ and $\omega_a = -0.72 \pm 0.21$. The significance of this deviation from $\Lambda$CDM is $3.2\sigma$, reduced from the earlier value of $4.2\sigma$ obtained in the DESY5 analysis. Motivated by these results and the recent DESI measurements, a number of subsequent studies have explored the implications of such dynamics of dark energy and deviations, and whether new physics in the dark energy sector can help with cosmological tensions \cite{park2024using,yin2024cosmic,shlivko2024assessing,lodha2025desi,carloni2025does,croker2024desi,mukherjee2024model,roy2025dynamical,wang2024dark,orchard2024probing,dinda2025model,jiang2024nonparametric,vagnozzi2020new,vagnozzi2023seven,pedrotti2025bao,rebouccas2025investigating,pang2025constraints,kessler2025one,gao2025null,borghetto2025bounded,wolf2025robustness,peng2025dark,tsedrik2025interacting,reeves2025tuning,odintsov2025modified,shim2025squeezing,gao2024evidence,luongo2025dark,giare2025dynamical,chakraborty2025desi,giare2025overview,wolf2025navigating,ye2025nec,lopez2025crosschecking,ye2025hints,silva2025new,park2024w_0w_a,fazzari2025cosmographic,zhou2025measuring,wu2025observational,van2025linear,van2509ii,van2025iii,barua2025cosmological,zhang2026prospects,braglia2025exotic,li2025testing,lee2025shape,li2025exploring,mishra2025braneworld,mazumdar2025constraint,liu2025torsion,giare2025neutrino,wang2025can,van2025compartmentalization,mukherjee2025new,moffat2025dynamical,giani2025matter,cheng2025pressure,ye2025tension,jhaveri2025turning,scherer2025challenging,mirpoorian2025dynamical,kou2025unified,dinda2025calibration,wolf2025cosmological,escudero2025sound,toomey2025theory,huang2025reionization,plaza2025probing,yang2025constraining,wang2025lensing,yang2025dark,petri2025dark,meetei2025interaction,arora2025dynamical,ishak2025fall,zapata2025holographic,goldstein2025monodromic,qiang2025new,li2025reconstructing,herold2025bayesian,an2025topological,wang2025model,Capozziello:2019cav,Capozziello:2020ctn,Piedipalumbo:2023dzg,Demianski:2012ra,Demianski:2018nov,Capozziello:2025kws,zhou2025dynamical,lopez2025non,colgain2025desi,afroz2025hint,dymnikova1998self,dymnikova2000decay,dymnikova2001decay,doroshkevich1984formation,ray2011phenomenology,doroshkevich1988cosmological,doroshkevich1989large,khlopov2012physical,doroshkevich1985fluctuations,novikov2016ultralight,novikov2016quantum,novikov2023gravitational}.

The purpose of this paper is to review and investigate the implications of recent cosmological observations on the status of the standard $\Lambda$CDM paradigm. There is convincing evidence that a simple cosmological constant has at least some difficulties in the successful fitting of the latest sets of data, and that an evolving dark energy maydescription  be better suited for the description of the data. However, a detailed analysis of all datasets is necessary to give a complete and convincing answer to this question. 

In the analysis of the observational data in the present study we use the CPL parametrization of the equation of state of dark energy, a simple but still powerful approximation describing the possible redshift evolution of the dark energy. The CPL dark energy parameter depends on two arbitrary constants $\left(\omega _0,\omega _a\right)$, and one of the main tasks of our data analysis is to convincingly determine the numerical values of these parameters. Moreover, in the present study we will also try to answer the question of the nature of dark energy (quintessence, phantom, quintom, etc.), a fundamental problem whose solution could give fundamental insights into the nature and physical origin of the dark energy. 

The present work is organized as follows. We review the theoretical basis and the classification of redshift dependent dark energy and dark matter models in Section~\ref{sec_2}. The observational datasets and the statistical methodology applied are presented in Section~\ref{sec_3}.  The results of the statistical analysis, including the determinations of the values of the CPL parameters, are presented and discussed in detail in Section~\ref{sec_4}. We discuss and conclude the results of our work in Section~\ref{sec_5}. The characteristics of the dark energy constraints are presented in Appendix~\ref{appendix_a1}. A comparison of the cosmological implications of the  DESI Data Release 1  and Data Release 2 is presented in the Appendix~\ref{appendix_a2}.  

\section{Theoretical Background}\label{sec_2}

In the present Section, we introduce the fundamentals of the $\omega(z)$CDM cosmological model, and discuss its theoretical formulation. In the next Section, the cosmological datasets and the methodology for the statistical analysis is presented. 

\subsection{The $\omega(z)$CDM model}

The standard $\Lambda$CDM model is described by the simple form of the Hubble function, which, as a function of the redshift, is given by
\begin{equation}\label{Lambda}
H_\Lambda(z)=H_{0\Lambda}\sqrt{(1+z)^3\Omega_{m0}+\Omega _{\Lambda 0}},
\end{equation}
where we have introduced the redshift variable $z$, defined as $1+z=1/a$, and $\Omega _{m0}$ and $\Omega_{\Lambda 0}={\rm constant}$ denote the matter and dark energy density parameters, with the dark energy density parameter assumed to be proportional to the cosmological constant $\Lambda$. $\Omega_{m0}$ and $\Omega _{\Lambda 0}$ satisfy the closure relation $\Omega_{m0}+\Omega _{\Lambda 0}=1$.  Eq.~(\ref{Lambda}) can be straightforwardly generalized to include a redshift varying dark energy term, so that 
\begin{equation}
H_\omega(z)=H_{0\omega}\sqrt{(1+z)^3\Omega_{m0}+\Omega _{DE0}f_{DE} (z)},
\end{equation}
where the function $f_{DE}(z)$ describes the possible variation with respect to the redshift of the dark energy, while $\Omega _{DE0}$ is a constant \cite{Benetti:2019gmo,Petreca:2023nhy}. A time-varying dark energy can be described by an effective pressure $p_{DE}$ and effective energy density $\rho_{DE}$, assumed to obey an equation of state of the form $p_{DE}(z)=\omega (z)\rho_{DE}(z)$. By assuming matter conservation, the effective dark energy pressure and density must satisfy the conservation equation
\begin{equation}
\frac{d}{dt}\rho_{DE}(t)+3H(t)\left[\rho_{DE}(t)+p_{DE}(t)\right]=0,
\end{equation}
which gives 
\begin{equation}
f_{DE}(z)=\frac{\rho_{DE}(z)}{\rho_{DE0}}=\exp\left[3\int_0^z{\frac{1+\omega (z)}{1+z}dz}\right],
\end{equation}   
where $\rho_{DE0}=\rho_{DE}(0)$,  and $d/dt=-(1+z)H(z)d/dz$. For the CPL parameterization $\omega (z)=\omega _0+\omega _az/(1+z)$, and the function $f_{DE}(z)$ is given by
\begin{equation}
f_{DE}(z)=(1+z)^{3\left(1+\omega _0+\omega _a\right)}\exp\left(-\frac{3\omega _az}{1+z}\right).
\end{equation}

To quantify the deviations of the $\Lambda$CDM model from an arbitrary $\omega (z)$CDM model, we can proceed as follows. By means of a simple algebraic transformation we can write the Hubble function of an arbitrary model as
\begin{equation}\label{16}
\begin{split}
\frac{H_{\omega}(z)}{H_{0\omega}} 
&= \frac{\sqrt{(1+z)^3\Omega_{m0}+\Omega_{DE0}f_{DE}(z)}}{\sqrt{(1+z)^3\Omega_{m0}+\Omega_{\Lambda0}}}\\
& \times \sqrt{(1+z)^3\Omega_{m0}+\Omega_{\Lambda0}} \\
&= A(z)\sqrt{(1+z)^3\Omega_{m0}+\Omega_{\Lambda0}} \\
&= A(z)\frac{H_{\Lambda}(z)}{H_{0\Lambda}} ,
\end{split}
\end{equation}
where 
\begin{equation}
A(z)=\frac{\sqrt{(1+z)^3\Omega_{m0}+\Omega _{DE0}f_{DE} (z)}}{\sqrt{(1+z)^3\Omega_{m0}+\Omega _{\Lambda 0}}}.
\end{equation}

Hence, we have $H_{\omega }(z)/H_{0\omega }=A(z)\left[ H_{\Lambda
}(z)/H_{0\Lambda }\right] $, $H_{\omega }(z)=\left( H_{0\omega }/H_{0\Lambda
}\right) A(z)H_{\Lambda }(z)$. If $A(z)\equiv H_{0\Lambda }/H_{0\omega }$, $%
\forall z\geq 0$, then the given cosmological model is fully equivalent to
the $\Lambda$CDM model. If $A(z)\equiv 1$, then $H_\omega (z)/H_{0\omega}=H_\Lambda (z)/H_{0\Lambda}$. The deviations from $\Lambda$CDM can therefore be described with the help of the function $A(z)$. We introduce the notation as
\begin{equation}
r_\omega (z)=\frac{\Omega _{DE0}f_{DE} (z)}{(1+z)^3\Omega_{m0}}, r_\Lambda (z)=\frac{\Omega _{\Lambda 0}}{(1+z)^3\Omega_{m0}},
\end{equation}
$A(z)$ can be written as
\begin{equation}
A(z)=\sqrt{\frac{1+r_\omega (z)}{1+r_\Lambda (z)}}.
\end{equation}
In the case of the CPL parameterization we obtain the following expression
\begin{equation}
r_{\omega }(z)=\frac{\Omega _{DE0}}{\Omega _{m0}}(1+z)^{3\left( \omega _{0}+\omega
_{a}\right) }\exp \left( -\frac{3\omega _{a}z}{1+z}\right) . 
\end{equation}

Hence, for the function $A(z)$  we find
\begin{equation}
A(z)=\sqrt{\frac{1+r_{DE0}(1+z)^{3\left( \omega _{0}+\omega _{a}\right) }\exp
\left( -\frac{3\omega _{a}z}{1+z}\right) }{1+r_{\Lambda 0}\left( 1+z\right)
^{-3}}},
\end{equation}
where we have denoted $r_{DE0}=\Omega _{DE0}/\Omega _{m0}$ and $r_{\Lambda 0}=\Omega _{\Lambda 0}/\Omega_{m0}$, respectively.

For the $\Lambda$CDM model the deceleration parameter can be determined as
\begin{equation}
\begin{split}
q_\Lambda(z)
&= \frac{d}{dt}\!\left(\frac{1}{H_\Lambda}\right) - 1 
= -(1+z)H_\Lambda(z)\frac{d}{dz}\!\left(\frac{1}{H_\Lambda}\right)-1 \\
&= \frac{\Omega_{m0}(1+z)^3 - 2\Omega_{\Lambda0}}
{2\left[\Omega_{m0}(1+z)^3+\Omega_{\Lambda0}\right]} \\
&= \frac{(1+z)^3 - 2r_{\Lambda0}}
{2\left[(1+z)^3 + r_{\Lambda0}\right]} .
\end{split}
\end{equation}
For the $\omega (z)$CDM model described by the Hubble function (\ref{16}), the deceleration parameter is defined as 
\begin{equation}
\begin{split}
q_\omega(z)
&= \frac{d}{dt}\!\left(\frac{1}{H_\omega}\right) - 1
= -(1+z)H_\omega(z)\frac{d}{dz}\!\left(\frac{1}{H_\omega}\right) - 1 \\
&= \frac{H_{0\Lambda}}{H_{0\omega}} \frac{d}{dt}\!\left(\frac{1}{A(t)H_\Lambda(t)}\right) - 1 \\
&= -(1+z)A(z)H_\Lambda(z)\frac{d}{dz}\!\left(\frac{1}{A(z)H_\Lambda(z)}\right) - 1 .
\end{split}
\end{equation}
Thus, for the relation between the deceleration parameter of the $\omega (z)$CDM and $\Lambda$CDM models we obtain the general relationship
\begin{equation}
q_\omega (z) =q_\Lambda (z) +\frac{(1+z)A'(z)}{A(z)}.
\end{equation}
For the CPL parameterization we obtain
\begin{eqnarray}
\frac{q_{\omega }(z)}{q_{\Lambda }(z)} &=&1+\frac{3r_{\Lambda 0}\left[
z(\omega _{0}+\omega _{a}+1)+\omega _{0}+1\right] }{(1+z)\left[
(1+z)^{3}-2r_{\Lambda 0}\right] } \nonumber\\
&&+\frac{3\left\{ (1+z)^{2}\left[ z(\omega _{0}+\omega _{a})+\omega _{0}%
\right] \right\} }{\left[ (1+z)^{3}-2r_{\Lambda 0}\right] } \nonumber\\
&&-\frac{3\left[ r_{\Lambda 0}+(1+z)^{3}\right] }{(1+z)\left[
(1+z)^{3}-2r_{\Lambda 0}\right] } \nonumber\\
&&\times \frac{e^{\frac{3\omega _{a}z}{1+z}}\left[ z(\omega _{0}+\omega
_{a})+\omega _{0}\right] }{\left[ r_{DE0}(1+z)^{3(\omega _{0}+\omega
_{a})}+e^{\frac{3\omega _{a}z}{1+z}}\right] }.
\end{eqnarray}
We may call the procedure described above the $A(z)$ diagnostic of modified cosmological models that deviate from $\Lambda$CDM.

\subsection{The $\Lambda$$w(z)$DM model}

We consider now the possibility that dark matter evolves on a cosmological scales in a way that deviates from the standard $\lambda$CDM evolution. This means that dark matter may not be "absolutely" called, but it may have some pressure, which justifies keeping in the model title only the letters DM. We assume thus that dark matter with energy density $\rho_{DM}$ has a pressure $p_{DM}=w(z)\rho _{DM}c^2$. We further assume that baryonic matter is cold and that there is no interaction between the three considered major components of the Universe, dark energy, dark matter, and baryonic matter, respectively. We consider now the case in which dark energy is the cosmological constant $\Lambda$, and thus it is not affected by the cosmological expansion. However, the differences in the dark matter properties may induce significant differences with respect to the $\Lambda$CDM model. 

By considering that dark matter and baryonic matter expand adiabatically, the first law of thermodynamics gives for the variation of the dark matter energy density the equation
\be
\frac{d\rho_{DM}}{\rho_{DM}+p_{DM}/c^2}=\frac{d\rho_{DM}}{\rho_{DM}\left(1+w(z)\right)}=-\frac{dV}{V}=-3\frac{da}{a},
\ee   
where $V=a^3$ is the comoving cosmological volume and $a$ is the scale factor. By taking into account the relation between $a$ and the redshift $z$, we obtain the equation
\be
\frac{d\rho_{DM}}{\rho_{DM}\left(1+w(z)\right)}=3\frac{dz}{1+z},
\ee 
giving 
\be
\rho_{DM}(z)=\rho_{DM0}(1+z)^3f_{DM}(z),
\ee
where 
\be
f_{DM}(z)=\exp\left[3\int_0^z{\frac{w(x)dx}{1+x}}\right].
\ee

Hence, we obtain the Hubble function of the $\Lambda$$w(z)$DM model as
\be
\frac{H_w(z)}{H_{0w}}=\sqrt{\Omega _{DM0}(1+z)^3f_{DM}(z)+\Omega _{b0}(1+z)^3+\Omega _{\Lambda0}},
\ee 
where $\Omega _{b0}$ is the density parameter of baryonic matter, $\Omega _{\Lambda0}={\rm constant}$ is the density parameter of the constant dark energy (the cosmological constant), and $H_{w0}$ is the present value of the Hubble function.

The simplest possible warm dark model corresponds to the case $w(z)=w_0={\rm constant}$, which gives $F(z)=(1+z)^{3w}$, giving for the Hubble function the expression
\be
\frac{H_w(z)}{H_{0w}}=\sqrt{\Omega _{DM0}(1+z)^{3(1+w_0)}+\Omega _{b0}(1+z)^3+\Omega _{\Lambda0}}.
\ee 

We can also consider a more general equation of state for dark matter of the form
\be
w(z)=w_0(1+z)^{\alpha},
\ee
where $w_0$ and $\alpha $ are constants. Hence,
\be
f_{DM}(z)=\exp\left[\frac{3w_0}{\alpha}(1+z)^\alpha\right],
\ee
giving for the Hubble function the expression
\bea
\frac{H_w(z)^2}{H_{0w}^2}&=&\Omega _{DM0}(1+z)^{3}\exp\left[\frac{3w_0}{\alpha}(1+z)^\alpha\right]\nonumber\\
&&+\Omega _{b0}(1+z)^3+\Omega _{\Lambda0}.
\eea

For the $\Lambda$$w(z)$DM model we define the $A(z)$ diagnostic function as
\begin{equation}
A(z)=\frac{\sqrt{\Omega _{DM0}(1+z)^3f_{DM}(z)+\Omega _{b0}(1+z)^3+\Omega _{\Lambda0}}}{\sqrt{(1+z)^3\Omega_{m0}+\Omega _{\Lambda 0}}}.
\end{equation}

By denoting $r_{w0}=\Omega _{DM0}/\Omega _{\Lambda0}$ and $r_{b0}=\Omega _{b0}/\Omega _{\Lambda 0}$, the $A(z)$ diagnostic function takes the form
\begin{equation}
A(z)=\sqrt{\frac{1+\left[ r_{w0}f_{DM}(z)+r_{b0}\right] (1+z)^{3}}{1+\left(
1/r_{\Lambda 0}\right) (1+z)^{3}}}.
\end{equation}

The study of the $A(z)$ diagnostic function could give significant information on the deviation of the $\Lambda$$w(z)$DM model with respect to the $\Lambda$CDM model. 

\subsection{The $\omega (z)$$w(z)$DM model}

For the sake of completeness we also mention the $\omega (z)$$w(z)$DM model, in which there is a simultaneous redshift evolution of both dark energy and dark matter, with the parameters of the equations of state denoted $\omega (z)$ and $w(z)$, respectively. The Hubble function for the model is defined according to

\begin{equation}
\begin{split}
\frac{H_{\omega w}(z)}{H_{0\omega w}} 
 =  \sqrt{\left[ \Omega
_{DM0}f_{DM}(z)+\Omega _{b0}\right] (1+z)^{3}+\Omega _{DE0}f_{DE}(z) }.
\end{split}
\end{equation}

The functional forms of $f_{DM}(z)$ and $f_{DE}(z)$ can be determined once the parameters of the equations of state are defined, under the assumption of the adiabatic, non-interacting evolution of dark energy, dark matter, and baryonic matter, respectively.  

\section{Dataset and Methodology}\label{sec_3}
In this work, we constrain the parameters of the dynamically dark energy models introduced in Section~\ref{sec_2}. Our primary objective is to obtain the parameter space using recent cosmological measurements and to evaluate its statistical performance. To efficiently sample the parameter space, we use the Metropolis–Hastings Markov Chain Monte Carlo (MCMC) algorithm~\cite{hastings1970monte}, implemented within the SimpleMC cosmological inference framework ~\cite{simplemc,aubourg2015}. The convergence of MCMC chains is carefully monitored using the Gelman–Rubin diagnostic~\cite{gelman1992inference}, quantified by the statistic $(R - 1)$. We adopt the standard convergence criterion of $R - 1 < 0.01$, ensuring that the chains have reached a statistically stable state and that the posterior distributions are well sampled. Post-processing and visualization of the MCMC outputs are carried out using the \texttt{GetDist} package~\cite{lewis2025getdist}, which provides marginalized one- and two-dimensional likelihood contours, as well as estimates of the derived parameters with their corresponding confidence intervals.

In addition to parameter estimation, we assess the evidence from the Bayesian model $\ln \mathcal{Z}$ for each dark energy model using \texttt{MCEvidence}~\cite{heavens2017marginal}. The Bayesian evidence represents the integrated likelihood of the model over its parameter priors and serves as a quantitative measure of how well a given model fits the observational data while penalizing model complexity. The comparison of the model between two competing cosmological scenarios, $a$ and $b$, is performed by the Bayes factor $B_{ab} = Z_a / Z_b$, or equivalently, its logarithmic form $\ln B_{ab} = \Delta \ln Z$. Following the revised Jeffreys scale~\cite{kass1995bayes}, we interpret the relative evidence as: $0 \leq |\Delta \ln Z| < 1 \Rightarrow \text{weak evidence,}$ $1 \leq |\Delta \ln Z| < 3 \Rightarrow \text{moderate evidence,}$ $3 \leq |\Delta \ln Z| < 5 \Rightarrow \text{strong evidence,}$ $|\Delta \ln Z| \geq 5 \Rightarrow \text{decisive evidence.}$ The model that produces the larger Bayesian evidence (or equivalently, smaller $|\ln Z|$) is statistically preferred.

Furthermore, we also use a simple frequentist approach based on the difference in the best-fit chi-square values, defined as $\Delta \chi^{2} = \chi^{2}_{\text{$\omega_0 \omega_a$CDM Model}} - \chi^{2}_{\text{$\Lambda$CDM Model}}.$ A negative $\Delta \chi^{2}$ implies that the $\omega_0 \omega_a$CDM model provides a better fit to the observational data compared to the standard $\Lambda$CDM model, whereas a positive value suggests a poorer fit. The joint use of $\Delta \ln Z$ and $\Delta \chi^{2}$ therefore offers a consistent and complementary framework to assess both the quality of fit and the overall statistical strength of each cosmological model.

During our analysis, we use several cosmological datasets to constrain the model parameters. These include the Lyman-$\alpha$ Baryon Acoustic Oscillation measurements from DESI~DR2, as well as the galaxy BAO measurements from DESI~DR2 and the Dark Energy Spectroscopic Instrument Data Release~1. We also use various Type~Ia supernova samples, including Pantheon$^+$, the Dark Energy Survey Supernova Program, and Union3, the CamSpec compressed Cosmic Microwave Background likelihood, as well as the Big Bang Nucleosynthesis prior to the baryon density parameter $\Omega_{\mathrm{b}}h^{2}$. In the following, we provide more details about each of the chosen observational datasets used in our analysis.

\begin{itemize}
\item \textbf{Baryon Acoustic Oscillation :} We use the Baryon Acoustic Oscillation (BAO) measurements from over 14 million galaxies and quasars obtained by the Dark Energy Spectroscopic Instrument. These measurements are reported at the effective redshift $z_{\mathrm{eff}}$ for each redshift bin (see Eq.~2.1 of \cite{adame2025desi} for details), where $z_{\mathrm{eff}}$ represents the redshift corresponding to the maximum statistical weight of the sample. For BGS, only the angle-averaged distance ratio $D_V/r_d$ is reported, corresponding to a purely isotropic BAO fit to the monopole. For the other tracers LRG1, LRG2, LRG3+ELG1, ELG2, and QSO the DESI DR2 analysis provides the Hubble distance ratio $D_H/r_d$ and the comoving angular diameter distance ratio $D_M/r_d$. The relevant distance measures are defined as follows: $D_H(z) = \frac{c}{H(z)}$, $D_M(z) = c \int_0^z \frac{dz'}{H(z')}$, and $D_V(z) = \left[ z , D_M^2(z) , D_H(z) \right]^{1/3}.$ Here, $r_d$ represents the sound horizon at the drag epoch, defined as $r_d = \int_{z_d}^{\infty} \frac{c_s(z)}{H(z)},dz,$
where $c_s(z)$ is the sound speed of the photon–baryon fluid. In the standard flat $\Lambda$CDM cosmological model, the sound horizon is given by $r_d = 147.09 \pm 0.26~\mathrm{Mpc}$ \cite{aghanim2020planck}.
    \begin{itemize}
    \item \textbf{DESI DR2 BAO :} In our analysis, we use the recent DESI~DR2 measurements consisting of 13 data points spanning the redshift range ( $0.1 < z < 4.2$ )~\cite{karim2025desi}.
    
    \item \textbf{DESI DR1 BAO :} We also use 12 measurements from the DESI~DR1 catalog, spanning the redshift range ( $0.1 < z < 4.16$ )~\cite{adame2025desi}.

    \item \textbf{Ly$\alpha$ Forest BAO :} We also use the BAO measurements from the Lyman-$\alpha$ forest, which provide constraints on $D_{\mathrm{H}}(z_{\mathrm{eff}})/r_{\mathrm{d}}$ and $ D_{\mathrm{M}}(z_{\mathrm{eff}})/r_{\mathrm{d}}$ at $ z_{\mathrm{eff}} = 2.33$, with values $D_{\mathrm{H}}(z_{\mathrm{eff}})/r_{\mathrm{d}} = 8.632 \pm 0.101~(\mathrm{stat{+}sys})$, $D_{\mathrm{M}}(z_{\mathrm{eff}})/r_{\mathrm{d}} = 38.99 \pm 0.53~(\mathrm{stat{+}sys})$, and $\rho(D_{\mathrm{H}}/r_{\mathrm{d}}, D_{\mathrm{M}}/r_{\mathrm{d}}) = -0.431~(\mathrm{stat{+}sys}).$\cite{karim2025alpha}
    \end{itemize}
   \item \textbf{Type Ia Supernovae:} We also use the Pantheon$^{+}$ (PP) Type Ia Supernovae (SNe~Ia) catalog, which contains 1,701 light curves from 1,550 SNe Ia spanning the redshift range ( 0.001 $\leq$ z $\leq$ 2.26 ) \cite{scolnic2022pantheon,brout2022pantheon}. In our analysis, we consider 1,590 light curves, excluding those with $z < 0.01$ due to significant systematic uncertainties arising from peculiar velocities. Further, we use the catalog of 1,829 photometric light curves of SNe~Ia collected over five years by the Dark Energy Survey Supernova Program (DES-SN5Y)~\cite{abbott2024dark}. This catalog includes 1,635 DES SNe Ia measurements spanning the redshift range $0.10 < z < 1.13$, along with 194 SNe Ia at $z < 0.1$ from the CfA/CSP Foundation sample~\cite{hicken2009cfa3,hicken2012cfa4,foley2017foundation}. Finally, we use the Union3 catalog~\cite{rubin2025union}, which includes 2,087 SNe~Ia. A large fraction of this dataset, comprising 1,363 SNe~Ia, overlaps with the Pantheon$^{+}$ sample but is analyzed using a different approach based on the Bayesian hierarchical modeling framework, Unity~1.5~\cite{rubin2015unity}. In all SNe~Ia catalogs, the absolute magnitude $\mathcal{M}$ is completely degenerate with the Hubble constant $h$ . Consequently, without an external calibration of either $\mathcal{M}$ or $h$, the distance moduli can be shifted by an arbitrary constant offset without altering the relative distances. To address this degeneracy and ensure unbiased cosmological inference, we marginalize over the nuisance parameter $\mathcal{M}$. For more details, see Eqs.~(A9–A12) of \cite{goliath2001supernovae}.
   
   \item \textbf{Cosmic Chronometers:}
   In this analysis, we use 15 measurements spanning the redshift range ( 0.17 $\leq z \leq$ 1.96 ) reported in \cite{moresco2012improved,moresco2015raising,moresco20166}, as these include the full covariance matrix accounting for both statistical and systematic uncertainties \cite{moresco2018setting,moresco2020setting}. These measurements are obtained using the differential age method, which relies on the study of massive, passively evolving galaxies formed at redshifts around ( z $\sim$ 2–3 ). By comparing the change in redshift with the change in age of these galaxies, the Hubble parameter can be directly determined without assuming any specific cosmological model \cite{jimenez2002constraining}.
   
   \item \textbf{CMB Compressed Likelihood:} Finally, following \cite{lemos2023cmb}, we use the CamSpec CMB compressed likelihood, with the information compressed into three parameters, $\boldsymbol{\mu} = (\omega_b, \omega_{bc}, D_M(1090)/r_d)$. The numerical details of the CamSpec likelihood implementation, including the mean vector $\boldsymbol{\mu}$ and covariance matrix $\mathbf{C}$, are described in Appendix~A of \cite{karim2025desi}. The full CamSpec likelihood \cite{rosenberg2022cmb} incorporates the temperature (TT), polarization (EE), and temperature–polarization cross (TE) power spectra for multipoles $\ell > 30$, derived from the Planck PR4 NPIPE CMB maps \cite{akrami2020planck}. We use the CMB compressed likelihood since the dynamical dark energy models considered mainly affect the late-time expansion and geometrical features of the CMB. The full CMB spectrum contains small non-geometrical anomalies, such as the lensing amplitude excess and the low-$\ell$ power deficit, which may introduce systematic biases. Planck data alone, for instance, show a $\gtrsim 2\sigma$ preference for phantom dark energy~\cite{escamilla2024state}, largely due to the low-$\ell$ deficit. To avoid such effects, we use the compressed CMB likelihood in our analysis. Throughout our paper, we use the CMB for the notation.

  \item \textbf{Big Bang Nucleosynthesis (BBN) prior on $\Omega_b h^2$} \\
  We use the $\Omega_b h^2$ determination from~\cite{schoneberg20242024}, derived using the \textsc{PRyMordial} code~\cite{burns2024prymordial}, which accounts for uncertainties in nuclear reaction rates. This yields $\Omega_b h^2 = 0.02218 \pm 0.00055$ under the standard $\Lambda$CDM assumption, and $\Omega_b h^2 = 0.02196 \pm 0.00063$ when $N_{\mathrm{eff}}$ is allowed to vary, in which case a covariance between $\Omega_b h^2$ and $N_{\mathrm{eff}}$ is included. We refer to this prior as BBN, and use it whenever a CMB independent calibration is required.
\end{itemize}

In our analysis, we also vary the sum of neutrino masses and the number of effective relativistic species. Below, you will find further details below

\subsection{Sum of neutrino masses and Number of effective relativistic species }
Neutrinos are an essential component of the thermal history of the Universe. 
The hot Big Bang framework predicts a relic background of cosmic neutrinos, analogous to the cosmic microwave background but consisting of weakly interacting fermions. In the early Universe, neutrinos behave as radiation, influencing the formation of acoustic peaks in the primordial plasma, while at later epochs they contribute as a non-relativistic matter component, affecting the growth of large-scale structure. Consequently, cosmological observations are sensitive to both the effective number of relativistic species and the total mass of all neutrinos, \(\sum m_\nu\), offering information complementary to laboratory measurements \cite{collaboration2016desi}.

In the standard cosmological model, the total neutrino mass is typically fixed to \(\sum m_\nu = 0.06 \, \mathrm{eV}\), corresponding to the minimal mass allowed by neutrino oscillation experiments, which measure only mass squared differences and not the absolute mass scale. These results imply lower bounds of \(\sum m_\nu \gtrsim 0.059 \, \mathrm{eV}\) for the normal hierarchy and \(\sum m_\nu \gtrsim 0.10 \, \mathrm{eV}\) for the inverted hierarchy, though the ordering itself remains unknown \cite{gonzalez2021nufit}.

Direct laboratory constraints, such as those from the KATRIN experiment measuring the endpoint of tritium \(\beta\)-decay, currently yield \(m_\beta < 0.8 \, \mathrm{eV}\) (90\% C.L.), implying \(\sum m_\nu \lesssim 2.4 \, \mathrm{eV}\), independent of cosmological assumptions \cite{bornschein2005katrin,eliasdottir2022next}. Together with oscillation data, this sets a broad allowed range from roughly \(0.06\) to a few eV.

Cosmological data, however, provide much tighter bounds. Since most cosmological observables depend mainly on the total mass rather than the detailed mass splittings, analyses generally assume three degenerate massive neutrinos a simplification that accurately reproduces both normal and inverted hierarchy effects \cite{lesgourgues2006massive}. 
In this picture, nonzero \(\sum m_\nu\) values or upper limits translate consistently across hierarchies \cite{di2018exploring,choudhury2020updated}.

Massive neutrinos impact structure formation primarily through two mechanisms: 
(i) their large thermal velocities cause free-streaming that suppresses clustering on small scales, reducing the amplitude of the matter power spectrum and slightly shifting the BAO scale; and 
(ii) as they become non-relativistic, they contribute to the total matter density, modifying the expansion history according to
\[
\Omega_\nu h^2 = \frac{\sum m_\nu}{93.14 \, \mathrm{eV}} .
\]
These effects alter both the timing of matter \(\Lambda\) equality and the late-time growth of structure \cite{lesgourgues2006massive}. While BAO measurements alone mainly constrain the expansion geometry, combining them with CMB and lensing data significantly tightens the limits on \(\sum m_\nu\) by breaking degeneracies with \(H_0\) and \(\omega_m\). Future surveys like DESI will further improve sensitivity, as its full-shape power spectrum analyses capture the scale-dependent suppression from neutrino mass in addition to precise geometric constraints \cite{desi2024desi}.

In addition to varying the total neutrino mass, we also treat the effective number of relativistic species, \(N_{\text{eff}}\), as a free parameter. 
Allowing \(N_{\text{eff}}\) to vary accounts for the potential existence of extra light relics beyond the three active neutrinos of the Standard Model such as sterile neutrinos or other forms of dark radiation. In the standard cosmological framework, the expected value is \(N_{\text{eff}} = 3.044\) \cite{froustey2020neutrino,bennett2021towards}; any significant deviation from this prediction would indicate the presence of new physics in the early Universe.

In our analysis, we use the following relation for the present-day radiation density: $\Omega_{\mathrm{rad}} = 2.469 \times 10^{-5}\, h^{-2} \left( 1 + 0.2271\, N_{\mathrm{eff}} \right)$~\cite{komatsu2009five}. In the standard cosmological scenario, the effective number of relativistic species is taken to be $N_{\mathrm{eff}} = 3.04$~\cite{Parkinson}. However, in the $\omega_0 \omega_a$CDM + $N_{\mathrm{eff}}$ model, we treat $N_{eff}$ as a free parameter. Assuming spatial flatness, the dark energy density parameter follows from the closure relation $\Omega_{\mathrm{de,0}} = 1 - \Omega_{\mathrm{rad,0}} - \Omega_{m0}.$ The choice of uniform priors for the $\omega_0\omega_a$CDM model is summarized in Table~\ref{tab_0}.

\begin{table}
\centering
\begin{tabular}{lll}
\hline
\textbf{Model} & \textbf{Parameter} & \textbf{Prior} \\
\hline
\multirow{4}{*}{\(\Lambda\)CDM} 
& \( h \) & \( \mathcal{U}[0.4, 0.9] \) \\
& \( \Omega_{m0} \) & \( \mathcal{U}[0.1, 0.5] \) \\
& \( \sum m_\nu \)eV & \(\mathcal{U}[0, 5] \) \\
& \( N_{eff }\) & \( \mathcal{U}[2, 5] \) \\
\hline
\multirow{2}{*}{$\omega_0\omega_a$CDM} 
& \( \omega_0 \) & \( \mathcal{U}[-3, 1] \) \\
& \( \omega_a \) & \( \mathcal{U}[-3, 2] \) \\
\hline
\end{tabular}
\caption{The table summarizes the chosen priors and the parameters of the $\Lambda$CDM and $\omega_0 \omega_a$CDM models used in our analysis. Note that the $\omega_0 \omega_a$CDM model is an extension of the $\Lambda$CDM model, so the parameters $h$, $\Omega_{m0}$, $\sum m_\nu\,(\mathrm{eV})$, and $N_{\mathrm{eff}}$ are common to both. Here, $\mathcal{U}$ denotes uniform priors, and $h \equiv H_0/100$.}\label{tab_0}
\end{table}
\begin{table*}
\setlength{\tabcolsep}{5pt}
\resizebox{\textwidth}{!}{%
\begin{tabular}{lccccccccccc}
\hline
\textbf{Dataset/Models} & $h$ & $\Omega_m$ & $\omega_0$ &  $\omega_a$ & $\sum{m_{\nu}}\,[\mathrm{eV}]$ &  $N_{eff}$ & $r_d$ (Mpc)& $|\Delta \ln \mathcal{Z}_{\Lambda\mathrm{CDM}, \mathrm{Model}}|$ &  $\Delta \chi^2$ & Significance \\ 
\hline
\textbf{$\Lambda$CDM} \\
DESI DR2 + BBN & $0.693{\pm 0.010}$ & $0.297{\pm 0.0088}$ & --- & --- & --- & --- & $147.26{\pm 1.57}$ & 0 & 0 & --- \\
DESI DR2 + BBN + Pantheon$^+$ & $0.692{\pm 0.010}$ & $0.303{\pm 0.0082}$ & --- & --- & --- & --- & $147.48{\pm 1.54}$ & 0 & 0 & --- \\
DESI DR2 + BBN + DES-SN5Y & $0.692{\pm 0.010}$ & $0.309{\pm 0.0082}$ & --- & --- & --- & --- & $145.80{\pm 1.51}$ & 0 & 0 & --- \\
DESI DR2 + BBN + DES-SN5Y ($z>0.5$) & $0.692{\pm 0.010}$ & $0.300{\pm 0.0082}$ & --- & --- & --- & --- & $146.94{\pm 1.52}$ & 0 & 0 & --- \\
DESI DR2 + BBN + Union3 & $0.692{\pm 0.010}$ & $0.303{\pm 0.0085}$ & --- & --- & --- & --- &  $146.50{\pm 1.55}$ & 0 & 0 & --- \\
\hline
\textbf{$\omega_0\omega_a$CDM} \\
DESI DR2 + BBN & $0.640_{-0.042}^{+0.032}$ & $0.361^{+0.044}_{-0.025}$ & $-0.40_{-0.23}^{+0.38}$ & $-1.95_{-1.3}^{+0.61}$ & --- & --- & $145.91{\pm 5.93}$ & 1.72 & -1.95 & 1.97 \\
DESI DR2 + BBN + Pantheon$^+$ & $0.672_{-0.023}^{+0.036}$ & $0.301^{+0.026}_{-0.012}$ & $-0.888_{-0.070}^{+0.058}$ & $-0.26_{-0.50}^{+0.58}$ & --- & --- & $149.07{\pm 6.12}$ & 0.02 & -1.91 & 1.75 \\
DESI DR2 + BBN + DES-SN5Y & $0.677_{-0.022}^{+0.030}$ & $0.319^{+0.022}_{-0.013}$ & $-0.783_{-0.092}^{+0.073}$ & $-0.80{\pm 0.64}$ & --- & --- & $146.44{\pm 5.19}$ & 4.10 & -5.88 & 2.63 \\
DESI DR2 + BBN + DES-SN5Y (z$>$0.01) & $0.680_{-0.022}^{+0.030}$ & $0.307^{+0.030}_{-0.018}$ & $-0.87_{-0.16}^{+0.12}$ & $-0.44_{-0.69}^{+0.84}$ & --- & --- & $146.81{\pm 7.64}$ & 0.49 & -0.37 & 0.93 \\
DESI DR2 + BBN + Union3 & $0.671_{-0.023}^{+0.026}$ & $0.333^{+0.020}_{-0.015}$ & $-0.67{\pm 0.12}$ & $-1.22{\pm 0.68}$ & --- & ---  & $145.49{\pm 4.31}$ & 3.40 & -4.75 & 2.75 \\
\hline
\textbf{$\Lambda$CDM} \\
DESI DR2 + CC & $0.692 \pm 0.010$ & $0.297 \pm 0.008$ & --- & --- & --- & --- & $147.38 \pm 1.58$ & 0 & 0 & ---  \\
DESI DR2 + CC + Pantheon$^+$ & $0.691 \pm 0.010$ & $0.304 \pm 0.008$ & --- & --- & --- & --- & $146.61 \pm 1.53$ & 0 & 0 & --- \\
DESI DR2 + CC + DES-SN5Y & $0.691 \pm 0.010$ & $0.309 \pm 0.008$ & --- & --- & --- & --- & $145.98 \pm 1.54$ & 0 & 0 & --- \\
DESI DR2 + CC + DES-SN5Y ($z > 0.1$) & $0.691 \pm 0.010$ & $0.300 \pm 0.008$ & --- & --- & --- & --- & $147.07 \pm 1.57$ & 0 & 0 & ---\\
DESI DR2 + CC + Union3 & $0.692 \pm 0.010$ & $0.303 \pm 0.008$ & --- & --- & --- & --- & $146.65 \pm 1.54$ & 0 & 0 & --- \\
\hline
\textbf{$\omega_0\omega_a$CDM} \\
DESI DR2 + CC & $0.652_{-0.041}^{+0.032}$ & $0.337_{-0.035}^{+0.048}$ & $-0.61_{-0.32}^{+0.37}$ & $-1.2_{-1.3}^{+1.0}$ & --- & --- & $147.50{\pm 4.18}$ & 0.94 & -1.44 & 1.13  \\
DESI DR2 + CC + Pantheon$^+$ & $0.676_{-0.021}^{+0.025}$ & $0.306_{-0.013}^{+0.020}$ & $-0.889_{-0.068}^{+0.059}$ & $-0.31_{-0.43}^{+0.51}$ & --- & --- & $148.13{\pm 4.40}$ & 0.25 & -1.87 & 1.75 \\
DESI DR2 + CC + DES-SN5Y & $0.677_{-0.019}^{+0.022}$ & $0.321_{-0.012}^{+0.017}$ & $-0.786_{-0.081}^{+0.068}$ & $-0.80_{-0.48}^{+0.55}$ & --- & --- & $146.21{\pm 3.71}$ & 3.63 & -5.71 & 2.87 \\
DESI DR2 + CC + DES-SN5Y ($z > 0.1$) & $0.679_{-0.021}^{+0.025}$ & $0.307_{-0.020}^{+0.026}$ & $-0.89_{-0.15}^{+0.12}$ & $-0.36_{-0.59}^{+0.78}$ & --- & --- & $147.61{\pm 4.62}$ & 0.97 & 0.64 & 0.81 \\
DESI DR2 + CC + Union3 & $0.668 \pm 0.022$ & $0.330_{-0.014}^{+0.019}$ & $-0.70 \pm 0.11$ & $-1.07 \pm 0.58$ & --- & --- & $146.11{\pm 3.66}$  & 3.02 & -4.53 & 2.73 \\
\hline
\textbf{$\Lambda$CDM} \\
DESI DR2 + CMB & $0.683{\pm 0.0043}$ & $0.304{\pm 0.0056}$ & --- & --- & --- & --- & $147.17{\pm 0.23}$ & 0 & 0 & --- \\
DESI DR2 + CMB + Pantheon$^+$ & $0.681{\pm 0.0041}$ & $0.306{\pm 0.0054}$ & --- & --- & --- & --- &  $147.09{\pm 0.22}$ & 0 & 0 & --- \\
DESI DR2 + CMB + DES-SN5Y & $0.679{\pm 0.0040}$ & $0.309{\pm 0.0054}$ & --- & --- & --- & --- &   $147.13{\pm 0.23}$ & 0 & 0 & --- \\
DESI DR2 + CMB + DES-SN5Y ($z > 0.1$) & $0.682{\pm 0.0042}$ & $0.305{\pm 0.0055}$ & --- & --- & --- & --- &   $147.00{\pm 0.23}$ & 0 & 0 & --- \\
DESI DR2 + CMB + Union3 & $0.681{\pm 0.0042}$ & $0.306{\pm 0.0056}$ & --- & --- & --- & --- & $147.09{\pm 0.22}$ & 0 & 0 & --- \\
\hline
\textbf{$\omega_0\omega_a$CDM}\\
DESI DR2 + CMB & $0.637_{-0.020}^{+0.018}$ & $0.356{\pm 0.022}$ & $-0.43{\pm 0.21}$ & $-1.72_{-0.61}^{+0.67}$ & --- & --- & $146.71{\pm 0.28}$ & 2.50 & -3.61 & 2.71  \\
DESI DR2 + CMB + Pantheon$^+$ & $0.676{\pm 0.006}$ & $0.312{\pm 0.006}$ & $-0.865{\pm 0.059}$ & $-0.49{\pm 0.22}$ & --- & --- & $146.94{\pm 0.27}$  & 0.37 & -2.85 & 2.29 \\
DESI DR2 + CMB + DES-SN5Y & $0.669{\pm 0.006}$ & $0.320{\pm 0.006}$ & $-0.782{\pm 0.058}$ & $-0.75{\pm 0.23}$ & --- & ---  & $146.87{\pm 0.26}$ & 2.83 & -7.03 & 3.76 \\
DESI DR2 + CMB + DES-SN5Y ($z>0.1$) & $0.670{\pm 0.009}$ & $0.318_{-0.011}^{+0.009}$ & $-0.803_{ -0.11}^{+0.092}$ & $-0.68_{-0.29}^{+0.39}$ & --- & ---  & $146.88{\pm 0.27}$ & 0.95 & -1.66 & 1.95 \\
DESI DR2 + CMB + Union3 & $0.660{\pm 0.008}$ & $0.329{\pm 0.009}$ & $-0.685{\pm 0.083}$ & $-1.02{\pm 0.29}$ & --- & --- & $146.82{\pm 0.26}$ & 3.54 & -6.16 & 3.80 \\
\hline
\textbf{$\Lambda$CDM + $\sum{m_{\nu}}\,[\mathrm{eV}]$} \\
DESI DR2 + CMB  & $0.682 \pm 0.0052$ & $0.302 \pm 0.0063$ & --- &  --- &  $<0.056$ & --- & $147.21{\pm 0.27}$ & 0 & 0 & --- \\ 
DESI DR2 + CMB + Pantheon$^+$ & $0.680 \pm 0.0049$ & $0.305 \pm 0.0059$ & --- &  --- & $<0.070$ & --- & $147.10{\pm 0.27}$ & 0 & 0 & --- \\ 
DESI DR2 + CMB + DES-SN5Y & $0.677 \pm 0.0051$ & $0.309 \pm 0.0061$ & --- &  --- & $<0.076$ & --- & $146.01{\pm 0.28}$ & 0 & 0 & --- \\ 
DESI DR2 + CMB + DES-SN5Y ($z>0.01$) & $0.681 \pm 0.0050$ & $0.304 \pm 0.0060$ & --- &  --- & $<0.065$ & --- & $147.17{\pm 0.27}$ & 0 & 0 & --- \\ 
DESI DR2 + CMB + Union3 & $0.679 \pm 0.0052$ & $0.306 \pm 0.0062$ & --- &  --- & $<0.073$ & --- & $147.09{\pm 0.28}$ & 0 & 0 & --- \\ 
\hline
\textbf{$\omega_0\omega_a$CDM + $\sum{m_{\nu}}\,[\mathrm{eV}]$} \\
DESI DR2 + CMB & $0.628_{-0.027}^{+0.022}$ & $0.366 \pm 0.029$ & $-0.36 \pm 0.27$ & $-1.98_{-0.91}^{+0.80}$ & $<0.136$ & --- & $146.42{\pm 0.48}$ & 4.15 & -3.56 & 2.37 \\ 
DESI DR2 + CMB + Pantheon$^+$ & $0.673{\pm 0.0066}$ & $0.314 \pm 0.0071$ & $-0.857 \pm 0.057$ & $-0.55_{-0.22}^{+0.25}$ & $<0.110$ & --- & $146.82{\pm 0.38}$ & 0.01 & -2.63 & 2.51 \\
DESI DR2 + CMB + DES-SN5Y & $0.665{\pm 0.0064}$ & $0.322 \pm 0.007$ & $-0.769 \pm 0.060$ & $-0.83_{-0.25}^{+0.27}$ & $<0.113$ & --- & $146.73{\pm 0.41}$ & 4.49 & -6.99 & 3.85 \\
DESI DR2 + CMB + DES-SN5Y ($z>0.01$) & $0.667{\pm 0.010}$ & $0.321 \pm 0.011$ & $-0.790 \pm 0.11$ & $-0.740{\pm 0.37}$ & $<0.109$ & --- & $146.78{\pm 0.39}$ & 3.43 & -1.30 & 1.91 \\
DESI DR2 + CMB + Union3 & $0.656{\pm 0.008}$ & $0.322 \pm 0.009$ & $-0.674 \pm 0.091$ & $-1.09{\pm 0.33}$ & $<0.125$ & --- & $146.67{\pm 0.40}$ & 4.12 & -5.97 & 3.58 \\
\hline
\textbf{$\Lambda$CDM + $N_{eff}$} \\
DESI DR2 + CMB & $0.698{\pm 0.009}$ & $0.293{\pm 0.007}$ & --- & --- & --- &  $3.22{\pm 0.13}$ & $145.58{\pm 0.85}$ & 0 & 0 & --- \\
DESI DR2 + CMB + Pantheon$^+$ & $0.692{\pm 0.008}$ & $0.298{\pm 0.006}$ & --- & --- & --- &  $3.27_{-0.13}^{+0.12}$ & $145.84{\pm 0.84}$ & 0 & 0 & --- \\
DESI DR2 + CMB + DES-SN5Y & $0.687{\pm 0.008}$ & $0.303{\pm 0.006}$ & --- & --- & --- &  $3.22{\pm 0.12}$ & $146.08{\pm 0.82}$ & 0 & 0 & --- \\
DESI DR2 + CMB + DES-SN5Y $(z > 0.1)$ & $0.696{\pm 0.008}$ & $0.295{\pm 0.006}$ & --- & --- & --- &  $3.30{\pm 0.13}$ & $145.70{\pm 0.83}$ & 0 & 0 & --- \\
DESI DR2 + CMB + Union3 & $0.692{\pm 0.008}$ & $0.298{\pm 0.007}$ & --- & --- & --- &  $3.26{\pm 0.13}$ & $145.88{\pm 0.83}$ & 0 & 0 & --- \\
\hline
\textbf{$\omega_0\omega_a$CDM + $N_{eff}$} \\
DESI DR2 + CMB & $0.646_{-0.040}^{+0.025}$ & $0.345_{-0.031}^{+0.040}$ & $-0.53_{-0.27}^{-0.34}$ & $-1.38_{-1.1}^{+0.88}$ & --- & $3.15_{-0.41}^{+0.30}$ & $146.44{\pm 1.90}$ & 2.24  & -2.74 & 1.54 \\
DESI DR2 + CMB + Pantheon$^+$ & $0.685{\pm 0.010}$ & $0.305 \pm 0.008$ & $-0.885 \pm 0.057$ & $-0.27{\pm 0.27}$ & --- & $3.33_{-0.30}^{+0.26}$ & $145.57{\pm 1.38}$ & 0.75 & -2.26 & 2.02 \\
DESI DR2 + CMB + DES-SN5Y & $0.673{\pm 0.009}$ & $0.315 \pm 0.008$ & $-0.796 \pm 0.064$ & $-0.62_{- 0.29}^{+0.33}$ & --- & $3.21{\pm 0.26}$ & $146.14{\pm 1.28}$ & 3.11 & -5.97 & 3.11 \\
DESI DR2 + CMB + DES-SN5Y ($z>0.01$) & $0.680{\pm 0.015}$ & $0.310 \pm 0.014$ & $-0.86 \pm 0.11$ & $-0.40_{- 0.40}^{+0.45}$ & --- & $3.26_{-0.30}^{+0.26}$ & $145.95{\pm 1.42}$ & 1.14 & -1.16 & 1.27 \\
DESI DR2 + CMB + Union3 & $0.664_{-0.014}^{+0.012}$ & $0.326 \pm 0.013$ & $-0.700 \pm 0.10$ & $-0.89{\pm 0.42}$ & --- & $3.18_{-0.28}^{+0.24}$ & $146.28{\pm 1.32}$ & 2.85 & -5.16 & 3.33 \\
\hline
\textbf{$\Lambda$CDM} \\
Ly$\alpha$ + CMB + Galaxy BAO & $0.683{\pm 0.004}$ & $0.304{\pm 0.0056}$ & --- & --- & --- & --- & $147.17{\pm 0.23}$ & 0 & 0 & --- \\
Ly$\alpha$ + CMB + Pantheon$^+$ & $0.673{\pm 0.005}$ & $0.318 \pm 0.007$ & --- & --- & --- & --- & $146.44{\pm 1.71}$ & 0 & 0 & --- \\
Ly$\alpha$ + CMB + DES-SN5Y & $0.670{\pm 0.005}$ & $0.322 \pm 0.007$ & --- & --- & --- & --- & $146.39{\pm 1.45}$ & 0 & 0 & --- \\
Ly$\alpha$ + CMB + DES-SN5Y ($z>0.1$) & $0.674{\pm 0.005}$ & $0.315 \pm 0.007$ & --- & --- & --- & --- & $146.51{\pm 1.72}$ & 0 & 0 & --- \\
Ly$\alpha$ + CMB + Union3 & $0.672{\pm 0.005}$ & $0.319 \pm 0.008$ & --- & --- & --- & --- & $146.44{\pm 1.58}$ & 0 & 0 & --- \\
\hline
\textbf{$\omega_0\omega_a$CDM} \\
Ly$\alpha$ + CMB + Galaxy BAO & $0.637_{-0.020}^{+0.018}$ & $0.356{\pm 0.022}$ & $-0.43{\pm 0.21}$ & $-1.72_{-0.61}^{+0.67}$ & --- & --- & $146.71{\pm 0.28}$ & 2.24 & -3.16 & 2.71 \\
Ly$\alpha$ + CMB + Pantheon$^{+}$ & $0.673{\pm 0.012}$ & $0.317_{-0.013}^{+0.011}$ & $-0.913{\pm 0.09}$ & $-0.34_{-0.44}^{+0.52}$ & --- & --- & $146.78{\pm 0.30}$ & 0.75 & -0.40 & 0.97 \\
Ly$\alpha$ + CMB + DES-SN5Y & $0.672_{-0.0095}^{+0.011}$ & $0.318_{-0.011}^{+0.0096}$ & $-0.77{\pm 0.10}$ & $-0.88{\pm 0.50}$ & --- & --- & $146.76{\pm 0.29}$ & 3.11 & -2.85 & 2.30 \\
Ly$\alpha$ + CMB + DES-SN5Y ($z>0.01$) & $0.675{\pm 0.012}$ & $0.316{\pm 0.012}$ & $-0.99{\pm 0.18}$ & $-0.08{\pm 0.75}$ & --- & --- & $146.77{\pm 0.31}$ & 1.14 & -0.15 & 0.06 \\
Ly$\alpha$ + CMB + Union3 & $0.692{\pm 0.007}$ & $0.323_{-0.014}^{+0.012}$ & $-0.66{\pm 0.14}$ & $-1.27{\pm0.65}$ & --- & --- & $146.76{\pm 0.29}$ & 2.85 & -2.44 & 2.43 \\
\hline
\end{tabular}
}
\caption{This table shows the marginalized posterior means and 68\% credible intervals for the cosmological parameters of the $\Lambda$CDM and $\omega_0\omega_a$CDM models, considering variations in the total neutrino mass $\sum m_{\nu}\,[\mathrm{eV}]$ and the effective number of relativistic species $N_{\mathrm{eff}}$. The constraints are derived using combinations of DESI DR2, DESI DR2 Ly$\alpha$, CMB, BBN, and CC data, together with different Type Ia supernova samples (Pantheon$^+$, DES-SN5Y, DES-SN5Y with $z > 0.1$, and Union3).}\label{tab_1}
\end{table*}


\begin{figure*}
\begin{subfigure}{.33\textwidth}
\includegraphics[width=\linewidth]{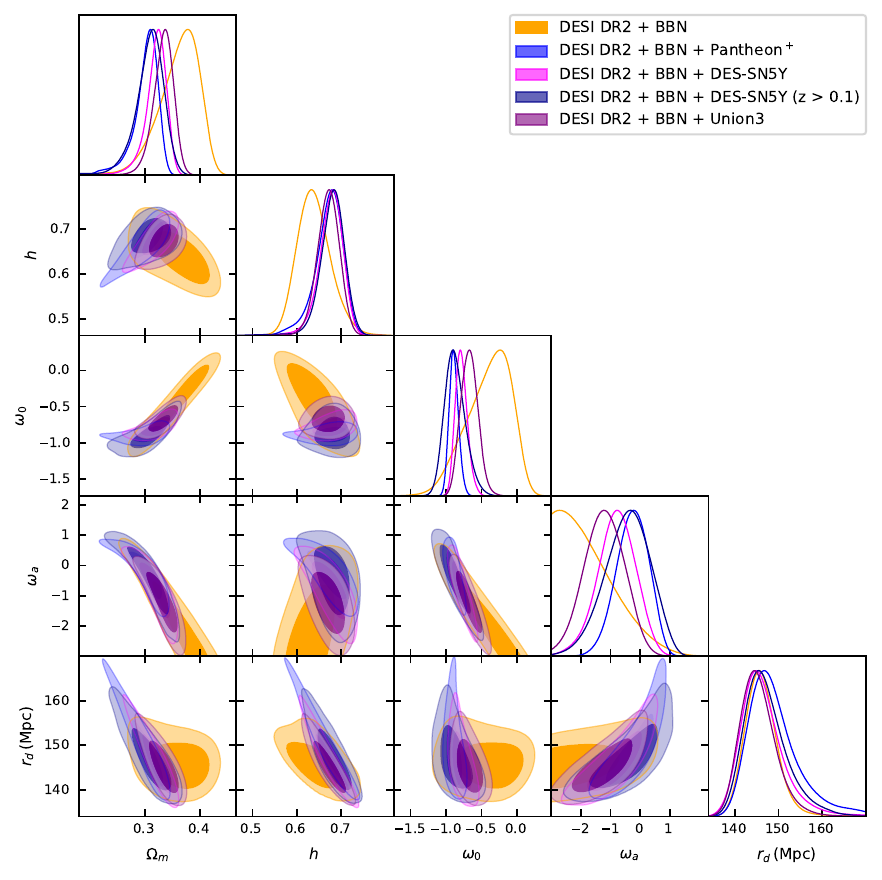}
\end{subfigure}
\hfil
\begin{subfigure}{.33\textwidth}
\includegraphics[width=\linewidth]{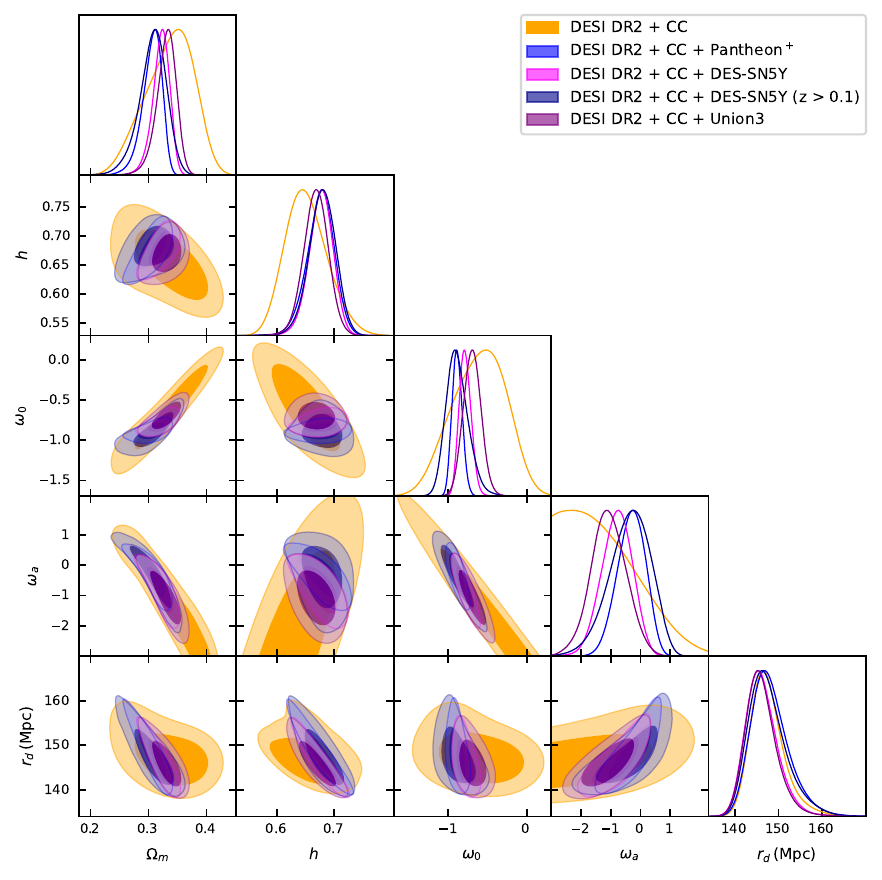}
\end{subfigure}
\hfil
\begin{subfigure}{.33\textwidth}
\includegraphics[width=\linewidth]{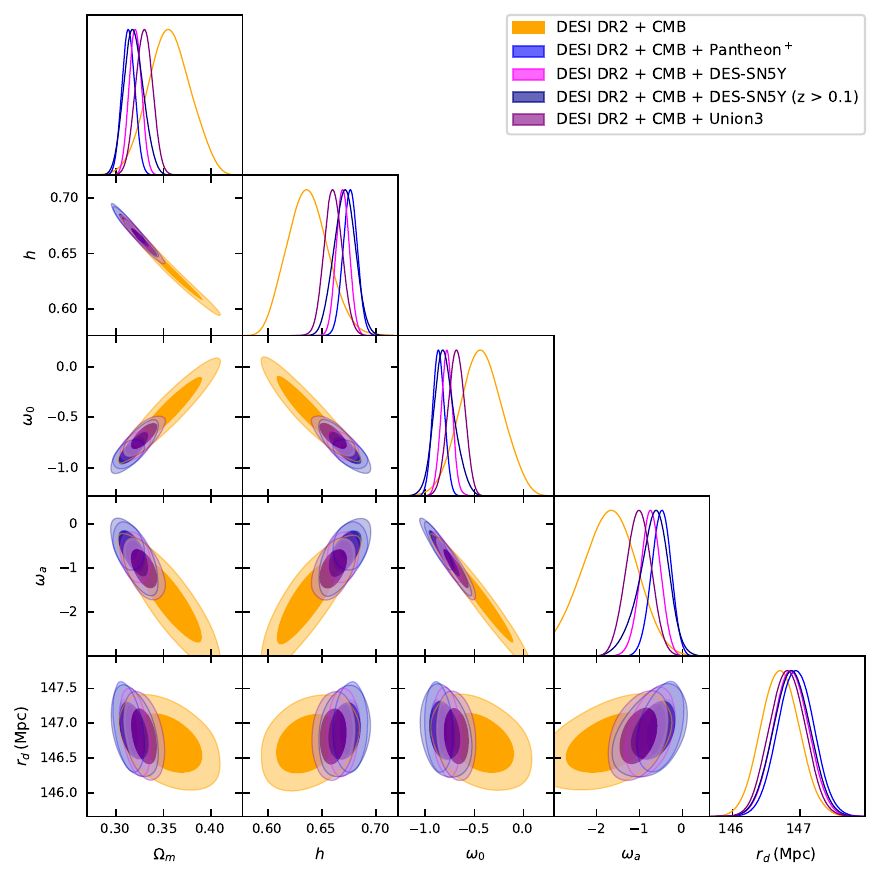}
\end{subfigure}
\begin{subfigure}{.33\textwidth}
\includegraphics[width=\linewidth]{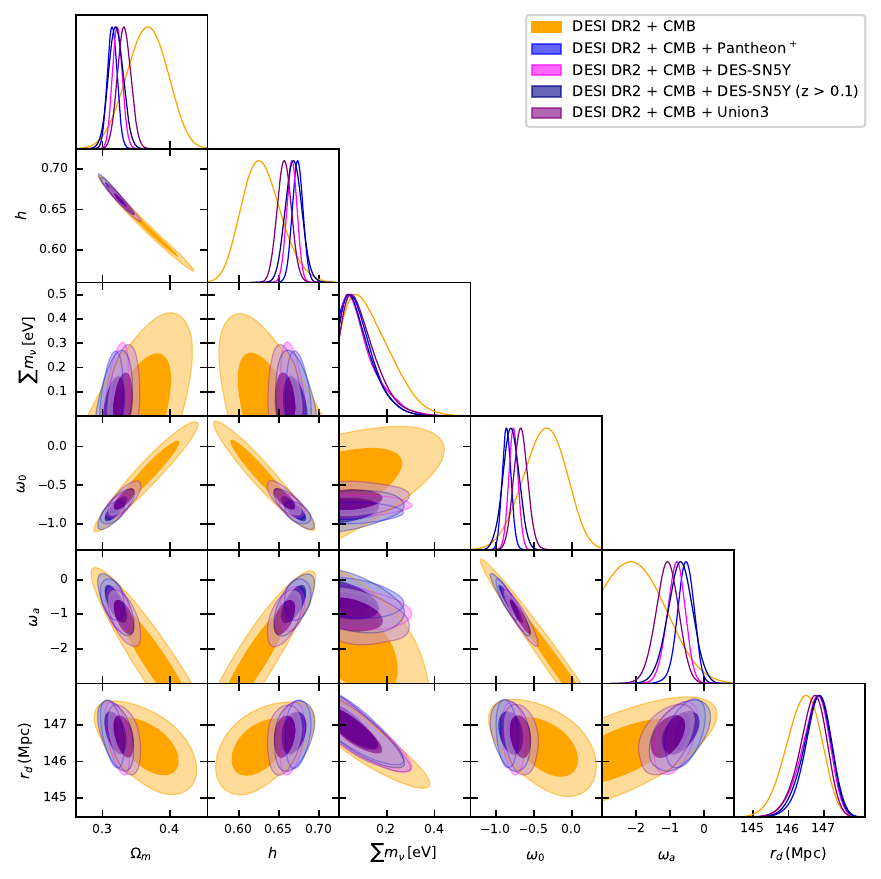}
\end{subfigure}
\hfil
\begin{subfigure}{.33\textwidth}
\includegraphics[width=\linewidth]{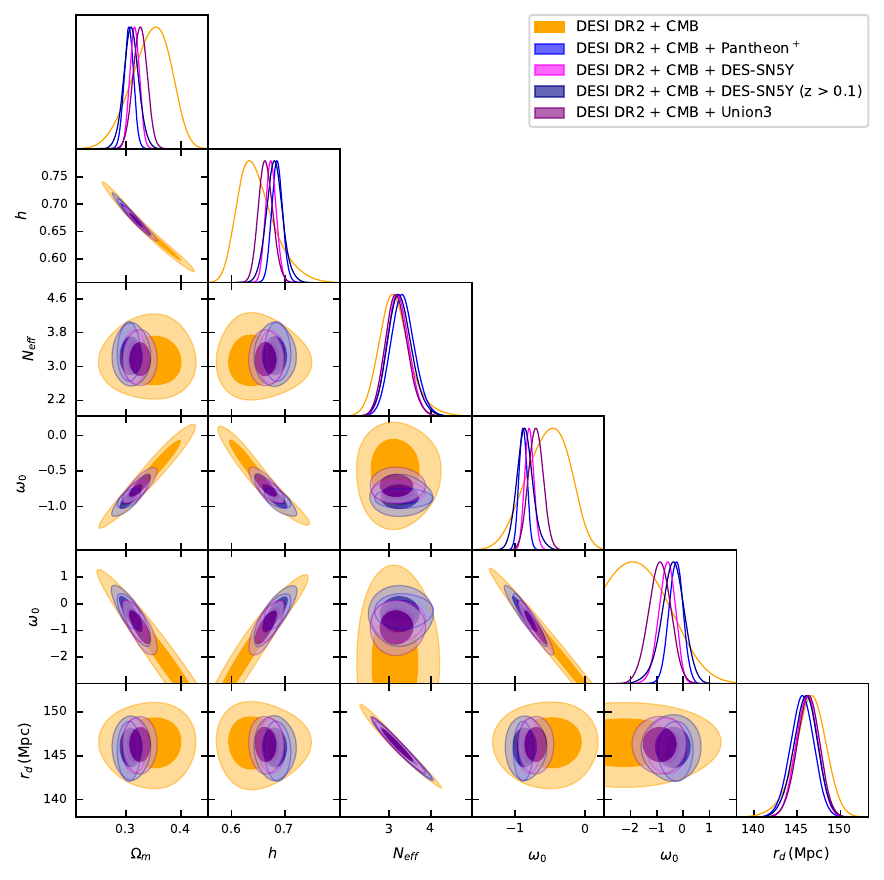}
\end{subfigure}
\hfil
\begin{subfigure}{.33\textwidth}
\includegraphics[width=\linewidth]{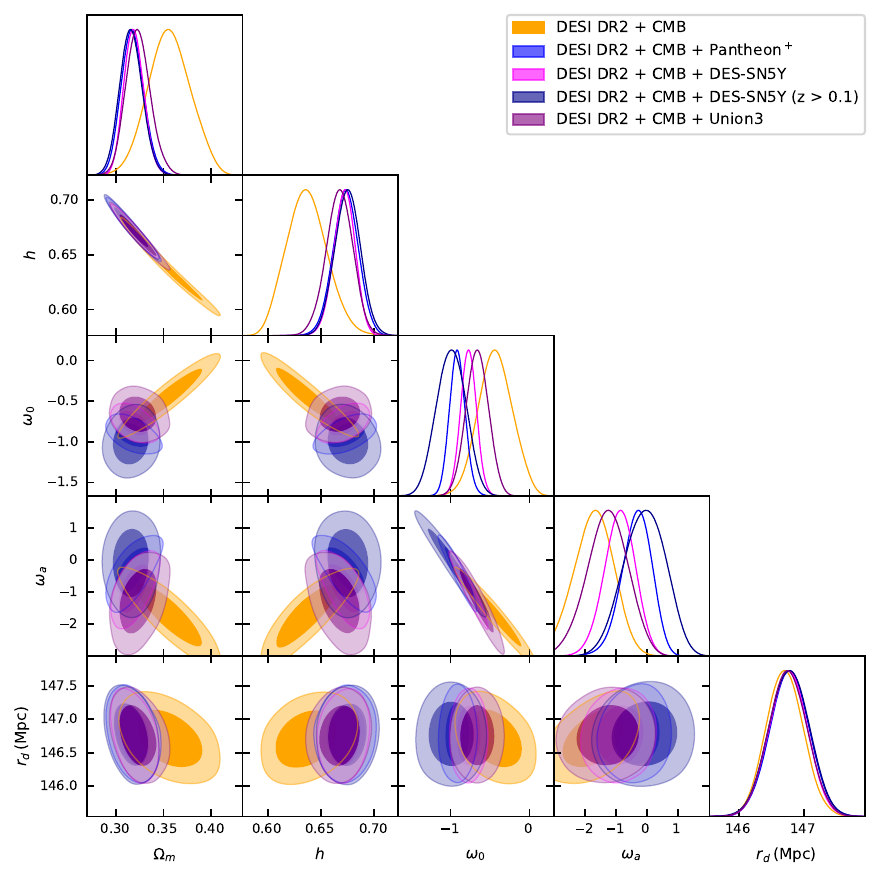}
\end{subfigure}
\caption{The figure shows contours of the cosmological parameters for the $\omega_0\omega_a$CDM model at 68\% (1$\sigma$) and 95\% (2$\sigma$) confidence levels. The constraints are derived using combinations of DESI DR2, DESI DR2 Ly$\alpha$, CMB, BBN, and CC data, together with different Type~Ia supernova samples (Pantheon$^+$, DES-SN5Y, DES-SN5Y with $z > 0.1$, and Union3).}\label{fig_1}
\end{figure*}

\section{Results}\label{sec_4}
In this section, we discuss our main results, which are organized into five subsections. We begin with the status of the $H_0$ tension after DESI DR2, and then examine the implications of DESI DR2 for dynamical dark energy. Next, we consider whether the data favor not only dynamical but also phantom-like behavior. We then explore whether these indications of 
dynamical dark energy could instead be driven by systematics. Finally, we present a brief statistical analysis supporting our findings.

Fig~\ref{fig_1} presents the corner plots of the $\omega_0\omega_a$CDM model using DESI DR2 data combined with various SN~Ia samples. The first-row panels show results with BBN, CC, and CMB (left to right). The second row shows the 
$\omega_0\omega_a$CDM+$\sum m_\nu$ and $\omega_0\omega_a$CDM+$N_{\mathrm{eff}}$ 
extensions, and finally the case including DESI DR2 Ly$\alpha$. The diagonal panels show the 1D posteriors, and the off-diagonal panels show the 2D contours at 68\% and 95\% C.L. Table~\ref{tab_1} shows the marginalized posterior means and the 68\% credible  intervals for each case.

\subsection{$H_0$ Tension after DESI DR2}\label{sec_4a}
In this subsection, we discuss one of the most important and open issues in modern cosmology the Hubble tension and examine the status of this problem in light of the recent DESI DR2 measurements. Before doing so, it is important to note that, in order to fully resolve the $H_0$ tension, one must not only increase the value of $H_0$ but also reduce the sound horizon scale $r_d$ at the epoch of baryon drag by approximately $7\%$. This balance is crucial because the angular size of the sound horizon, precisely measured by the CMB, depends on the ratio $r_d / D_A(z_*)$. Hence, increasing $H_0$ without a corresponding modification of $r_d$ would lead to inconsistencies with BAO observations.

\begin{figure*}
\centering
\includegraphics[scale=0.60]{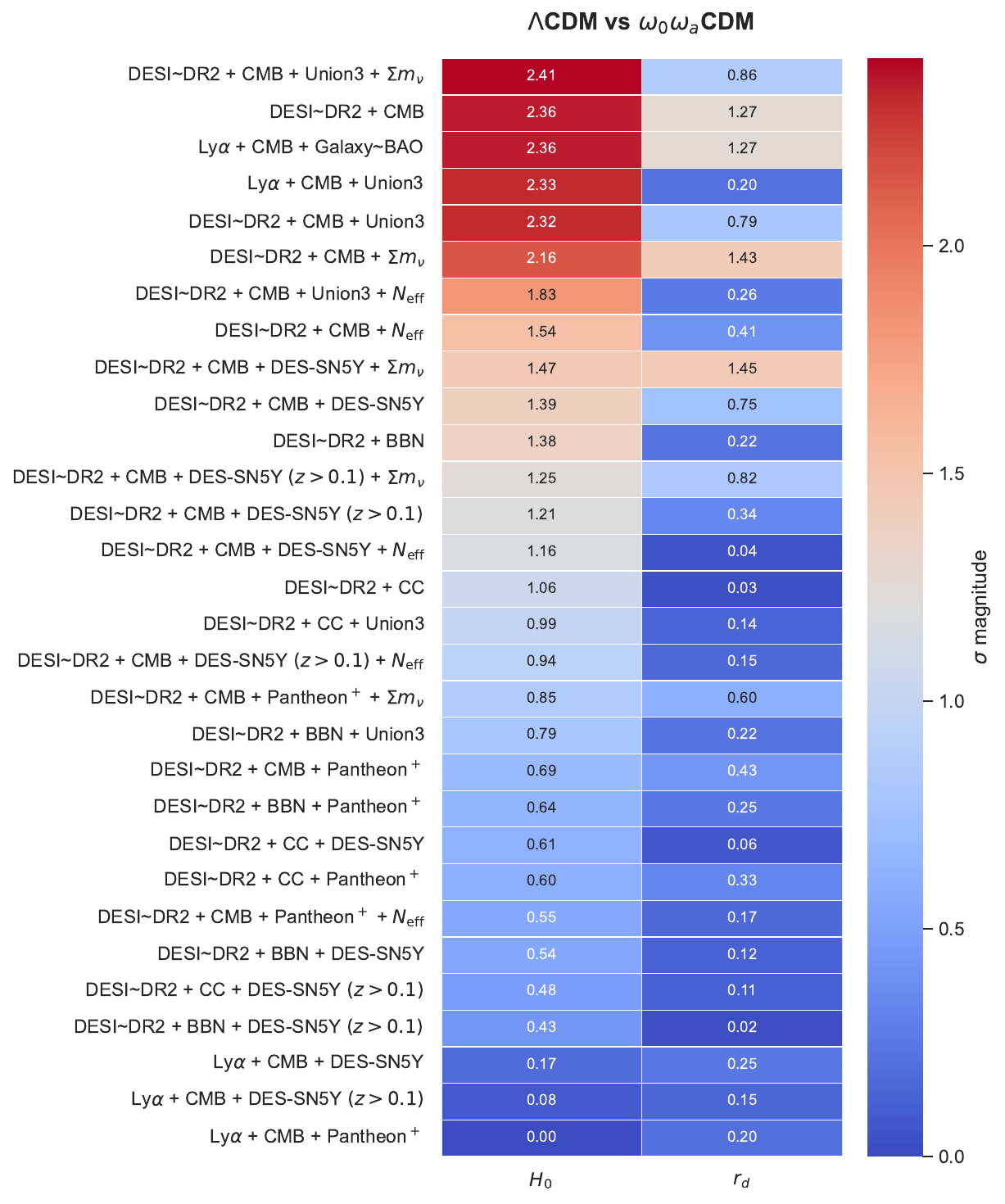}
\caption{This figure shows the comparative analysis of the predicted values of the Hubble parameter ($h$) and the sound horizon scale ($r_d$) between the standard $\Lambda$CDM and $\omega_0\omega_a$CDM models. Each row corresponds to a specific combination of datasets, including DESI~DR2, DESI~DR2~Ly$\alpha$, CMB, BBN, and CC data, combined with different Type~Ia supernova samples (Pantheon$^+$, DES-SN5Y, DES-SN5Y with $z > 0.1$, and Union3). The color scale indicates the statistical significance (in units of $\sigma$) of the deviation between the two models, with warmer tones representing larger differences.}\label{fig_2}
\end{figure*}

We begin with a comparative analysis between the $\Lambda$CDM and $\omega_0\omega_a$CDM models, focusing on the predicted values of the Hubble parameter ($h$) and the sound horizon scale ($r_d$) across various observational combinations. When we combine the DESI~DR2 data with BBN and CC measurements, the tension in the inferred value of $h$ remains below $1.5\sigma$, while that in $r_d$ is within $0.3\sigma$, indicating consistency between the two models. However, when CMB data are included, the tension between the models becomes more pronounced, reaching up to $\sim2.5\sigma$ for $H_0$ in combinations such as DESI~DR2 + CMB and DESI~DR2 + CMB + $\Sigma m_\nu$. Including $\Sigma m_\nu$ and $N_{\mathrm{eff}}$ as free parameters slightly changes this behavior, sometimes reducing or increasing the level of tension depending on the chosen combination of datasets \cite{Capozziello:2023ewq,Capozziello:2024stm}. In contrast, when DESI~DR2 Ly$\alpha$ forest measurements are added, the differences in both $h$ and $r_d$ drop below $1\sigma$, showing the strong agreement between the $\Lambda$CDM and $\omega_0\omega_a$CDM models. A visual summary of these results is shown in Fig.~\ref{fig_2}, which presents the statistical deviations of the $h$ and $r_d$ parameters for the $\Lambda$CDM and $\omega_0\omega_a$CDM models corresponding to each dataset.

Before discussing the $h$ tension, we note that for this specific analysis we 
restrict our SNe Ia dataset to the Pantheon$^+$ sample only. This choice is 
motivated by the fact that Pantheon$^+$ is the most internally consistent of the currently available compilations and avoids the additional calibration systematics and internal tensions present in DESY5 and Union3. As discussed in Sec.IV~A~2 of Ref.~\cite{pedrotti2025bao}, these issues can bias late-time cosmological inferences and complicate the interpretation of the Hubble tension. Therefore, using Pantheon$^+$ ensures a clean and constructive comparison.

\begin{figure*}
\centering
\includegraphics[scale=0.55]{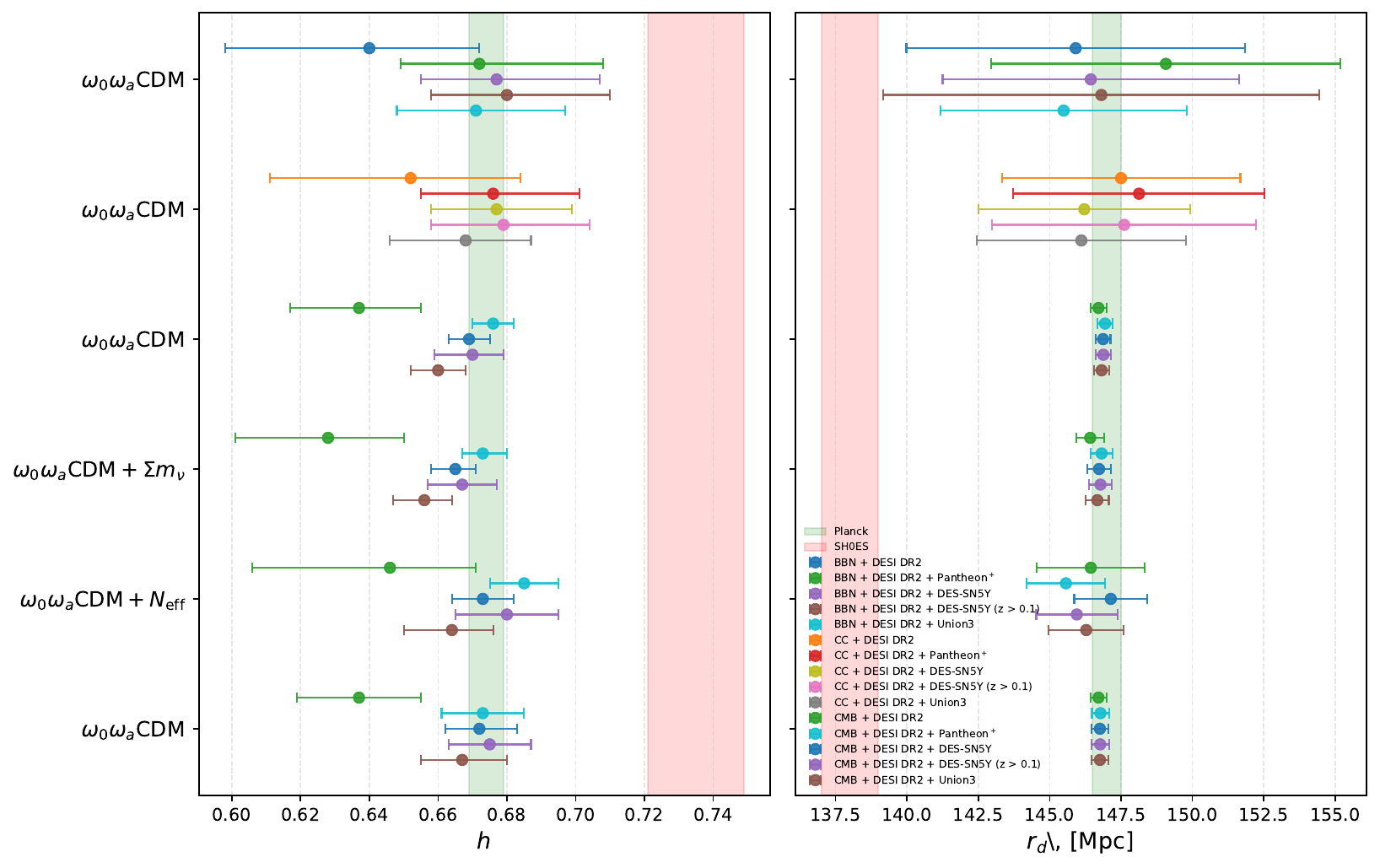}
\caption{This figure shows the whisker plots of the Hubble constant $h$ (left panel) and the sound horizon at the drag epoch $r_d$ (right panel). Each point represents the marginalized mean value corresponding to each dataset combination, as indicated in the legend at the lower left of the right panel, with horizontal error bars showing the 68\% (1$\sigma$) uncertainties. The green and red shaded vertical bands indicate the $1\sigma$ ranges of the \textit{Planck} ($h = 0.674 \pm 0.005$, $r_d = 147.0 \pm 0.5$~Mpc) and \textit{SH0ES} ($h = 0.735 \pm 0.014$, $r_d = 138.0 \pm 1.0$~Mpc) measurements, respectively.}\label{fig_3}
\end{figure*}

Let us return to our point, the Hubble tension. As is well known, the $\omega_{0}\omega_{a}$CDM model is a dynamical dark energy model, and we know that dark energy is completely sub-dominant at recombination; indeed, dynamical dark energy cannot reduce the sound horizon. This feature can be observed in Fig.~\ref{fig_3} on the $r_d$-plane, where we find that, focusing only on the Pantheon$^{+}$ combinations, the inferred sound-horizon values lie in the range $r_d \simeq 147$-$149$~Mpc. Relative to the reference value $r_d = 138.0 \pm 1.0$~Mpc, this corresponds to deviations of $\sim 2\sigma$ for the BBN and CC combinations, which increase up to $6$-$9\sigma$ whenever CMB data are included. Indeed, the combination of DESI~DR2 with the other datasets does not reduce the sound horizon $r_d$.

Further, the inferred value of the Hubble parameter lies in the range $h \simeq 0.672$-$0.676$. Compared with the local measurement $h = 0.735 \pm 0.014$~\cite{riess2022comprehensive}, the BBN and CC combinations show a tension at the level of $\sim 1.9$-$2\sigma$. This tension increases to about $3.8$-$3.9\sigma$ when CMB data are included. None of the Pantheon$^{+}$ combinations bring the inferred value of $h$ into agreement with the local measurement; indeed, the Hubble tension remains 
unresolved.

In \cite{verde2024tale} it is shown that, in $\Lambda$CDM, the DESI+Planck combination predicts $H_{0} = 68.17 \pm 0.28~\mathrm{km\,s^{-1}\,Mpc^{-1}}$, 
corresponding to an $\sim 5\sigma$ tension with the local measurements. Meanwhile, dynamical dark energy models predict values between $H_{0} = 63.6^{+1.6}_{-2.1}$ and $H_{0} = 67.51 \pm 0.59~\mathrm{km\,s^{-1}\,Mpc^{-1}}$, 
depending on the choice of dataset. None of these values are close to the local measurement, showing that dynamical dark energy fails to fit BAO + CMB + SNe Ia + local $H_{0}$, even though dynamical dark energy can describe BAO + CMB + SNe Ia well.

Mathematically speaking, as we know, for dynamical dark energy models the energy 
density evolves according to the following relation: $\frac{\rho_{\mathrm{DE}}(z)}{\rho_{\mathrm{DE},0}}
= \exp\!\left[ 3 \int_{0}^{z} \frac{1 + \omega(z')}{1 + z'} \, dz' \right].$ For the $\Lambda$CDM model, the dark energy density remains constant with redshift, i.e., $\frac{\rho_{\mathrm{DE}}(z)}{\rho_{\mathrm{DE},0}} = 1.$ However, in dynamical dark energy scenarios, two possibilities arise: either $\frac{\rho_{\mathrm{DE}}(z)}{\rho_{\mathrm{DE},0}} > 1,$
indicating an increasing dark energy density with redshift, or $\frac{\rho_{\mathrm{DE}}(z)}{\rho_{\mathrm{DE},0}} < 1,$
corresponding to a decreasing dark energy density as the Universe evolves. In general, for dynamical dark energy models, the evolution of $H(z)$ can be 
given by the following expression: $H^{2}(z) = H_{0}^{2}(1 -\Omega_{m})\,\frac{\rho_{\mathrm{DE}}(z)}{\rho_{\mathrm{DE},0}} + H_{0}^{2}\Omega_{m}(1+z)^{3}.$ As discussed earlier, dark energy is completely sub-dominant at recombination and becomes more relevant at low redshifts (late time). Since the matter component evolves in the same way in both $\Lambda$CDM and in dynamical dark energy models, the quantity $H_{0}^{2}\Omega_{m}$ ends up being constrained to nearly the same value in both cases, as a consequence, if $\frac{\rho_{\mathrm{DE}}(z)}{\rho_{\mathrm{DE},0}}$ increases, then $H_{0}$ must decrease, and vice versa.

\begin{figure}
\centering
\includegraphics[scale=0.40]{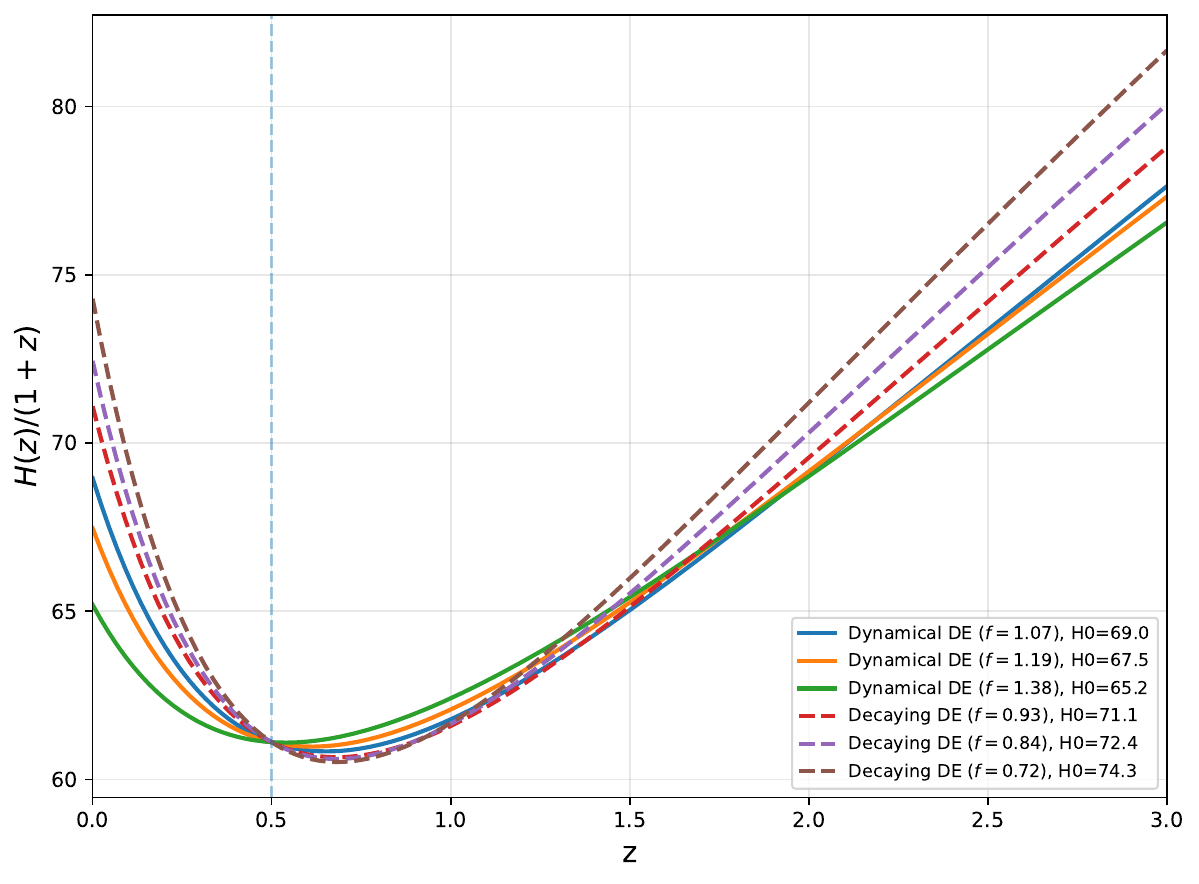}
\caption{This figure shows how different choices for the dark-energy evolution, 
characterized by the ratio 
$f(z_p)=\rho_{\mathrm{DE}}(z_p)/\rho_{\mathrm{DE}}(0)$, affect 
the expansion history $H(z)/(1+z)$. Models with ($f(z_p)>1$, $\omega(z)>-1$) are shown as solid lines and predict a lower value of $H_0$, whereas models with ($f(z_p)<1$, $\omega(z)<-1$) are shown as dotted lines and predict a higher value of $H_0$.}\label{fig_4}
\end{figure}

As a simple test, in Fig.~\ref{fig_4} we show how variations in the dark energy evolution, modeled as $f(z) = \rho_{\mathrm{DE}}(z)/\rho_{\mathrm{DE},0} = 1 + A z + B z^2$, affect the inferred Hubble parameter. We consider both dynamical dark-energy scenarios, shown as solid lines ($f>1$), 
and decaying dark-energy scenarios, shown as dotted lines ($f<1$), adjusting $H_0$ to maintain consistency at a pivot redshift of $z=0.5$, while keeping the matter density parameter fixed at $\Omega_{m0} = 0.301$. The resulting behavior of $H(z)/(1+z)$ clearly show the negative correlation correlation between $f(z)$ and $H_0$: models with increasing dark energy density ($f>1$) lead to lower values of $H_0$, and vice versa. Indeed, both DESI~DR1 and DESI~DR2 prefer $\omega(z) > -1$, which implies $\rho_{\mathrm{DE}}(z)/\rho_{\mathrm{DE},0} > 1$. This suggests that resolving the $H_0$ tension after the DESI release is highly unlikely, as DESI results indicate evidence for dynamically evolving dark energy. In \cite{vagnozzi2018constraints,vagnozzi2020new,alestas2020h,banerjee2021hubble,lee2022local,colgain2025much, Colgain:2024xqj} also shows that if \(\omega > -1\), this lowers \(H_0\)

\begin{figure*}
\centering
\includegraphics[scale=0.30]{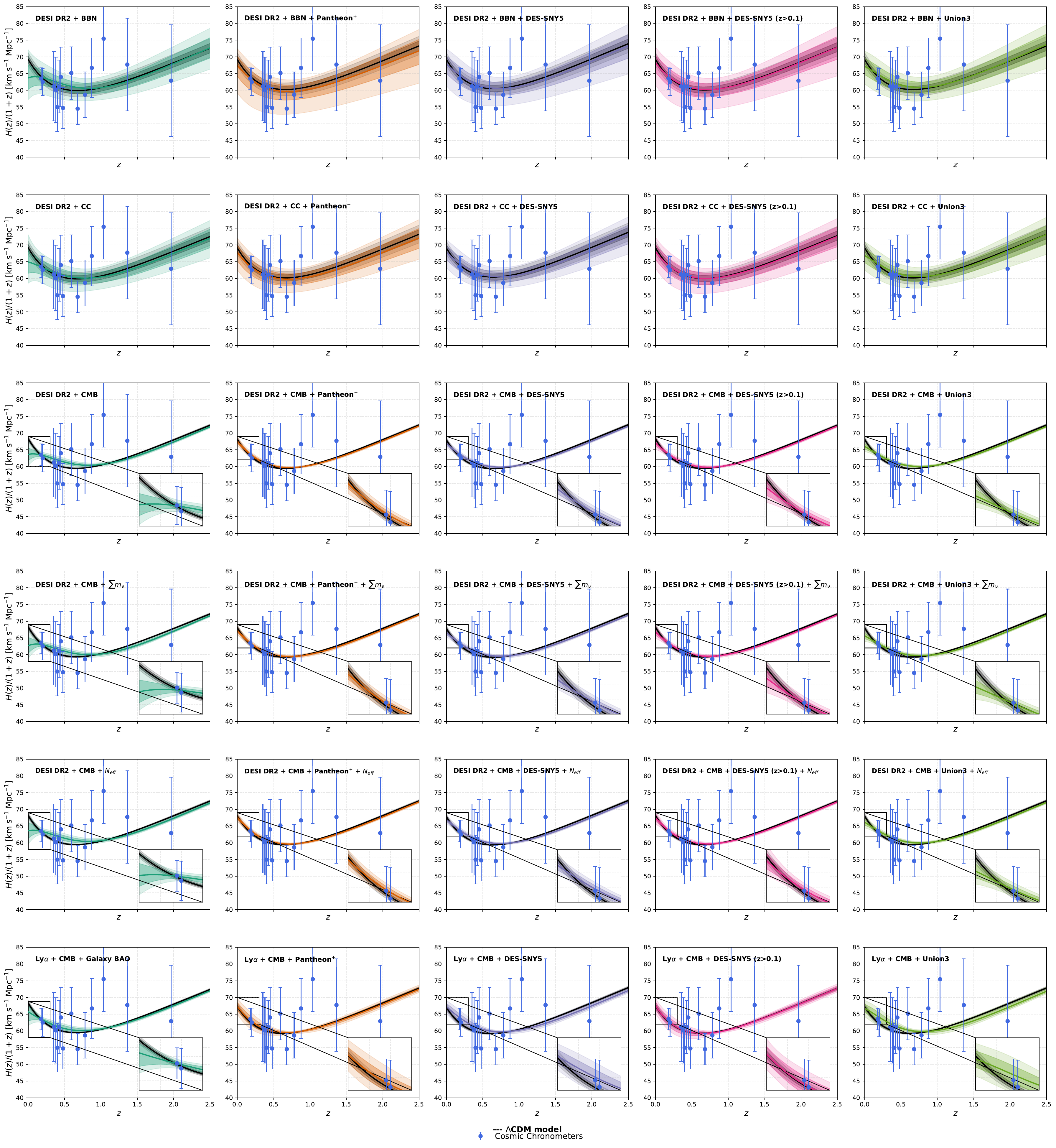}
\caption{This figure shows the evolution of $H(z)/(1+z)$ as a function of redshift $z$ for different cosmological dataset combinations, including DESI~DR2, DESI~DR2~Ly$\alpha$, CMB, BBN, and CC data, together with various Type~Ia supernova samples (Pantheon$^+$, DES-SN5Y, DES-SN5Y with $z > 0.1$, and Union3) The black dashed line represents the predicted o $\Lambda$CDM model, shown with its corresponding 1$\sigma$ and 2$\sigma$ confidence regions. The colored lines and shaded bands represent the predictions of the $\omega_0\omega_a$CDM model, also propagated with their respective 1$\sigma$ and 2$\sigma$ uncertainties for each dataset combination.}\label{fig_5}
\end{figure*}

Fig~\ref{fig_5} shows the evolution of $H(z)/(1+z)$ as a function of redshift for the set of dataset combinations considered in this work.  Each panel corresponds to a specific combination of DESI~DR2 data (BAO and Ly$\alpha$), CMB, BBN, CC, and different Type~Ia supernova samples (Pantheon$^+$, DES SN5Y, DES SN5Y with $z>0.1$, and Union3). The colored curves represent the mean expansion history, and the shaded regions denote the corresponding 68\% credible intervals. The blue points with error bars show the CC measurements of $H(z)/(1+z)$, allowing for direct visual comparison with the reconstructed expansion rate. The $\Lambda$CDM prediction is shown as a black solid line for reference. A zoom-in subplot is included in the lower-right corner to highlight the  behavior of both the  expansion history and the $\Lambda$CDM 
prediction at low redshifts. This inset makes it easier to see how each model shows different evolution and predicts a different present-day value of $H_0$ compared to the $\Lambda$CDM 
model at $z=0$.

\subsection{Dynamical dark energy after DESI DR2 ?}\label{sec_4b}
\begin{figure*}
\begin{subfigure}{.33\textwidth}
\includegraphics[width=\linewidth]{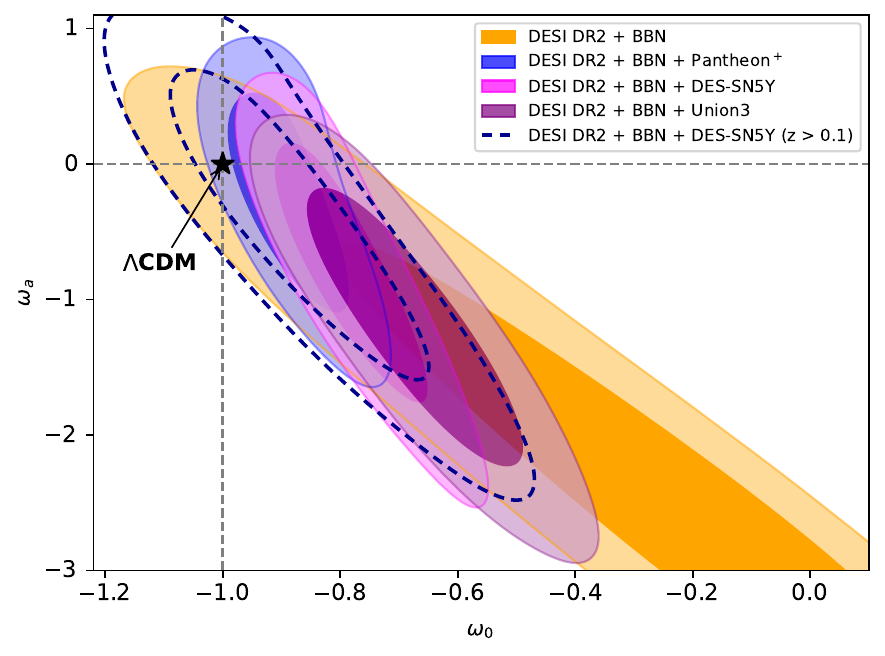}
\end{subfigure}
\hfil
\begin{subfigure}{.33\textwidth}
\includegraphics[width=\linewidth]{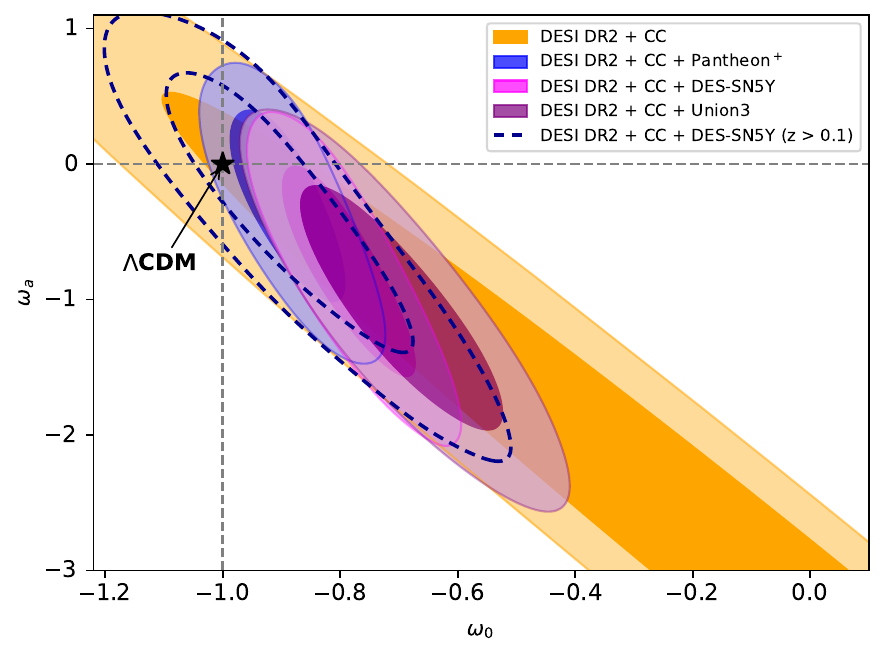}
\end{subfigure}
\hfil
\begin{subfigure}{.33\textwidth}
\includegraphics[width=\linewidth]{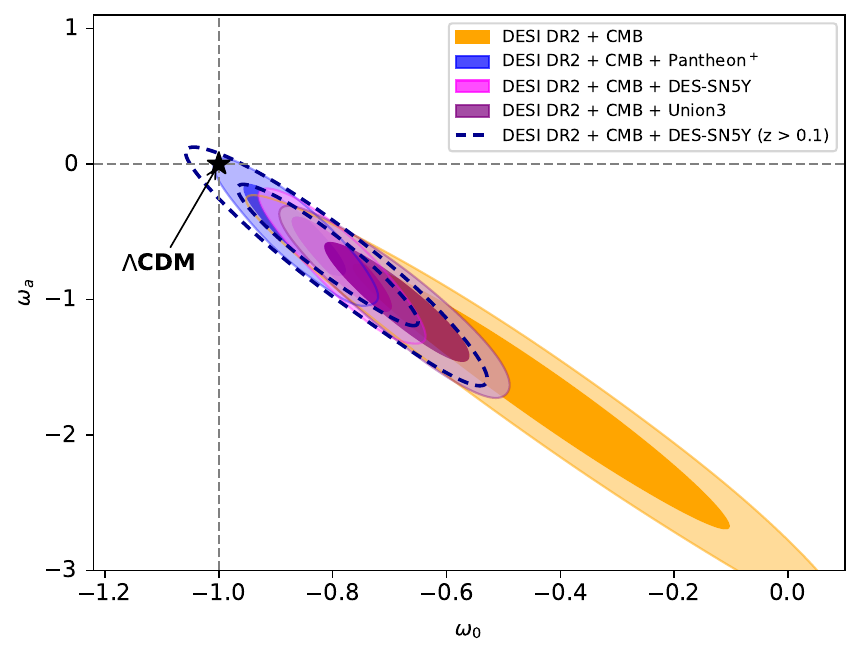}
\end{subfigure}
\begin{subfigure}{.33\textwidth}
\includegraphics[width=\linewidth]{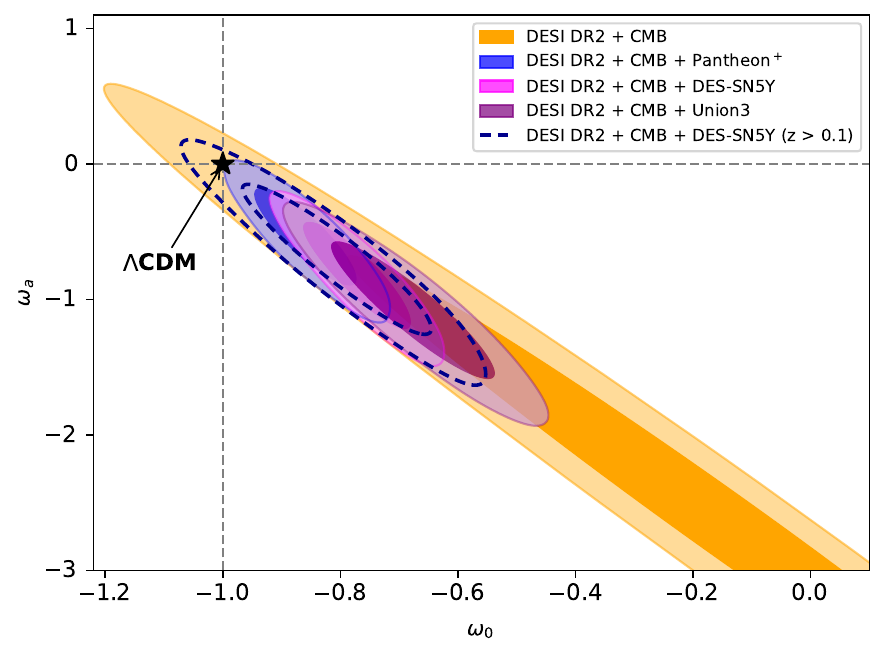}
\end{subfigure}
\hfil
\begin{subfigure}{.33\textwidth}
\includegraphics[width=\linewidth]{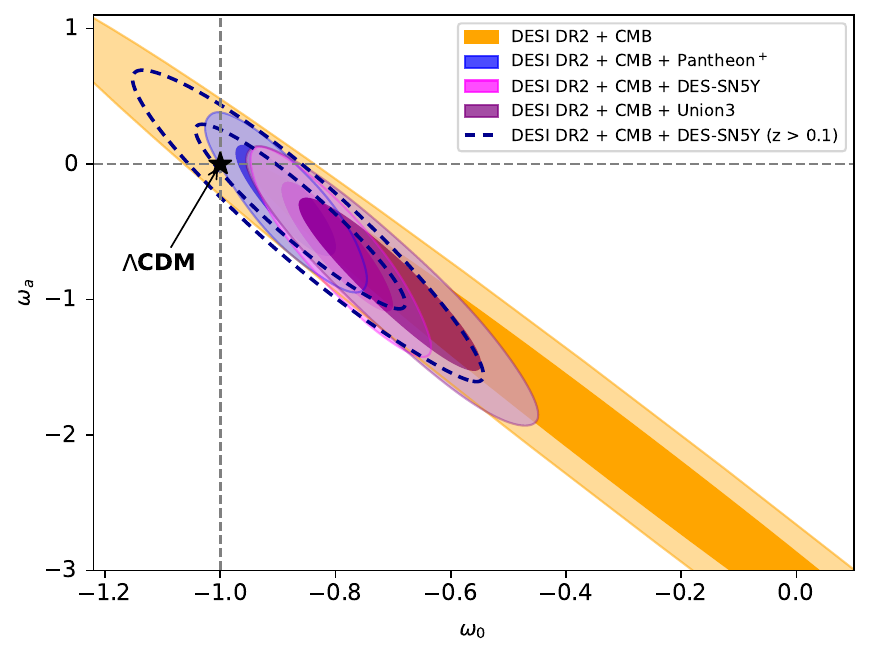}
\end{subfigure}
\hfil
\begin{subfigure}{.33\textwidth}
\includegraphics[width=\linewidth]{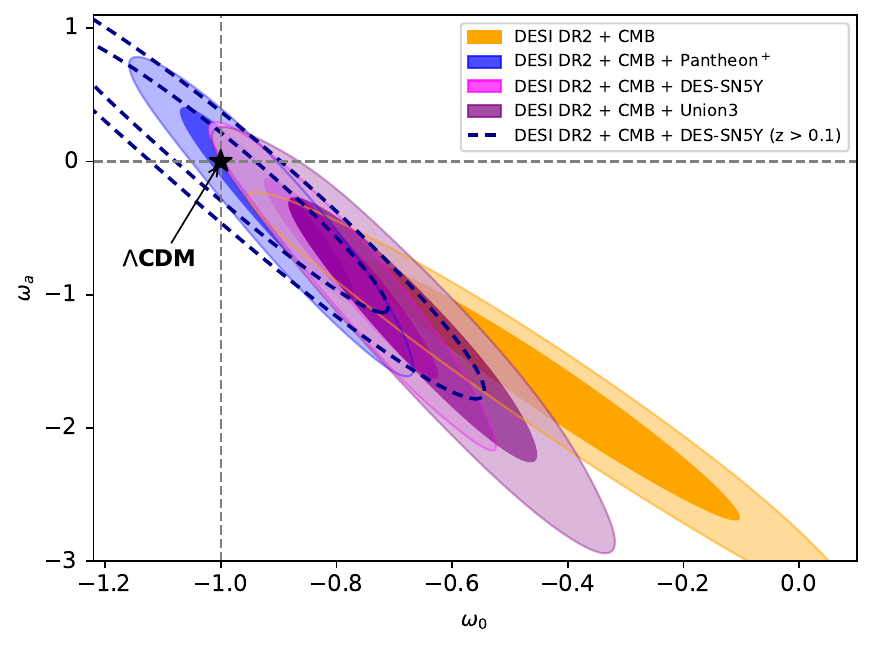}
\end{subfigure}
\caption{The figure shows the $\omega_0$–$\omega_a$ plane obtained from the $\omega_0\omega_a$CDM, $\omega_0\omega_a$CDM + $\sum m_{\nu}$, and $\omega_0\omega_a$CDM + $N_{\mathrm{eff}}$ models using different dataset combinations. The contours correspond to the 68\% (1$\sigma$) and 95\% (2$\sigma$) confidence regions derived from DESI~DR2, DESI~DR2~Ly$\alpha$, CMB, BBN, and CC data, together with various Type~Ia supernova samples (Pantheon$^+$, DES-SN5Y, DES-SN5Y with $z > 0.1$, and Union3). The black star marks the $\Lambda$CDM prediction ($\omega_0 = -1$, $\omega_a = 0$).}\label{fig_6}
\end{figure*}

We choose the $\omega_0\omega_a$CDM model as our main framework to study the nature of dark energy using the DESI DR2 measurements. An important feature of this parametrization is that it includes the standard $\Lambda$CDM model as a special case, corresponding to $\omega_0 = -1$ and $\omega_a = 0$. By mapping our results in the $(\omega_0, \omega_a)$ plane, one can directly compare DESI DR2 constraints with $\Lambda$CDM and investigate any possible preference for dynamical dark energy over the cosmological constant ($\Lambda$). In Fig.~\ref{fig_6}, we show the $(\omega_0, \omega_a)$ plane for the $\omega_0\omega_a$CDM, $\omega_0\omega_a$CDM + $\sum m_{\nu}\,[\mathrm{eV}]$, and $\omega_0\omega_a$CDM + $N_{\mathrm{eff}}$ model with different dataset combinations. The first row shows the results with BBN, CC, and CMB  (left to right), while the second row shows the $\omega_0\omega_a$CDM 
extensions with $\sum m_{\nu}$, $N_{\mathrm{eff}}$, and the DESI DR2 Ly$\alpha$ combination (left to right). In our analysis, we first consider the combination of DESI DR2 with BBN and different SNe~Ia datasets (Pantheon$^+$, DES-SN5Y, and Union3). Using the DESI DR2 data with BBN priors, we found
\begin{equation}
\begin{aligned}
\omega_0 &= -0.40^{+0.38}_{-0.23} \\
\omega_a &= -1.95^{+0.61}_{-1.3}
\end{aligned}
\qquad \Bigg\} \text{DESI DR2 + BBN}
\end{equation}
The results favor the $\omega_0 > -1$, $\omega_a < 0$ quadrant, showing a mild preference for dynamical dark energy up to $1.97\sigma$. It can be observed that the posteriors obtained using this combination are poorly constrained, as they go beyond the prior limits to accommodate $\omega_0 > -1$ (see \cite{wang2024self}). Furthermore, we combine the BBN and DESI DR2 data with different SNe~Ia samples. We find
\begin{equation}
\left.
\begin{aligned}
\omega_0 &= -0.888^{+0.058}_{-0.070} \\
\omega_a &= -0.26^{+0.58}_{-0.50}
\end{aligned}
\right\}
\begin{array}{l}
\text{DESI DR2+BBN+} \\
\text{Pantheon$^+$,}
\end{array}
\end{equation}
when combined with the Pantheon$^+$ sample

\begin{equation}
\left.
\begin{aligned}
\omega_0 &= -0.783^{+0.073}_{-0.092} \\
\omega_a &= -0.80 \pm 0.64
\end{aligned}
\right\}
\begin{array}{l}
\text{DESI DR2+BBN+} \\
\text{DES-SN5Y,}
\end{array}
\end{equation}
when the DES-SN5Y sample is included

\begin{equation}
\left.
\begin{aligned}
\omega_0 &= -0.87^{+0.12}_{-0.16} \\
\omega_a &= -0.44^{+0.84}_{-0.69}
\end{aligned}
\right\}
\begin{array}{l}
\text{DESI DR2+BBN+} \\
\text{DES-SN5Y ($z>0.1$),}
\end{array}
\end{equation}
when the DES-SN5Y sample is included

\begin{equation}
\left.
\begin{aligned}
\omega_0 &= -0.67 \pm 0.12 \\
\omega_a &= -1.22 \pm 0.68
\end{aligned}
\right\}
\begin{array}{l}
\text{DESI DR2+BBN+} \\
\text{Union3,}
\end{array}
\end{equation}
When the Union3 sample is included, these combinations show a preference for dynamical dark energy of up to $1.75\sigma$, $2.63\sigma$, $0.93\sigma$, and $2.75\sigma$ for the combinations with Pantheon$^+$, DES-Y5, and Union3, respectively.

Next, we extend our analysis by combining DESI DR2 data with CC measurements. The CC dataset provides direct estimates of the Hubble parameter $H(z)$ at different redshifts, and when combined with DESI DR2, it helps to constrain the late-time expansion history of the Universe. Using the DESI DR2 + CC combination, we find

\begin{equation}
\begin{aligned}
\omega_0 &= -0.61^{+0.37}_{-0.32} \\
\omega_a &= -1.2^{+1.0}_{-1.3}
\end{aligned}
\qquad \Bigg\} \text{DESI DR2 + CC,}
\end{equation}

which shows a preference for dynamical dark energy up to $1.13\sigma$, predicting values in the $\omega_0 > -1$, $\omega_a < 0$ quadrant. We then combine the DESI~DR2 and CC data with different Type~Ia supernova samples. For the Pantheon$^+$ compilation, the best-fit parameters are

\begin{equation}
\left.
\begin{aligned}
\omega_0 &= -0.889^{+0.059}_{-0.068} \\
\omega_a &= -0.31^{+0.51}_{-0.43}
\end{aligned}
\right\}
\begin{array}{l}
\text{DESI DR2+CC+} \\
\text{Pantheon$^+$,}
\end{array}
\end{equation}
showing a preference for dynamical dark energy up to $1.75\sigma$.

\begin{equation}
\left.
\begin{aligned}
\omega_0 &= -0.786^{+0.068}_{-0.081} \\
\omega_a &= -0.80^{+0.55}_{-0.48}
\end{aligned}
\right\}
\begin{array}{l}
\text{DESI DR2+CC+} \\
\text{DES-SN5Y,}
\end{array}
\end{equation}
when the DES-SN5Y sample is included, the  preference for dynamical dark energy increases to $2.87\sigma$.

\begin{equation}
\left.
\begin{aligned}
\omega_0 &= -0.89^{+0.12}_{-0.15} \\
\omega_a &= -0.36^{+0.78}_{-0.59}
\end{aligned}
\right\}
\begin{array}{l}
\text{DESI DR2+CC+} \\
\text{DES-SN5Y ($z>0.1$),}
\end{array}
\end{equation}
when the low-$z$ sample is excluded from the DES-SN5Y sample, the preference for dynamical dark energy decreases to $0.81\sigma$.

\begin{equation}
\left.
\begin{aligned}
\omega_0 &= -0.70 \pm 0.11 \\
\omega_a &= -1.07 \pm 0.58
\end{aligned}
\right\}
\begin{array}{l}
\text{DESI DR2+CC+} \\
\text{Union3,}
\end{array}
\end{equation}
when the Union3 sample is included, it shows a preference for dynamical dark energy up to $2.73\sigma$. These combinations consistently favor the $\omega_0 > -1$, $\omega_a < 0$ quadrant, suggesting a tendency toward dynamical dark energy when SNe~Ia and CC data are jointly considered with DESI~DR2. Although the preference for dynamical dark energy remains below the $3\sigma$ level.

We next combine DESI DR2 with the CamSpec CMB compressed likelihood. The inclusion of this information provides a robust early-Universe anchor and tightens the overall parameter constraints compared to the BBN and CC combinations. For the DESI DR2 + CMB case, we obtain

\begin{equation}\label{desi_cmb}
\begin{aligned}
\omega_0 &= -0.43 \pm 0.21 \\
\omega_a &= -1.72^{+0.67}_{-0.61}
\end{aligned}
\qquad \Bigg\} \text{DESI DR2 + CMB,}
\end{equation}

showing a preference for dynamical dark energy at about $2.71\sigma$. This combination predicts values in the $\omega_0 > -1$, $\omega_a < 0$ quadrant. When additional SNe~Ia datasets are included, the results become more tightly constrained. For the Pantheon$^+$ compilation, we find

\begin{equation}
\left.
\begin{aligned}
\omega_0 &= -0.865 \pm 0.059 \\
\omega_a &= -0.49 \pm 0.22
\end{aligned}
\right\}
\begin{array}{l}
\text{DESI DR2+CMB+} \\
\text{Pantheon$^+$,}
\end{array}
\end{equation}
showing a preference for dynamical dark energy at about $2.29\sigma$.

\begin{equation}
\left.
\begin{aligned}
\omega_0 &= -0.782 \pm 0.058 \\
\omega_a &= -0.75 \pm 0.23
\end{aligned}
\right\}
\begin{array}{l}
\text{DESI DR2+CMB+} \\
\text{DES-SN5Y,}
\end{array}
\end{equation}
When the DES-SN5Y sample is included, the preference for dynamical dark energy increases up to $3.76\sigma$.

\begin{equation}
\left.
\begin{aligned}
\omega_0 &= -0.803_{-0.11}^{+0.092} \\
\omega_a &= -0.68_{-0.29}^{+0.39}
\end{aligned}
\right\}
\begin{array}{l}
\text{DESI DR2+CMB+} \\
\text{DES-SN5Y ($z>0.1$),}
\end{array}
\end{equation}
when the low-$z$ sample is excluded from the DES-SN5Y sample, the preference for dynamical dark energy decreases to $1.95\sigma$.

\begin{equation}
\left.
\begin{aligned}
\omega_0 &= -0.685 \pm 0.083 \\
\omega_a &= -1.02 \pm 0.29
\end{aligned}
\right\}
\begin{array}{l}
\text{DESI DR2+CMB+} \\
\text{Union3,}
\end{array}
\end{equation}

When the Union3 sample is included, it shows a preference for dynamical dark energy up to $3.80\sigma$. The inclusion of CMB data significantly improves the constraints on $\omega_0$ and $\omega_a$, yielding consistent results across different SNe~Ia compilations. The combined DESI~DR2, CMB, and SNe~Ia datasets show the strongest preference for dynamical dark energy over the cosmological constant ($\Lambda$), although the statistical significance remains below the $4\sigma$ level.

We also consider the combination of Ly$\alpha$ forest measurements from DESI DR2 with galaxy BAO and CMB data. The Ly-$\alpha$ forest is a key probe of the high-redshift Universe, tracing large-scale structure through hydrogen absorption in quasar spectra. It provides precise constraints on the cosmic expansion rate at $2 < z < 4$ \cite{adame2025desi,cuceu2023constraints} and extends BAO measurements beyond the reach of galaxy surveys \cite{delubac2013baryon,slosar2013measurement,des2020completed}. With recent improvements from DESI, the Ly$\alpha$ forest offers an independent and robust test of the $\Lambda$CDM model and the possible dynamics of dark energy. Since the combination of the Ly$\alpha$ + galaxy BAO corresponds to the same dataset as the DESI DR2 + CMB case, the resulting constraints are identical, showing a preference for dynamical dark energy at about $2.71\sigma$. When different SNe~Ia samples are added to the Ly$\alpha$ and CMB combination, the results become more tightly constrained. For the Pantheon$^+$ sample, we find

\begin{equation}
\left.
\begin{aligned}
\omega_0 &= -0.913 \pm 0.090 \\
\omega_a &= -0.34^{+0.52}_{-0.44}
\end{aligned}
\right\}
\begin{array}{l}
\text{Ly$\alpha$+CMB+} \\
\text{Pantheon$^+$,}
\end{array}
\end{equation}
showing a preference for dynamical dark energy up to $0.97\sigma$.

\begin{equation}
\left.
\begin{aligned}
\omega_0 &= -0.77 \pm 0.10 \\
\omega_a &= -0.88 \pm 0.50
\end{aligned}
\right\}
\begin{array}{l}
\text{Ly$\alpha$+CMB+} \\
\text{DES-SN5Y,}
\end{array}
\end{equation}
when the DES-SN5Y sample is included, the preference for dynamical dark energy increases to $2.30\sigma$.

\begin{equation}
\left.
\begin{aligned}
\omega_0 &= -0.99 \pm 0.18 \\
\omega_a &= -0.08 \pm 0.75
\end{aligned}
\right\}
\begin{array}{l}
\text{Ly$\alpha$+CMB+} \\
\text{DES-SN5Y ($z>0.1$),}
\end{array}
\end{equation}
when the low-$z$ sample is excluded from the DES-SN5Y sample, the preference for dynamical dark energy decreases to $0.06\sigma$.

\begin{equation}
\left.
\begin{aligned}
\omega_0 &= -0.66 \pm 0.14 \\
\omega_a &= -1.27 \pm 0.65
\end{aligned}
\right\}
\begin{array}{l}
\text{Ly$\alpha$+CMB+} \\
\text{Union3,}
\end{array}
\end{equation}
when the Union3 sample is included, it shows a preference for dynamical dark energy up to $2.43\sigma$. The Ly$\alpha$ combination with different datasets also favors values in the $\omega_0 > -1$, $\omega_a < 0$ quadrant with moderate significance. Although the preference for dynamical dark energy remains below the $3\sigma$ level, the results still indicate a mild deviation from the cosmological constant scenario. Next, we discuss the impact of the DESI DR2 measurements on the $\omega_0\omega_a$CDM + $\sum m_{\nu}\,[\mathrm{eV}]$ and $\omega_0\omega_a$CDM + $N_{\mathrm{eff}}$ models.

\begin{figure*}
\begin{subfigure}{.48\textwidth}
\includegraphics[width=\linewidth]{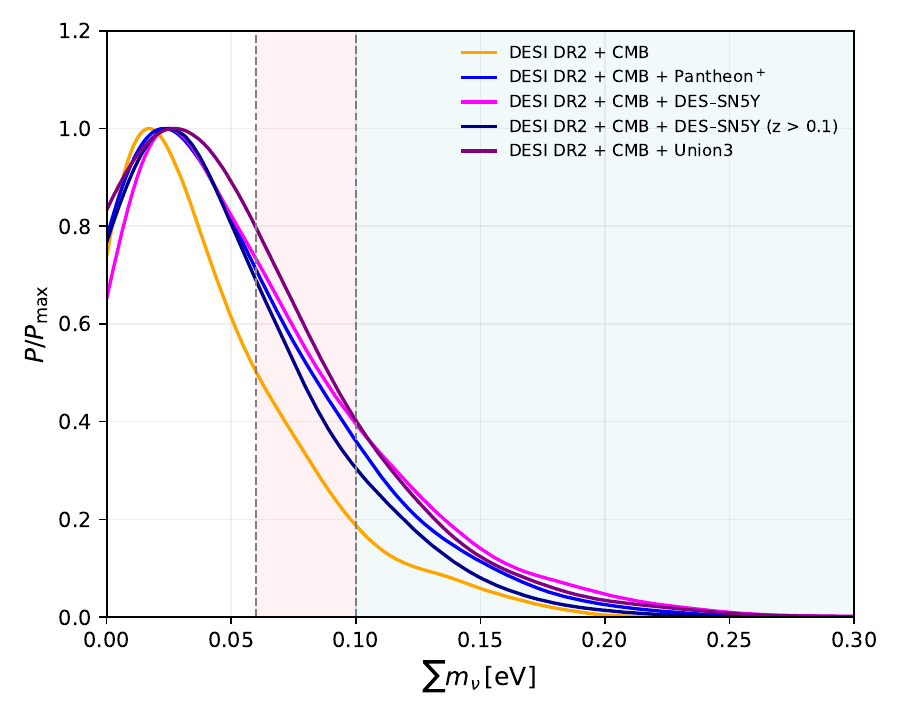}%
\end{subfigure}
\hfil
\begin{subfigure}{.48\textwidth}
\includegraphics[width=\linewidth]{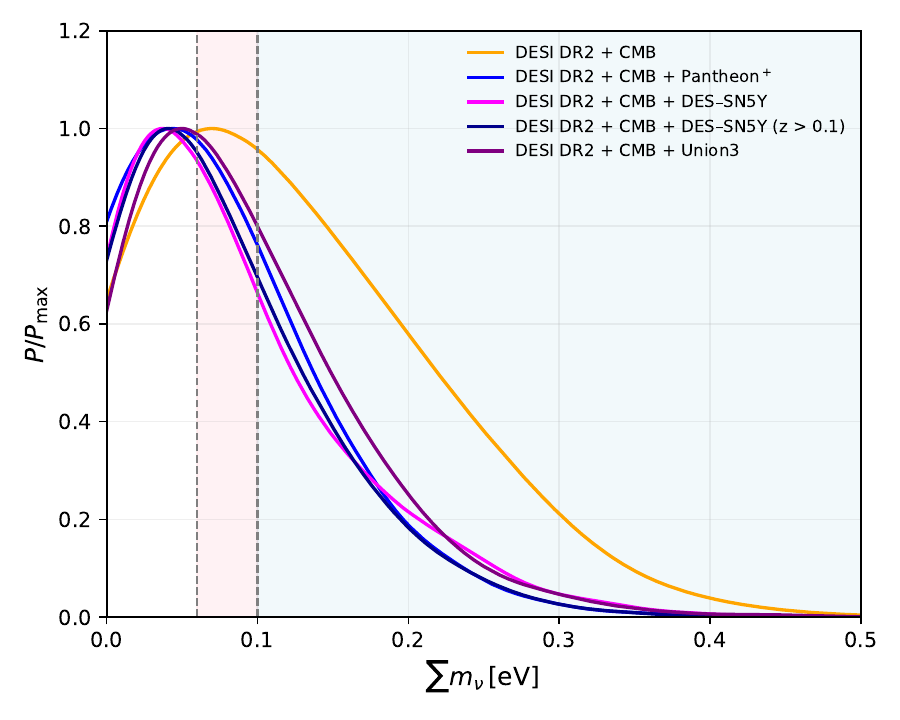}
\end{subfigure}
\caption{The figure shows the 1D marginalized posterior constraints on $\sum m_{\nu}\,[\mathrm{eV}]$ for the $\Lambda$CDM~+~$\sum m_{\nu}$ (left panel) and $\omega_0\omega_a$CDM~+~$\sum m_{\nu}$ (right panel) models, using DESI~DR2 and CMB data combined with various Type~Ia supernova samples (Pantheon$^+$, DES-SN5Y, and Union3).}\label{fig_7}
\end{figure*}

Before moving further, Fig.~\ref{fig_7} shows the normalized probability density of the total neutrino mass, \(\sum m_{\nu}\), in the \(\Lambda\)CDM + \(\sum m_{\nu}\) model (left panel) and the \(\omega_0\omega_a\)CDM + \(\sum m_{\nu}\) model (right panel) using DESI DR2 + CMB with different SNe~Ia samples. For the normal hierarchy (NH), the total neutrino mass must satisfy \(\sum m_\nu \ge 0.059 \, \mathrm{eV}\), while for the inverted hierarchy (IH) one requires \(\sum m_\nu \ge 0.10 \, \mathrm{eV}\). In the $\Lambda$CDM + $\sum m_{\nu},[\mathrm{eV}]$ model, we determine upper limits on the total neutrino mass at the 68\% confidence level using various dataset combinations with DESI DR2. For the DESI DR2 + CMB combination, we obtain $\sum m_{\nu} < 0.056,\mathrm{eV}$ (68\% C.L).

Including additional SNe~Ia samples slightly relaxes the constraints: $\sum m_{\nu} < 0.070,\mathrm{eV}$ for DESI DR2 + CMB + Pantheon$^+$, $\sum m_{\nu} < 0.076,\mathrm{eV}$ for DESI DR2 + CMB + DES-SN5Y, $\sum m_{\nu} < 0.065,\mathrm{eV}$ for DESI DR2 + CMB + DES-SN5Y ($z > 0.1$), and $\sum m_{\nu} < 0.073,\mathrm{eV}$ for DESI DR2 + CMB + Union3. There is no significant $2\sigma$+ detection of a non-zero neutrino mass; however, the posterior distributions still peak in the $\sum m_{\nu} > 0$ region. In addition to the $68\%$ confidence level upper limits, we also report the $95\%$ confidence level upper limits for the different dataset combinations. We find $\sum m_{\nu} < 0.125,\mathrm{eV}$ for DESI DR2 + CMB, $\sum m_{\nu} < 0.144,\mathrm{eV}$ for DESI DR2 + CMB + Pantheon$^+$, $\sum m_{\nu} < 0.158,\mathrm{eV}$ for DESI DR2 + CMB + DES-SN5Y, $\sum m_{\nu} < 0.132,\mathrm{eV}$ for DESI DR2 + CMB + DES-SN5Y ($z>0.1$), and $\sum m_{\nu} < 0.149,\mathrm{eV}$ for DESI DR2 + CMB + Union3. We find detection non zero neutrino masses about 1.6 $\sigma$+ at 68\% confidence limit. If the prior were extended to unphysical (negative) values, the posteriors would take the form of truncated distributions that would otherwise peak at negative neutrino masses similar behavior has been reported in \cite{green2025cosmological,elbers2025negative,noriega2024unveiling,naredo2024critical,elbers2025constraints}.
\begin{figure}
\centering
\includegraphics[scale=0.58]{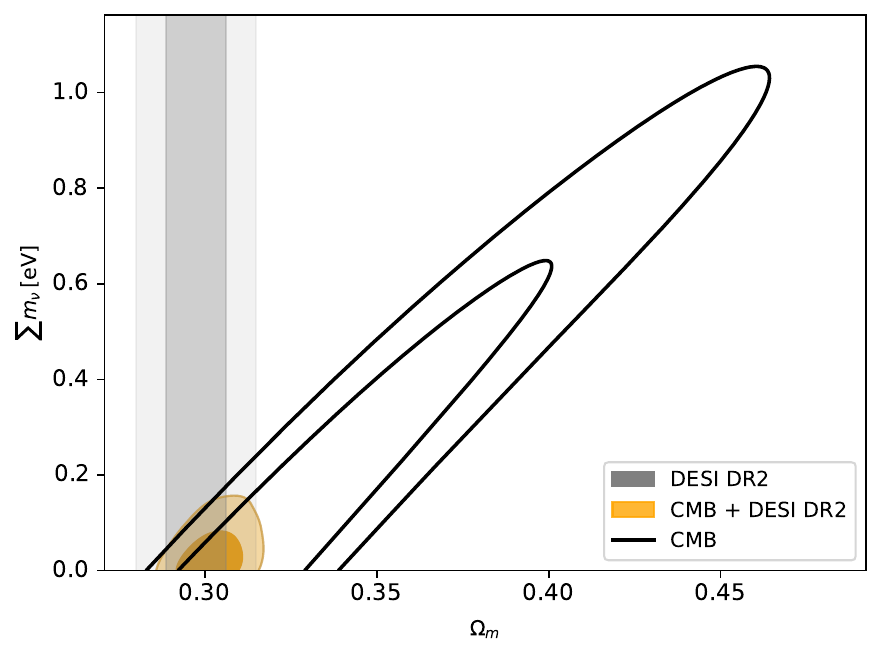}
\caption{This figure shows the 68\% and 95\% confidence contours for the $\Lambda$CDM~+~$\sum m_{\nu}$ model. These constraints assume $\sum m_{\nu} > 0$~eV and are obtained using DESI~DR2 alone, CMB alone, and DESI~DR2 combined with CMB data. The gray band represents the constraint from DESI~DR2 alone, showing that the results are largely insensitive to the neutrino mass parameter when external data are not included. It also shows the negative correlation between $\sum m_{\nu}$ and $\Omega_m$. The combination of DESI~DR2 with CMB data shows the tightest constraints, indicating that DESI~DR2 prefers lower $\Omega_m$ values.}\label{fig_8}
\end{figure}

Fig~\ref{fig_8} shows the $\Omega_m - \sum m_\nu$ parameter plane for the $\Lambda$CDM model, comparing the results from the DESI DR2 and CMB combination with those from the CMB data alone. It can be seen that when only the CMB dataset is considered, $\sum m_\nu$ and $\Omega_m$ show a positive correlation. In contrast, the inclusion of DESI DR2 data significantly tightens the constraint on $\Omega_m$, as shown by the narrow vertical contour. The preferred $\Omega_m$ value lies near the lower end of the CMB only posterior distribution, which in turn restricts the allowed range of $\sum m_\nu$ to very small values. This behavior reflects DESI’s tendency to favor slightly lower matter densities compared to the CMB results alone.

We further extend our analysis by allowing the total neutrino mass, $\sum m_{\nu}$, to vary within the $\omega_0\omega_a$CDM model. This enables us to examine the impact of varying neutrino mass on the preference for dynamical dark energy and the predicted values in the $\omega_0$–$\omega_a$ plane as well. In our analysis, when combining the CMB with DESI DR2 and assuming a $\sum m_{\nu} > 0$ prior, we find

\begin{equation}
\left.
\begin{aligned}
\omega_0 &= -0.36 \pm 0.27 \\
\omega_a &= -1.98^{+0.80}_{-0.91} \\
\sum m_{\nu} &< 0.136~\mathrm{eV}\ (68\%\ \mathrm{C.L.})
\end{aligned}
\right\}
\begin{array}{l}
\text{DESI DR2+CMB,}
\end{array}
\end{equation}
showing a preference for dynamical dark energy at $2.37\sigma$ from the $\Lambda$CDM point, and predicting values in the $\omega_0 > -1$ and $\omega_a < 0$ quadrant. When SNe~Ia data are included, the constraints tighten further. For the Pantheon$^+$ sample, we find

\begin{equation}
\left.
\begin{aligned}
\omega_0 &= -0.857 \pm 0.057 \\
\omega_a &= -0.55^{+0.25}_{-0.22} \\
\sum m_{\nu} &< 0.110~\mathrm{eV}\ (68\%\ \mathrm{C.L.})
\end{aligned}
\right\}
\begin{array}{l}
\text{DESI DR2+CMB+} \\
\text{Pantheon$^+$,}
\end{array}
\end{equation}
showing a preference for dynamical dark energy up to $2.51\sigma$.

\begin{equation}
\left.
\begin{aligned}
\omega_0 &= -0.769 \pm 0.060 \\
\omega_a &= -0.83^{+0.27}_{-0.25} \\
\sum m_{\nu} &< 0.113~\mathrm{eV}\ (68\%\ \mathrm{C.L.})
\end{aligned}
\right\}
\begin{array}{l}
\text{DESI DR2+CMB+} \\
\text{DES-SN5Y,}
\end{array}
\end{equation}
when the DES-SN5Y sample is included, the  preference for dynamical dark energy increases to $3.85\sigma$.

\begin{equation}
\left.
\begin{aligned}
\omega_0 &= -0.790 \pm 0.11 \\
\omega_a &= -0.740{\pm 0.37} \\
\sum m_{\nu} &< 0.109~\mathrm{eV}\ (68\%\ \mathrm{C.L.})
\end{aligned}
\right\}
\begin{array}{l}
\text{DESI DR2+CMB+} \\
\text{DES-SN5Y ($z>0.1$),}
\end{array}
\end{equation}
when the low-$z$ sample is excluded from the DES-SN5Y sample, the preference for dynamical dark energy decreases to $1.91\sigma$.

\begin{equation}
\left.
\begin{aligned}
\omega_0 &= -0.674 \pm 0.091 \\
\omega_a &= -1.09 \pm 0.33 \\
\sum m_{\nu} &< 0.125~\mathrm{eV}\ (68\%\ \mathrm{C.L.})
\end{aligned}
\right\}
\begin{array}{l}
\text{DESI DR2+CMB+} \\
\text{Union3,}
\end{array}
\tag{39}
\end{equation}
When the Union3 sample is included, the preference for dynamical dark energy increases to \(3.58\sigma\). These results indicate that allowing \(\sum m_{\nu}\) to vary also leads to a deviation from the cosmological constant. It is also worth noting that, for the \(\omega_0\omega_a\)CDM model, there is no \(2\sigma\) detection of non-zero neutrino masses, similar to the case of the \(\Lambda\)CDM model. As can be seen in Fig.~\ref{fig_7} (right panel), the posterior distributions peak in the \(\sum m_{\nu} > 0\) region, consistent with the 68\% confidence level. At the 2\(\sigma\) level, the marginalized values of the total neutrino mass are 
\(\sum m_{\nu} < 0.298\,\mathrm{eV}\), \(< 0.209\,\mathrm{eV}\), \(< 0.227\,\mathrm{eV}\), \(< 0.231\,\mathrm{eV}\), and \(< 0.247\,\mathrm{eV}\), for the DESI DR2 + CMB, DESI DR2 + CMB + Pantheon$^+$, DESI DR2 + CMB + DES–SN5Y, DESI DR2 + CMB + DES–SN5Y ($z>0.1$), and DESI DR2 + CMB + Union3 combinations, respectively.  These correspond to detection of non-zero neutrino masses about \(1.58\sigma\), \(1.64\sigma\), \(1.64\sigma\), \(1.62\sigma\), and \(1.58\sigma\). These results show evidence of 1.5$\sigma$+ nonzero Neutrino Masses.

As reported in \cite{choudhury2025cosmology} (see Fig.~2), when combining Planck PR4, lensing, and DESI DR2 data without Weak Lensing (WL), there is only about \(1\sigma\) indication of non zero neutrino masses. After including the WL data, the \(\sum m_{\nu}\) posterior shows a clear peak, corresponding to a \(2.1\sigma\) detection of non zero neutrino masses for Planck PR4+lensing+DESI2+DESY5+WL at \(\sum m_{\nu} = 0.19^{+0.15}_{-0.18}\,\mathrm{eV}\), and a \(1.9\sigma\) detection for Planck PR4+lensing+DESI2+Pantheon$^{+}$+WL. In \cite{choudhury2024updated} (see Fig.~3), Planck PR4+lensing shows a detection of non-zero neutrino masses at about \(1.56\sigma\), while Planck PR4+lensing+DESI DR1 yields about \(1.68\sigma\), and Planck PR4+lensing+DESI DR1+DESY5 gives around \(1.55\sigma\). Thus, the DESI DR1 data show about \(1.5\sigma\)+ detection of non zero neutrino masses. In \cite{du2025cosmological}, the authors report a preference for a nonzero neutrino mass at the $2.7\sigma$ level when CMB data are combined with DESI~DR2, DES~Y5, and DES~Y1 datasets.
\begin{figure*}
\centering
\includegraphics[scale=0.85]{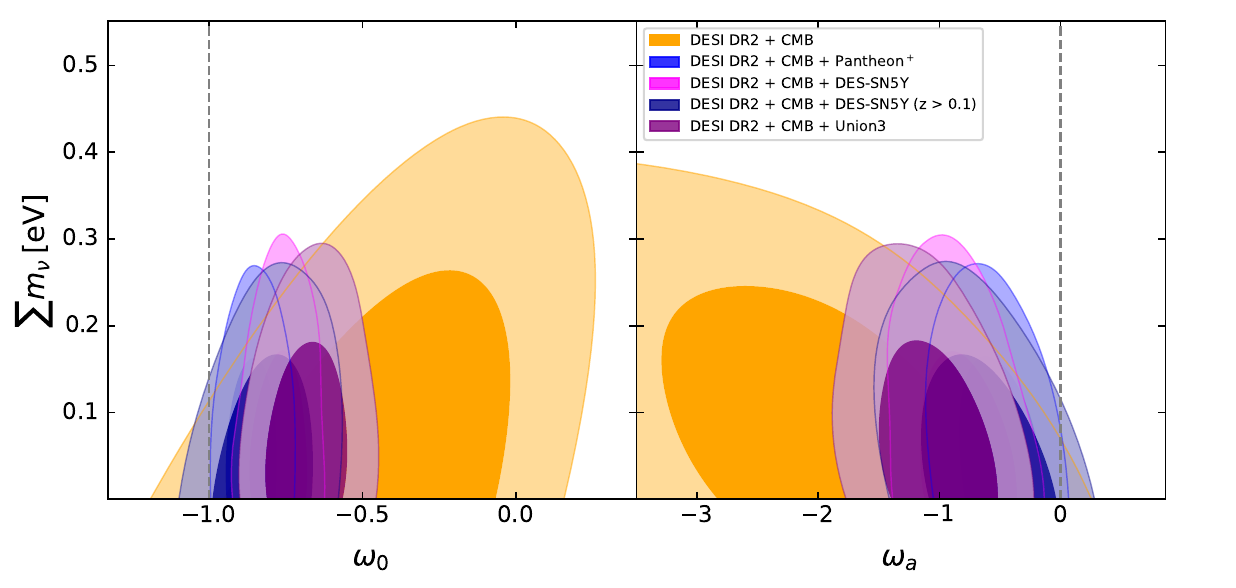}
\caption{The figure shows the ($\omega_0$–$\omega_a$)–$\sum m_{\nu}$~(eV) plane for the parameters of the $\omega_0\omega_a$CDM~+~$\sum m_{\nu}$~(eV) model, using DESI~DR2 and CMB data combined with various Type~Ia supernova samples (Pantheon$^+$, DES-SN5Y, and Union3). These constraints assume $\sum m_{\nu} > 0$~eV.}
\label{fig_9}
\end{figure*}

Fig.~\ref{fig_9} shows the posterior distributions of the parameters ($\omega_0$, $\omega_a$, and $\sum m_{\nu}$). It can be observed that adding DESI and SNe data to the CMB significantly tightens the constraints on ($\omega_0$, $\omega_a$)-$\sum m_{\nu}$. This improvement occurs because geometric measurements help break the parameter degeneracies present in the CMB-only analysis, thereby reducing the allowed range of $\sum m_{\nu}$. Finally, we extend our analysis by allowing the effective number of relativistic species to vary within the $\omega_0\omega_a$CDM model. This also allows us to examine the effect of DESI measurements on the standard value of $N_{\mathrm{eff}} = 3.046$ \cite{froustey2020neutrino,bennett2021towards}. For the DESI DR2 + CMB combination, we find

\begin{equation}
\begin{aligned}
\omega_0 &= -0.53_{-0.27}^{+0.34} \\
\omega_a &= -1.38_{-1.1}^{-0.88} \\
N_{\mathrm{eff}} &= 3.15^{+0.30}_{-0.41}
\end{aligned}
\qquad \Bigg\} \text{DESI DR2 + CMB,}
\end{equation}

showing preference for dynamical dark energy upto $1.5\sigma$ from the $\Lambda$CDM point. The inferred value of $N_{\mathrm{eff}}$ is consistent with the Standard Model expectation, indicating no significant evidence for extra relativistic degrees of freedom. When additional SNe~Ia data are included, we find

\begin{equation}
\left.
\begin{aligned}
\omega_0 &= -0.885 \pm 0.057 \\
\omega_a &= -0.27 \pm 0.27 \\
N_{\mathrm{eff}} &= 3.33^{+0.26}_{-0.30}
\end{aligned}
\right\}
\begin{array}{l}
\text{DESI DR2+CMB+} \\
\text{Pantheon$^+$,}
\end{array}
\end{equation}
show a preference for dynamical dark energy up to $2.02\sigma$.

\begin{equation}
\left.
\begin{aligned}
\omega_0 &= -0.796 \pm 0.064 \\
\omega_a &= -0.62^{+0.28}_{-0.33} \\
N_{\mathrm{eff}} &= 3.21 \pm 0.26
\end{aligned}
\right\}
\begin{array}{l}
\text{DESI DR2+CMB+} \\
\text{DES-SN5Y,}
\end{array}
\end{equation}
when the DES-SN5Y sample is included, it shows a preference for dynamical dark energy up to $3.11\sigma$.

\begin{equation}
\left.
\begin{aligned}
\omega_0 &= -0.860 \pm 0.11 \\
\omega_a &= -0.40^{+0.45}_{-0.40} \\
N_{\mathrm{eff}} &= 3.26_{-0.28}^{+0.24}
\end{aligned}
\right\}
\begin{array}{l}
\text{DESI DR2+CMB+} \\
\text{DES-SN5Y ($z>0.1$),}
\end{array}
\end{equation}
when the low-$z$ sample is excluded from the DES-SN5Y sample, the preference for dynamical dark energy decreases to $1.27\sigma$.

\begin{equation}
\left.
\begin{aligned}
\omega_0 &= -0.700 \pm 0.010 \\
\omega_a &= -0.89 {\pm 0.42}\\
N_{\mathrm{eff}} &= 3.18^{+0.24}_{-0.28}
\end{aligned}
\right\}
\begin{array}{l}
\text{DESI DR2+CMB+} \\
\text{Union3,}
\end{array}
\end{equation}
when the Union3 sample is included, it shows a preference for dynamical dark energy up to $3.33\sigma$. Overall, allowing $N_{\mathrm{eff}}$ to vary as a free parameter does not significantly alter the preference for a dynamical form of dark energy. Indeed, all combinations favor values in the $\omega_0 > -1$ and $\omega_a < 0$ quadrant. These results consistently show that dynamical dark energy models are preferred over the $\Lambda$CDM model when DESI~DR2 measurements are combined with other datasets. An important observation is that, in most data combinations such as DESI DR2 + BBN and DESI DR2 + CC the parameter $\omega_a$ tends to be pushed toward large negative values, often exceeding the prior limits, in order to accommodate values of $\omega_0 > -1$ (see also \cite{cortes2024interpreting,wang2024self}).

\begin{figure*}
\centering
\includegraphics[scale=0.55]{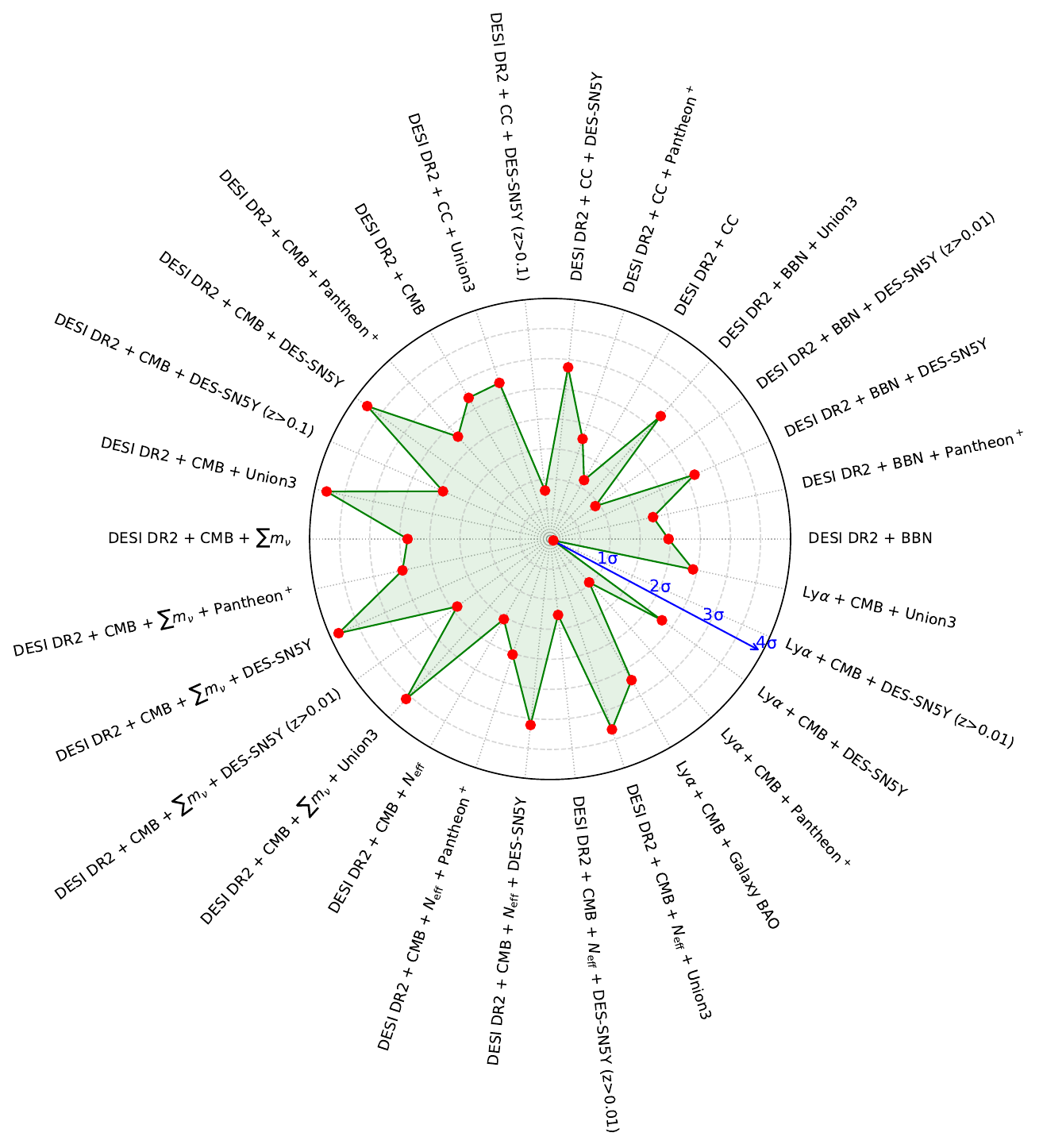}
\caption{This figure shows a radar plot showing the preference for dynamical dark energy in terms of $\sigma$. Each radial distance point, ending at a red dot, represents the significance level (in $\sigma$) of the preference for dynamical dark energy for a given dataset combination. Each radial line corresponds to a specific combination of datasets, as labeled around the outer edge of the plot.}\label{fig_10}
\end{figure*}
Fig.~\ref{fig_10} shows a radar plot summarizing the preference for dynamical dark energy across various dataset combinations. Each radial axis represents a different observational combination, including DESI DR2, CMB, BAO, BBN, and several supernova samples such as Pantheon$^+$, DES-SN5Y, and Union3. The radial distance from the center corresponds to the preference level, expressed in units of $\sigma$, with reference circles at 1$\sigma$, 2$\sigma$, 3$\sigma$, and 4$\sigma$  highlighted in blue for comparison.

As shown, most datasets show a preference for dynamical dark energy below the $3\sigma$ level, indicating that the cosmological constant ($\Lambda$) continues to provide a statistically consistent fit to the current cosmological data. However, combinations involving DESI DR2 and SNe Ia (particularly DES-SN5Y and Union3) show slightly higher significance levels, reaching up to $\sim 3.8\sigma$. These results show that the inclusion of DESI~DR2 with these SNe~Ia samples exhibits a notable preference for dynamical dark energy, indicating a departure from the cosmological constant ($\Lambda$).

These results do not discard the $\Lambda$CDM model; however, it is important to note that cosmology relies on observation based inference, where it is not possible to repeat measurements under identical conditions to achieve the same degree of precision as in laboratory experiments. For this reason, evidence at the level of 2–4$\sigma$ is generally regarded as significant in cosmology. This is precisely why several well known anomalies in the field such as the Hubble tension, the $S_8$ tension, the $M_B$ calibration tension, and the CMB lensing anomaly are consistently discussed as “tensions,” since they typically appear at the 2–4$\sigma$ significance level. Our findings fall within this range, suggesting that the evidence for dynamical dark energy is intriguing but not yet sufficient to rule out the $\Lambda$CDM model.


\begin{figure}
\centering
\includegraphics[scale=0.48]{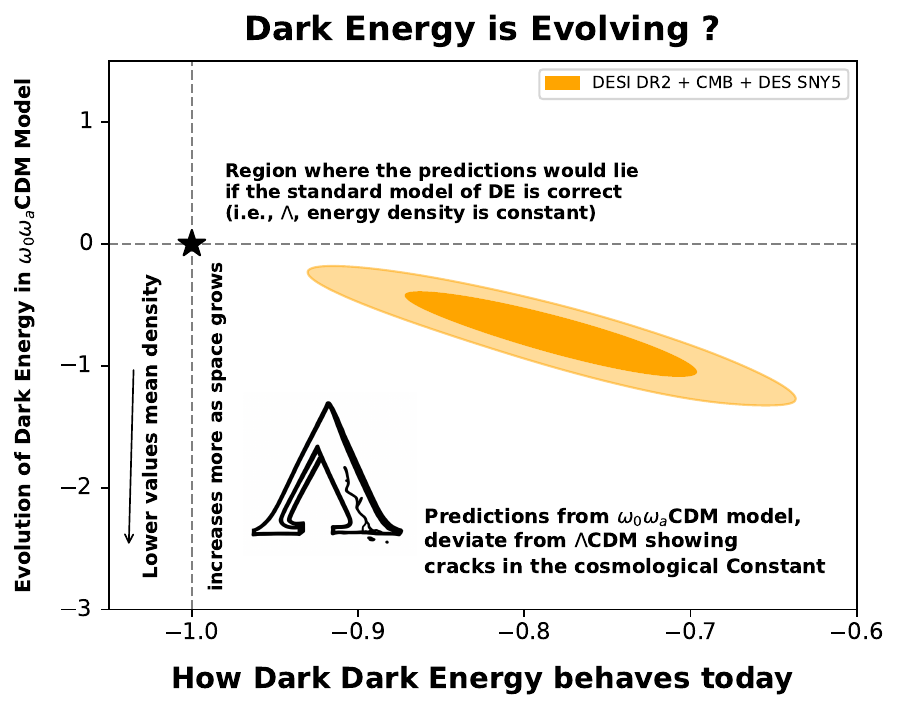}
\caption{The figure shows the $\omega_0$–$\omega_a$ quadrant of the $\omega_0\omega_a$CDM model. The contours represent the 68\% and 95\% confidence regions from the combined \textsc{DESI}~DR2 + CMB + DES-SN5Y datasets, which lie in the $\omega_0 > -1$ and $\omega_a < 0$ quadrant. The black star marks the $\Lambda$CDM prediction ($\omega_0 = -1$, $\omega_a = 0$).}\label{fig_11}
\end{figure}

Fig~\ref{fig_11} shows a summary of our results in the $(\omega_0, \omega_a)$ quadrant of the $\omega_0\omega_a$CDM model. Among all dataset combinations, the DES-SN5Y sample shows the most significant deviation when combined with CMB and DESI DR2 data. The black star marks the $\Lambda$CDM point ($\omega_0 = -1$, $\omega_a = 0$), corresponding to a constant dark energy. This can be observed clearly as the posterior shifts away from the cosmological‐constant point and lies in the quadrant $\omega_0 > -1$ and $\omega_a < 0$, suggesting the preference for dynamically dark energy rather than the cosmological constant.

Indeed, the combination of DESI DR2 with other cosmological measurements provides an intriguing perspective on the nature of dark energy. Unless there exists an unrecognized systematic error within one or more datasets, the results suggest that the $\Lambda$CDM model is being challenged by the combined DESI BAO, CMB, and SNe analyses. We may indeed be witnessing the first cracks in the cosmological constant; however, it's still too early to draw definitive conclusions. Upcoming Stage IV surveys particularly DESI DR3 will be crucial in providing deeper insight into the possible dynamical nature of dark energy.

\subsection{Not only dynamic but also phantom}\label{sec_4c}
\begin{figure*}
\centering
\includegraphics[scale=0.30]{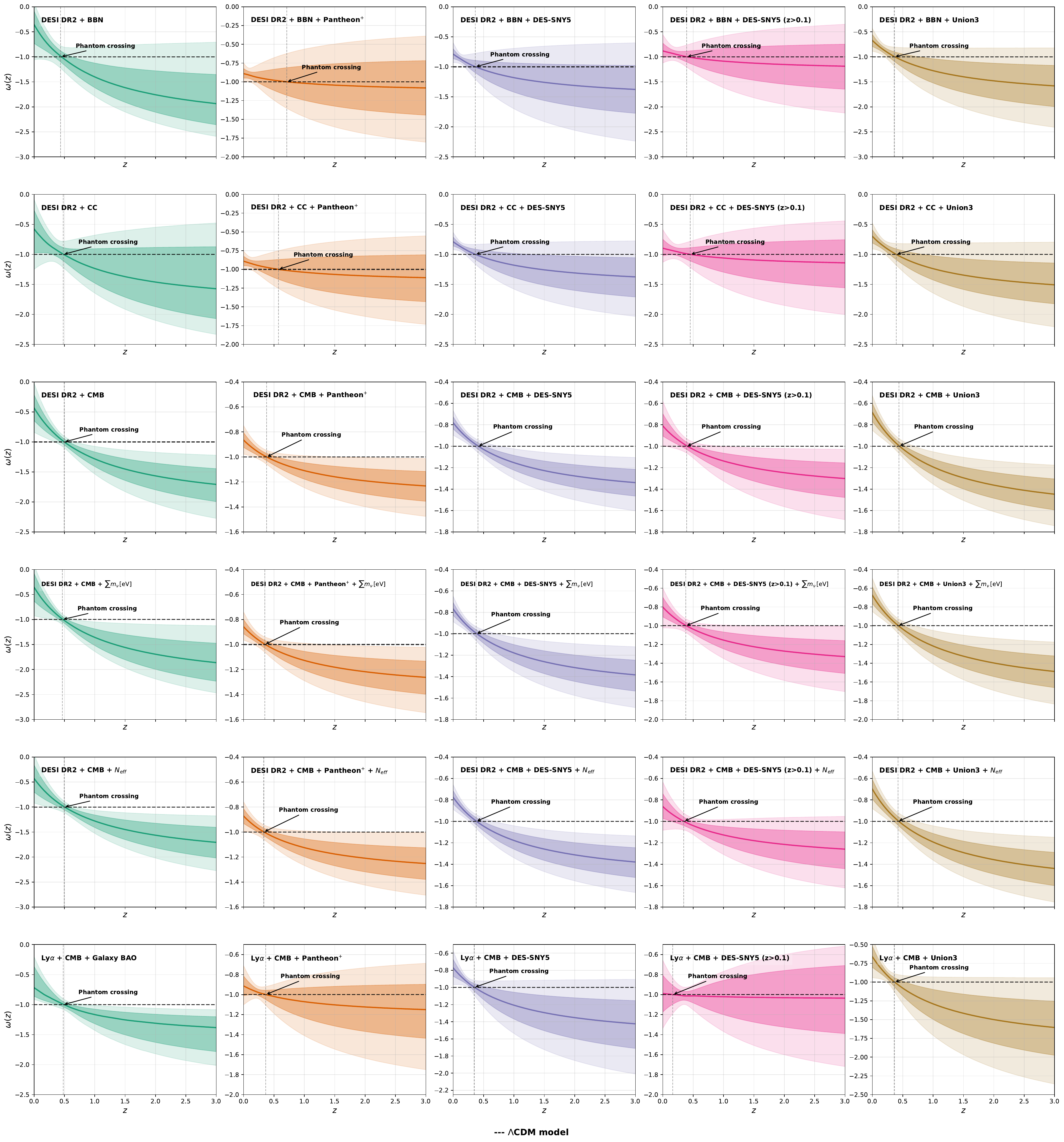}
\caption{This figure shows the reconstructed evolution of $\omega(z)$ as a function of redshift, obtained from the combination of DESI DR2, DESI DR2 Ly$\alpha$, CMB, BBN, and CC data, together with different Type~Ia supernova samples (Pantheon$^+$, DES-SN5Y, DES-SN5Y with $z > 0.1$, and Union3). The solid line represents the mean reconstruction, with shaded regions showing the $1\sigma$ and $2\sigma$ confidence intervals. The dashed horizontal line marks the cosmological-constant value $\omega = -1$, while the vertical line indicates the redshift at which a \textit{phantom crossing} occurs ($\omega < -1$).}\label{fig_12}
\end{figure*}

This subsection shows the evolution of the dark energy equation of state, $\omega(z)$, and the corresponding fractional dark energy density, $f_{\mathrm{DE}}(z)$, as functions of redshift, and illustrates their possible deviations from the cosmological-constant expectations, $\omega = -1$ and $f_{\mathrm{DE}}(z) = 1$. In Fig.~\ref{fig_12}, we show the evolution of $\omega(z)$ as a function of redshift, using the $\omega_0 \omega_a$CDM model with and without allowing $\sum m_{\nu}$ and $N_{\mathrm{eff}}$ to vary as free parameters. The analysis is performed using DESI~DR2, DESI~DR2 Ly$\alpha$, CMB, BBN, and CC data, combined with different Type~Ia supernova samples (Pantheon$^+$, DES-SN5Y, DES-SN5Y with $z > 0.1$, and Union3).

We observe that, in each dataset combination, the mean values of $\omega(z)$ fall below $\omega = -1$ at redshifts $z \gtrsim 0.5$, indicating a phantom regime ($\omega < -1$). At lower redshifts, around $z \lesssim 0.5$, $\omega(z)$ rises back above $-1$, entering the quintessence-like regime ($\omega > -1$). The points where the curves transition from the phantom region to the quintessence region are referred to as the \textit{phantom crossing}. This behaviour—phantom at high redshift and quintessence at low redshift is characteristic of the Quintom-B class of models~\cite{cai2025quintom,yang2025gaussian}. Moreover, several recent studies have reported similar phantom-crossing behaviour in analyses incorporating DESI measurements~\cite{ozulker2025dark,chen2025quintessential,silva2025testing,roy2025phantom,guedezounme2025phantom,hu2005crossing,yao2025general}.

Since the presence of a phantom crossing generally points toward dynamical dark energy rather than a cosmological constant, a wide range of theoretical models have been proposed to account for such transitions. Well-studied scalar-field scenarios include quintessence~\cite{tada2024quintessential,bhattacharya2024cosmological,berghaus2024quantifying,ramadan2024desi,gialamas2025quintessence,bayat2025examining,ratra1988cosmological,wetterich1988cosmology,wolf2024scant,payeur2025observations,dinda2025physical}, phantom~\cite{caldwell2002phantom}, Quintom~\cite{feng2006oscillating,guo2005cosmological,wolf2025matching,wang2025resolving,adam2025comparing,vazquez2024coupled}, and k-essence models~\cite{armendariz2000dynamical,chiba2000kinetically,csillag2025geometric}.

\begin{figure*}
\centering
\includegraphics[scale=0.30]{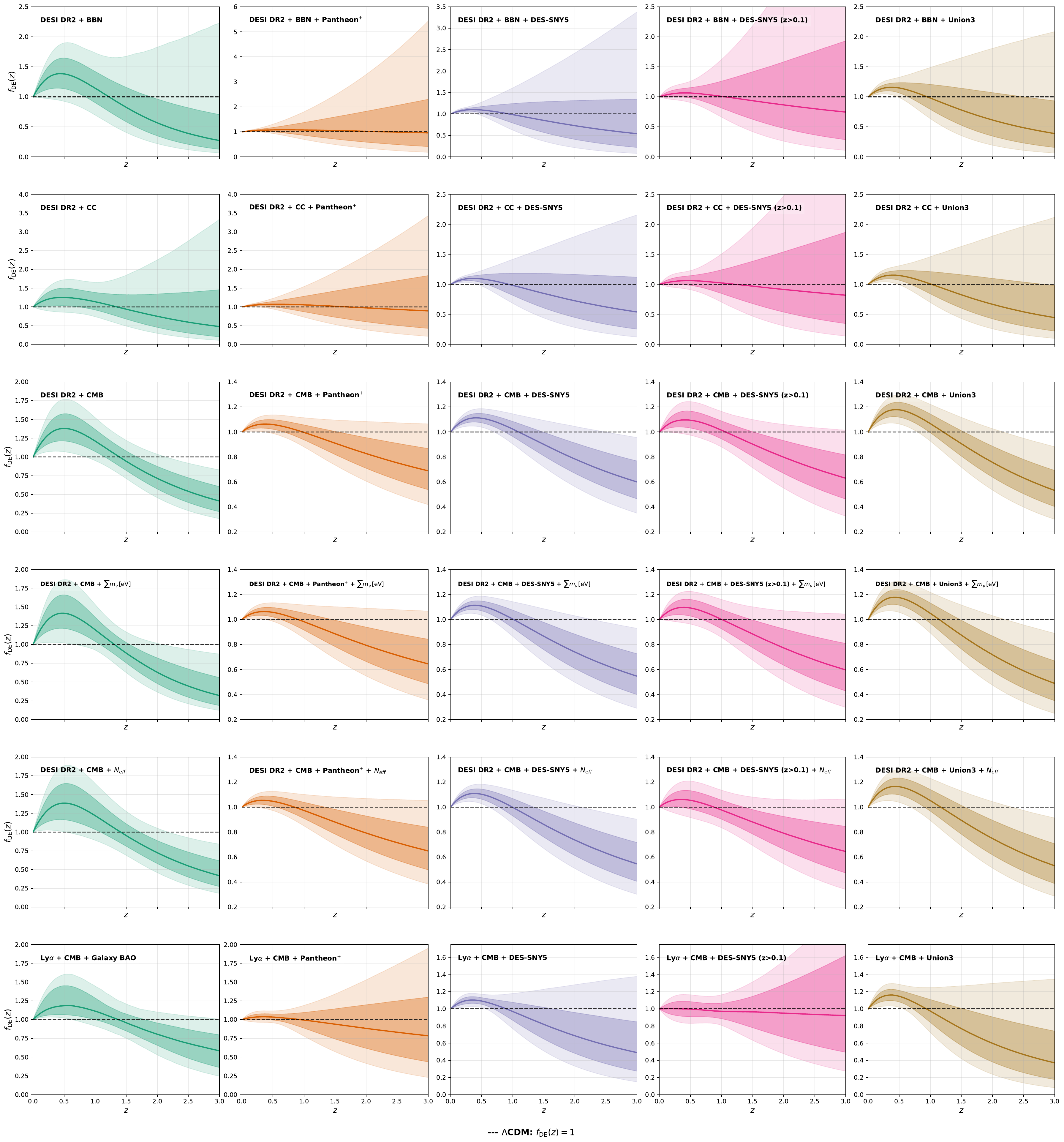}
\caption{This figure shows the reconstructed evolution of the fractional dark-energy contribution $f_{\mathrm{DE}}(z)$ as a function of redshift, obtained from the combination of DESI DR2, DESI DR2 Ly$\alpha$, CMB, BBN, and CC data, together with different Type~Ia supernova samples (Pantheon$^+$, DES-SN5Y, DES-SN5Y with $z > 0.1$, and Union3). The solid line represents the mean reconstruction, with shaded regions showing the $1\sigma$ and $2\sigma$ confidence intervals. The dashed horizontal line marks the reference value $f_{\mathrm{DE}} = 1$, corresponding to a cosmological constant. Departures from this constant value indicate possible dynamical dark-energy behavior.}\label{fig_13}
\end{figure*}

Similarly, Fig.~\ref{fig_13} shows the evolution of $f_{\mathrm{DE}}(z)$, which describes how much the dark sector contributes to the total energy budget as a function of redshift, or equivalently, how the effective dark-energy density changes over time. Any deviation from a constant $f_{\mathrm{DE}}(z)$ shows the preference of dynamical dark energy in the background. Each panel of the $f_{\mathrm{DE}}(z)$ shows that the dark-energy fraction increases toward low redshift and becomes dominant at $z \lesssim 1$, while it steadily decreases at higher redshift, where its contribution becomes negligible. Mild enhancements around intermediate redshifts in some datasets indicate small transient departures from a constant dark-energy density, consistent with the phantom-like features seen in $\omega(z)$.

\begin{figure}
\centering
\includegraphics[scale=0.48]{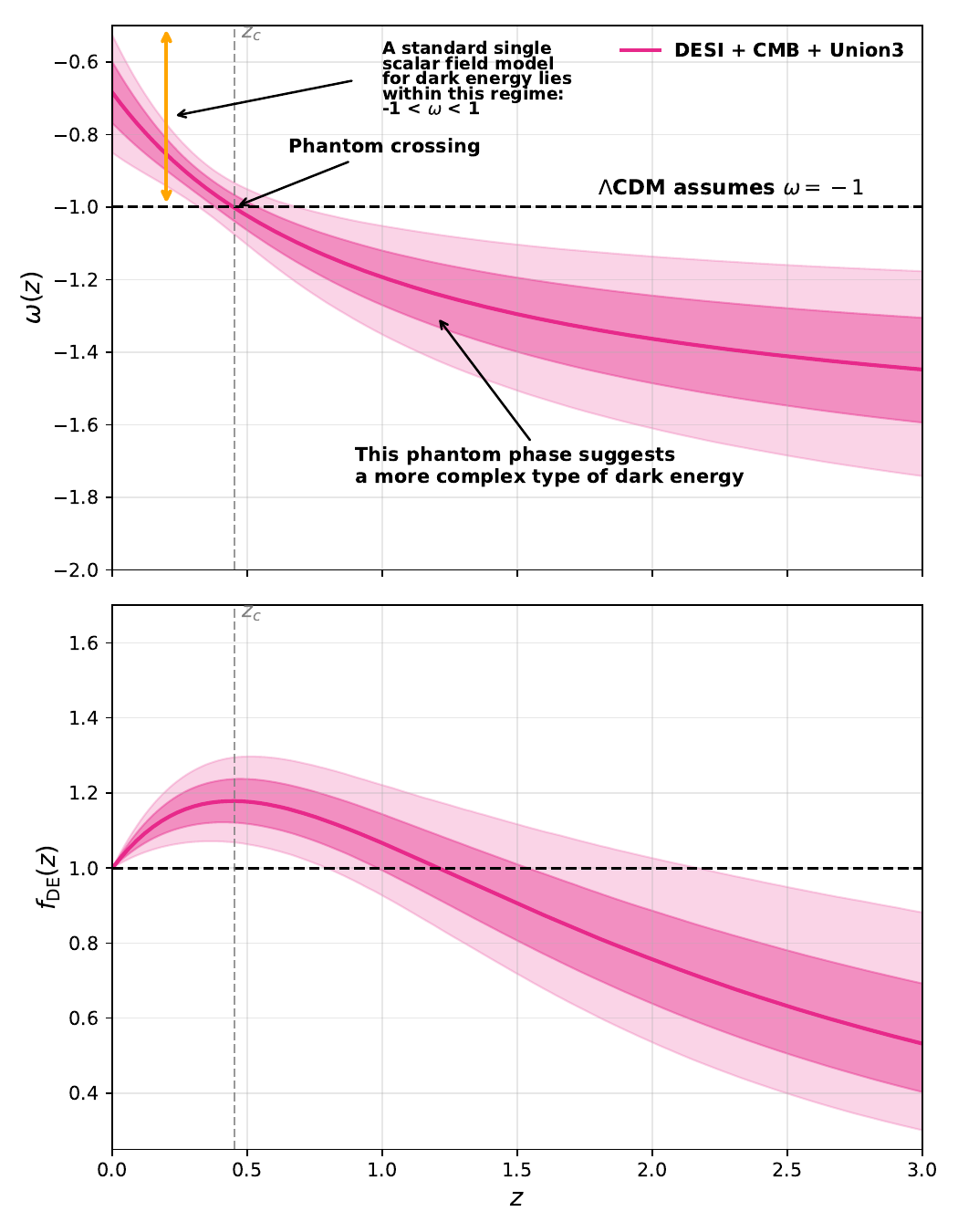}
\caption{This figure shows an annotated version of $\omega(z)$ (upper panel) and $f_{\mathrm{DE}}(z)$ (lower panel) as a function of redshift, obtained from the combined DESI DR2 + CMB + Union3 dataset. The pink solid line shows the mean evolution, with shaded bands representing the $1\sigma$ and $2\sigma$ confidence regions. The horizontal dashed line corresponds to the $\Lambda$CDM limit ($\omega = -1$), while the vertical line indicates the redshift $z_c$ where the \textit{phantom crossing} occurs. The region $-1 < \omega < 1$ corresponds to standard single-field dark energy, whereas the phantom phase suggests a more complex or non-canonical dark-energy behaviour.}\label{fig_14}
\end{figure}

As a summary, Fig.~\ref{fig_14} shows the evolution of the dark energy equation of state, $\omega(z)$, and the corresponding fractional dark-energy density, $f_{\mathrm{DE}}(z)$, using the DESI + CMB + Union3 combination within the $\omega_0\omega_a$CDM model. The upper panel shows that $\omega(z)$ crosses the $\omega=-1$ line at redshift $z_c$, entering a quintessence-like phase ($\omega>-1$) at lower redshifts after coming from a phantom regime at higher redshifts ($\omega<-1$). The point where the model transitions from the phantom region to the quintessence region is known as the \textit{phantom crossing}, indicated by the arrow. At $z=0$, the model lies within the range $-1 \leq \omega \leq 1$, as expected for a standard single scalar field dark energy model minimally coupled to gravity (e.g.\ quintessence), which is restricted to this parameter space. The lower panel shows the evolution of $f_{\mathrm{DE}}(z)$, which indicates how the effective dark-energy density changes with redshift. A small rise above the $\Lambda$CDM value ($f_{\mathrm{DE}}=1$) at intermediate redshifts aligns with the phantom-like behaviour seen in $\omega(z)$. At higher redshifts, $f_{\mathrm{DE}}(z)$ steadily decreases, reflecting that dark energy becomes less important in the early Universe. Taken together, the two panels show a consistent picture of a dynamical dark-energy component that moves across the $\omega=-1$ boundary.

\subsection{Dynamical dark energy or Systematics ?}\label{sec_4d}

\begin{figure*}
\begin{subfigure}{.33\textwidth}
\includegraphics[width=\linewidth]{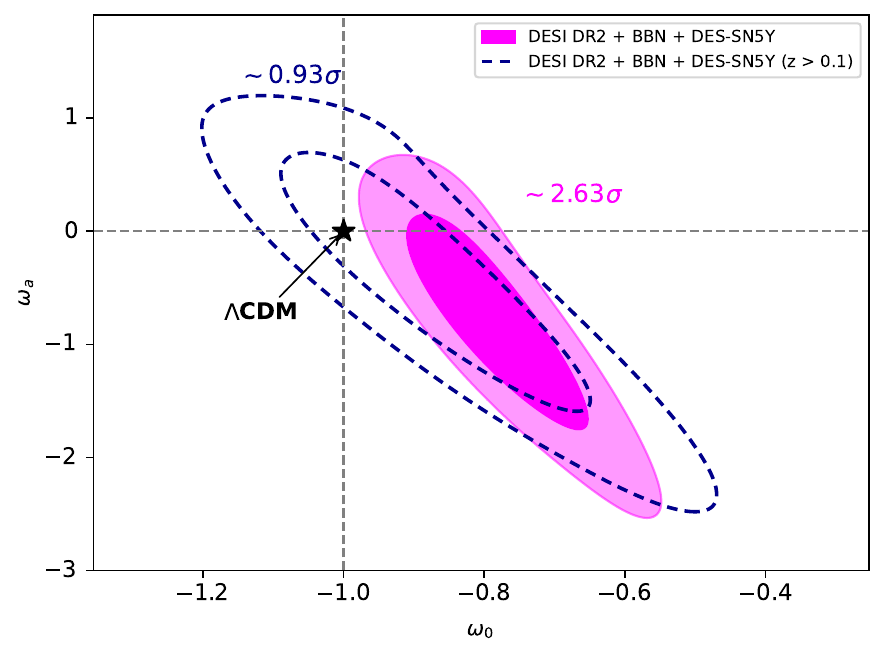}
\end{subfigure}
\hfil
\begin{subfigure}{.33\textwidth}
\includegraphics[width=\linewidth]{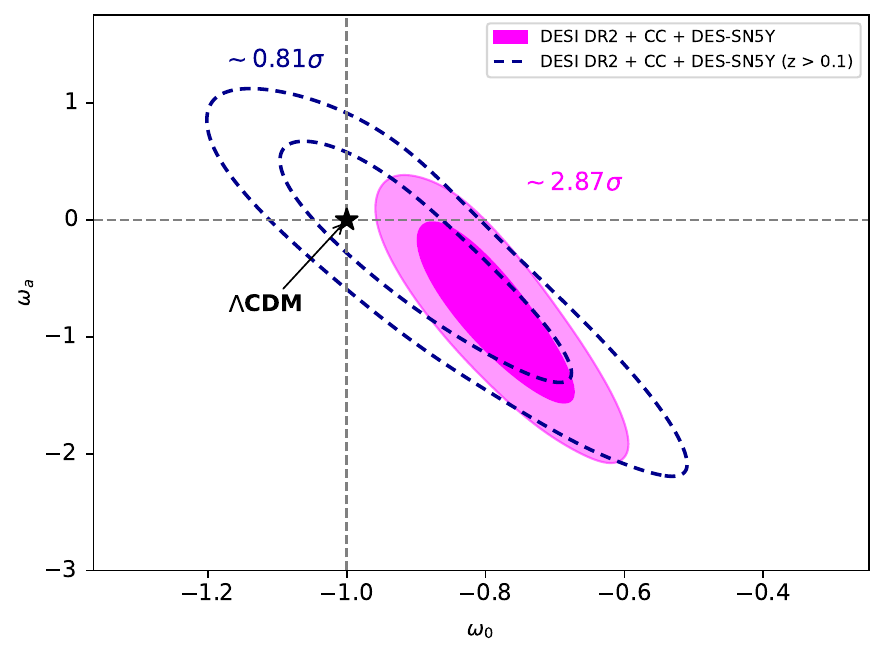}
\end{subfigure}
\hfil
\begin{subfigure}{.33\textwidth}
\includegraphics[width=\linewidth]{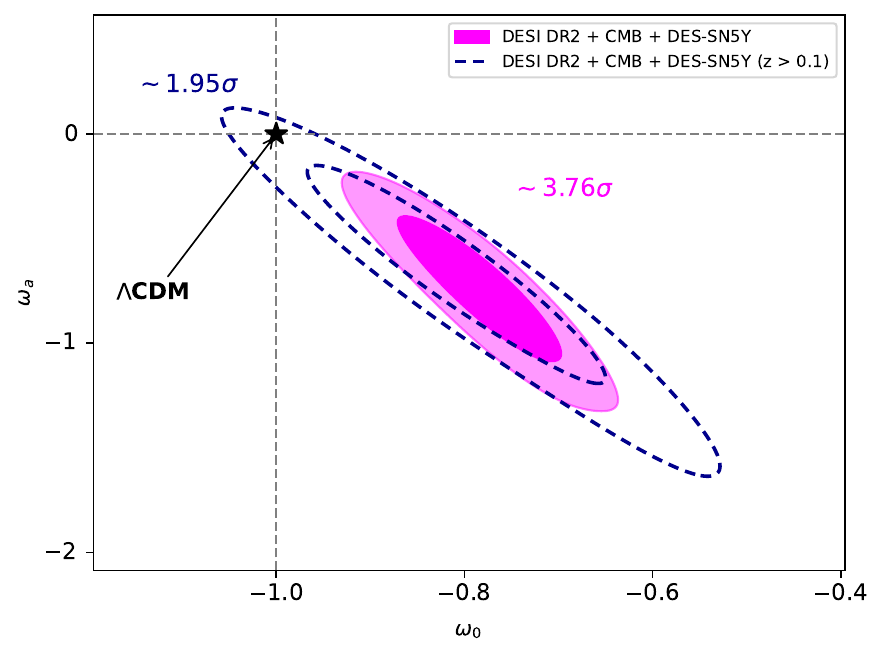}
\end{subfigure}
\begin{subfigure}{.33\textwidth}
\includegraphics[width=\linewidth]{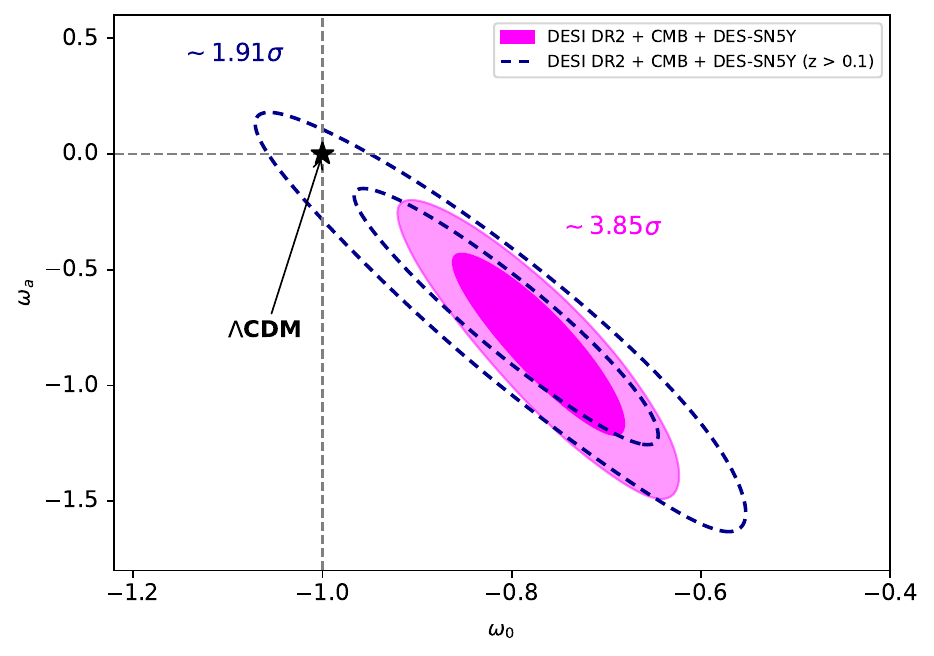}
\end{subfigure}
\hfil
\begin{subfigure}{.33\textwidth}
\includegraphics[width=\linewidth]{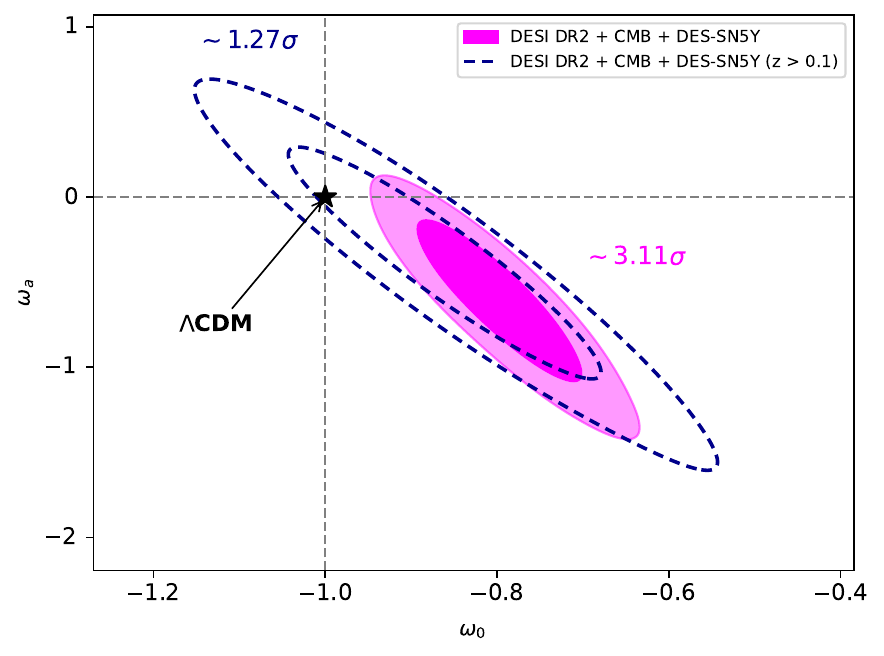}
\end{subfigure}
\hfil
\begin{subfigure}{.33\textwidth}
\includegraphics[width=\linewidth]{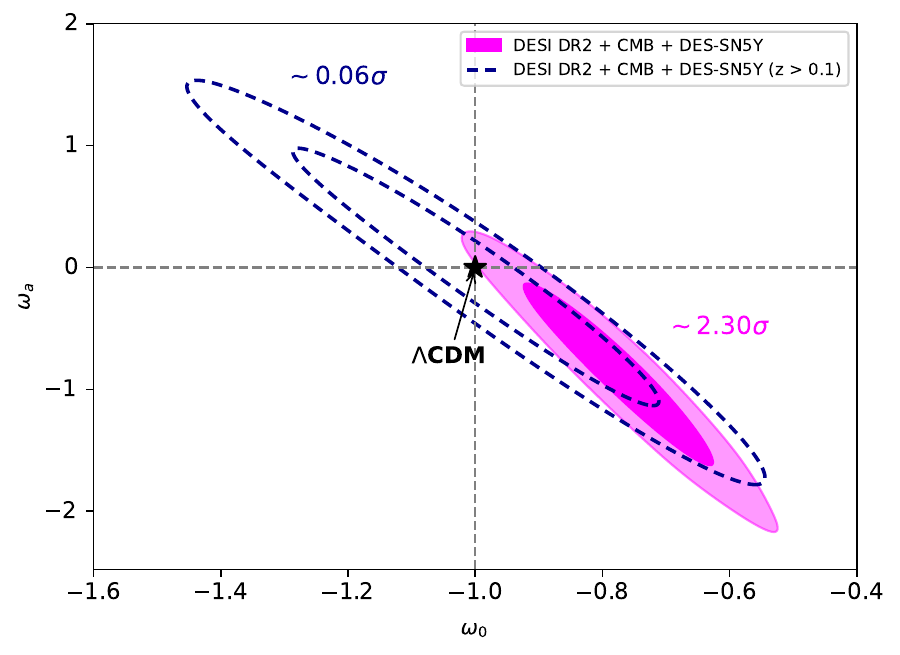}
\end{subfigure}
\caption{The figure shows the $\omega_0$–$\omega_a$ parameter plane obtained from the $\omega_0\omega_a$CDM, $\omega_0\omega_a$CDM + $\sum m_{\nu}$, and $\omega_0\omega_a$CDM + $N_{\mathrm{eff}}$ models using different dataset combinations. The contours correspond to the 68\% ($1\sigma$) and 95\% ($2\sigma$) confidence regions derived from DESI~DR2 and CMB combined with the DES-SN5Y sample (shown by the filled contours), and the DES-SN5Y sample with $z > 0.1$ (shown by the dotted contours). The black star marks the $\Lambda$CDM prediction ($\omega_0 = -1$, $\omega_a = 0$).}\label{fig_15}
\end{figure*}

As observed, the DES-SN5Y samples provide stronger evidence in favor of dynamical dark energy. In this subsection, we provide a detailed discussion about the bias caused by low-redshift SNe Ia in the evidence for dynamical dark energy. To do this, we consider the DES-SN5Y dataset and remove the 194 SNe Ia at $z < 0.01$, extracted from the CfA/CSP Foundation sample~\cite{hicken2009cfa3,hicken2012cfa4,foley2017foundation}. This leaves us with two datasets: the complete DES-SN5Y sample and another sample where the low-redshift $z < 0.01$ SNe Ia measurements have been removed, leaving a total of 1635 SNe Ia. In Fig.~\ref{fig_15}, we show the posterior distribution of the $\omega_0$-$\omega_a$ quadrant for the $\omega_0\omega_a$CDM model using different dataset combinations: DESI~DR2, CMB, BBN, CC, DES-SN5Y, and DES-SN5Y ($z>0.01$). The filled magenta contours correspond to the full DES-SN5Y sample, while the dotted blue contours represent the case where the low-$z$ ($z<0.01$) SNe~Ia sample is excluded. The standard $\Lambda$CDM model is represented by the black star in $\omega_0 = -1, , \omega_a = 0$ in the $\omega_0–\omega_a$ quadrant.

We find that the inclusion of the full DES-SN5Y sample shows a strong preference for a dynamical dark energy scenario. Specifically, the combination of DESI~DR2 + BBN + DES-SN5Y shows a preference for dynamical dark energy at the $\sim 2.6\sigma$ level, which decreases to $\sim 0.9\sigma$ when low-$z$ SNe~Ia are excluded. A similar pattern is observed for DESI~DR2 + CC + DES-SN5Y, where the preference decreases from $\sim2.9\sigma$ to $\sim0.8\sigma$ after excluding the low-$z$ sample.

When CMB data are included, the preference for dynamical dark energy becomes more pronounced. For example, the combination of DESI~DR2 + CMB + DES-SN5Y shows a preference of approximately $\sim 3.7\sigma$, which decreases to $\sim 1.9\sigma$ when the low-$z$ SNe~Ia are excluded. Extending the $\omega_0\omega_a$CDM model to include the total neutrino mass, $\sum m_{\nu}$, we find that the combination of DESI~DR2 + CMB + DES-SN5Y shows a preference for dynamical dark energy up to the $\sim 3.8\sigma$ level, which decreases to $\sim 1.9\sigma$ when low-$z$ SNe~Ia are excluded.

When varying the effective number of relativistic species, $N_{\mathrm{eff}}$, we find a similar behavior. The combination of DESI~DR2 + CMB + DES-SN5Y shows a preference for a dynamical dark energy of approximately $\sim 3.7\sigma$, which decreases to $\sim 1.2\sigma$ when the low-$z$ SNe~Ia are excluded. Furthermore, by including the Ly$\alpha$ forest measurements together with the CMB and DES-SN5Y data, we find that inclusion of the DES-SN5Y sample shows a preference of about $\sim 2.3\sigma$, which decreases to $\sim 0.06\sigma$ when the low-$z$ SNe~Ia sample is removed.

These results consistently show that the preference for dynamical dark energy is largely driven by the inclusion of low-redshift supernovae. When the nearby ($z < 0.1$) SNe~Ia are excluded, the results no longer require a dynamical dark energy component and thus fully restore $\Lambda$CDM concordance. The results clearly indicate that the apparent preference for a dynamical dark energy scenario originates primarily from the low-$z$ data. Therefore, these findings imply that either our Universe underwent a dramatic change very recently or, more plausibly, that we do not yet fully understand the astrophysical or observational systematics affecting our local Universe within a radius of roughly $300\,h^{-1}\,\mathrm{Mpc}$.
\begin{figure*}
\begin{subfigure}{.80\textwidth}
\includegraphics[width=\linewidth]{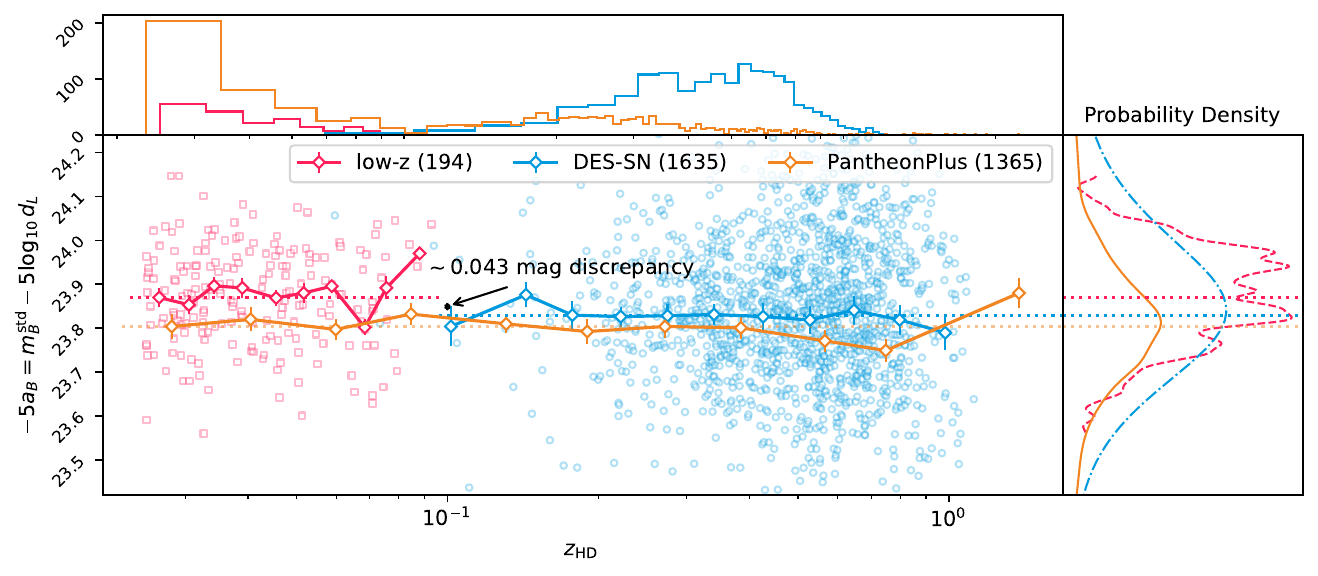}
\end{subfigure}
\hfil
\begin{subfigure}{.80\textwidth}
\includegraphics[width=\linewidth]{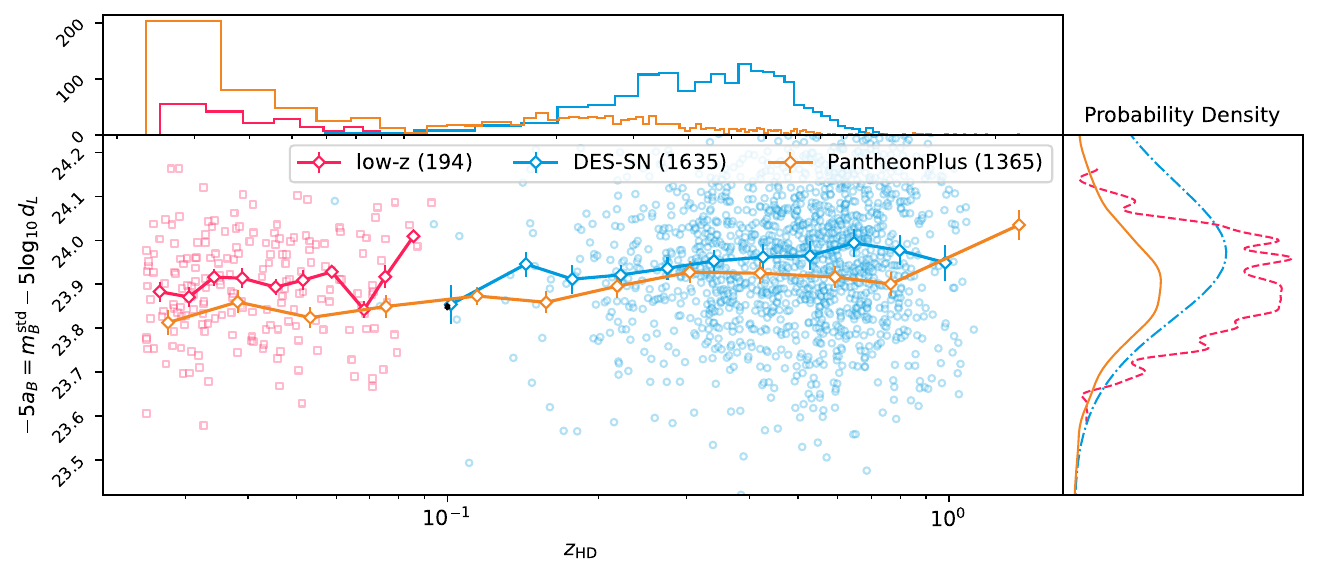}
\end{subfigure}
\caption{The figure shows the $-5a_B$ diagnosis for the low- and high-$z$ SNe~Ia samples. The analysis includes 194 SNe~Ia from the low-$z$ sample (red line), 1635 SNe~Ia from the DES-SN5Y sample (blue line), and 1365 SNe~Ia from the Pantheon$^+$ sample (orange line; SNe~Ia with $z \ge 0.0233$ are considered). The top and bottom-right panels show the probability density distributions for each sample. The observed $\sim 0.043$~mag discrepancy between the low and high-$z$ DES-SN5Y samples indicates a potential systematic effect in the DES-SN5Y sample.}\label{fig_16}
\end{figure*}


The above results motivate us to perform a systematic diagnosis to identify potential biases in low-$z$ SNe~Ia. We reconstruct the logarithmic magnitude distance relation, which connects the standardized apparent magnitude $m_{B,i}^{\mathrm{std}}$ of each SNe Ia to its dimensionless luminosity distance $\hat{d}_L(z_i)$ as $m_{B,i}^{\mathrm{std}} = 5 \, \log_{10}\!\left(\hat{d}_L(z_i)\right) - 5\,a_B$, where the intercept term $a_B$ represents the degeneracy between the absolute magnitude $M_B$ and the Hubble constant $H_0$, defined as: $-5a_B = M_B + 5\,\log_{10}\!\left(\frac{c/H_0}{\mathrm{Mpc}}\right) + 25.$ Here, the dimensionless luminosity distance $\hat{d}_L(z_i)$ is expressed as $\hat{d}_L(z_i) = (1 + z_{\mathrm{hel}}) 
\int_0^{z_i} \frac{dz'}{E(z')},$ where $\left(\frac{E(z)}{H_0}\right)^{2} = \Omega_m(1 + z)^3 + (1 - \Omega_m)$ follows the fiducial flat $\Lambda$CDM cosmology.

In this framework, the standardized magnitude $m_{B,i}^{\mathrm{std}}$ is corrected for the light-curve parameters, allowing $a_B$ to serve as a sensitive diagnostic of systematic effects. If the inferred intercept $-5a_B$ remains consistent between different samples of SNe Ia, it suggests the absence of major systematics. In contrast, any deviations in $-5a_B$ could indicate observational or modeling biases such as uncorrected peculiar velocity effects, uncertainties in redshift measurements, or inaccuracies in the modeling of $\hat{d}_L(z)$. By examining the redshift dependence of $-5a_B$, one can identify redshift ranges where potential systematics or signatures of new physics may emerge.

In Fig.~\ref{fig_16}, we show the diagnosis $a_B$. The upper panel presents the analysis for the $\Lambda$CDM model, while the lower panel shows the results for the best-fit $\omega_0\omega_a$CDM model. As we discussed earlier in this subsection, the dynamical DE in DES-SN5Y is mainly driven by the low-$z$ SNe~Ia. For this reason, we consider the DES-SN5Y dataset in two subsets: one excluding the low-$z$ sample, which contains 1635 SNe~Ia, and the other consisting of the low-$z$ sample with 194 measurements.

It can be observed that the high-quality DES-SN5Y sample (1635 SNe~Ia) shows a stable distribution $-5a_B$, while the low-$z$ sample shows irregular fluctuations in $-5a_B$. The weighted average for the low-$z$ sample is approximately 0.043~mag, Which is very close to the findings of \cite{efstathiou2025evolving}, but with 0.04~mag. The magnitude offset is estimated as $\Delta m_{\mathrm{offset}} := (m^{\mathrm{low-}z}_{\mathrm{\text{Pantheon$^+$}}} - m^{\mathrm{low-}z}_{\mathrm{\text{DES-SN5Y}}}) - (m^{\mathrm{high-}z}_{\mathrm{\text{Pantheon$^+$}}} - m^{\mathrm{high-}z}_{\mathrm{\text{DES-SN5Y}}}) \approx (-0.05) - (-0.01) = -0.04~\mathrm{mag},$ between Pantheon$^+$ and DES-SN5Y compilations at low and high redshifts, assuming a constant $M_B$. In \cite{vincenzi2025comparing}, the authors show that part of this offset can be corrected by refining the intrinsic scatter model and the mass-step estimate.

This discrepancy about $\sim 0.043$~mag in $-5a_B$ appears in the DES-SN5Y compilation around $z \approx 0.1$. This redshift range corresponds to a part of the Universe that is already homogeneous, so local inhomogeneities can be safely ruled out as the cause. In addition, such a sharp change is unlikely to result from any large scale homogeneous physical effect, since the Pantheon$^+$ compilation, which contains many more well-calibrated, low-$z$ Type Ia supernovae, shows no sign of a similar feature at low redshift.

In addition, we consider the $\omega_0\omega_a$CDM model. Under the $\omega_0\omega_a$CDM model, we found that the Pantheon$^+$ compilation also shows inconsistency in the $-5a_B$ values between its low-$z$ and high-$z$ samples, which can be observed at the bottom of Fig.~\ref{fig_16}. We also show the probability density, and it can be seen that the red dashed line represents the 1D probability density of the low intercept $z$ $-5a_B$ for DESY5. This distribution is non-Gaussian, it is irregular, meaning that those low-$z$ supernovae are scattered more than they should be.

In contrast, the other two cases (high-$z$ DESY5 and Pantheon$^+$) have smooth Gaussian distributions, which means that their SNe Ia behave consistently with the expected brightness–distance relations. So, the low-$z$ DESY5 supernovae are the odd ones, they scatter too much and do not align with a consistent distance magnitude relation. Pantheon$^+$, on the other hand, contains even more subsamples (18 total, from many surveys) but still shows a smooth Gaussian distribution around the mean intercept value. That tells us that Pantheon$^+$’ low-$z$ data are well-calibrated and internally consistent, although it is more complex. Meanwhile, DESY5’s low-$z$ data seem to suffer from systematic calibration errors.

It is important to note that we also consider the Pantheon$^+$ sample, which contains a larger number of low-$z$ well-calibrated SNe~Ia. For details on the preference for dynamical dark energy discussed in Sec.~\ref{sec_4b}, we examine the $\omega_0\omega_a$CDM model by considering all scenarios with and without varying $\sum m_{\nu}$ and $N_{\mathrm{eff}}$. The results of DESI~DR2 combined with BBN, CMB, and Pantheon$^+$ show a preference for dynamical dark energy ranging from $0.97\sigma$ to $2.51\sigma$, depending on the choice of the combination.

Hence, when the full DES–SN5Y sample is included, the preference for dynamical dark energy increases to $3.85\sigma$. However, when low-$z$ SNe~Ia are excluded from DES–SN5Y, this preference drops below $2\sigma$. The Pantheon$^+$ sample, which contains a larger number of low-$z$ SNe~Ia, also shows evidence of dynamical dark energy below $2.52\sigma$, indicating that the apparent signal is not statistically significant. Therefore, the success of the standard $\Lambda$CDM model remains unchallenged. In \cite{gialamas2025interpreting,efstathiou2025evolving,cortes2025desi,huang2025desi}, it is also shown that the evidence for dynamical dark energy is biased by the low-$z$ SNe~Ia sample in DES-SN5Y.

\subsection{Statistical Analysis}\label{sec_4e}
In this subsection, we present a detailed analysis based on the changes in the goodness of fit and Bayesian evidence. First, the changes in the goodness of fit, quantified by $\Delta \chi^2$, show that the inclusion of the entire DES-SN5Y sample significantly improves the fit in favor of the $\omega_0\omega_a$CDM model over the $\Lambda$CDM model, with values $\Delta \chi^2$ typically in the range of $-5$ to $-7$, corresponding to a level of significance $2.5$-$4\sigma$.  This improvement is most evident when DESI~DR2 is combined with CMB or BBN data, whereas the exclusion of the low-redshift sample ($z < 0.1$) SNe~Ia of DES-SN5Y results in $\Delta \chi^2 \approx 0$, indicating little to no statistical preference for the $\omega_0\omega_a$CDM model.

\begin{figure*}
\centering
\includegraphics[scale=0.55]{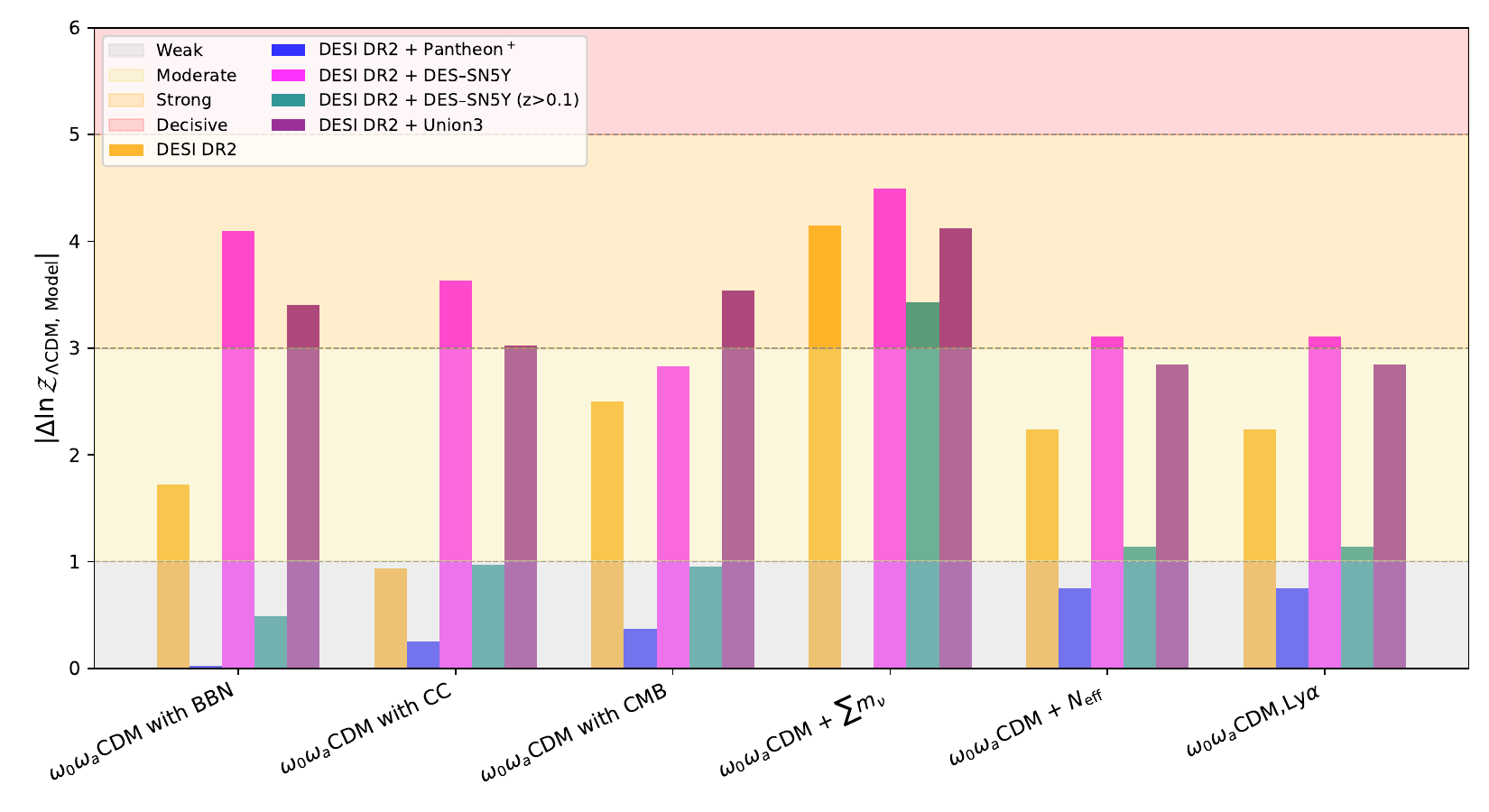}
\caption{This figure shows the Bayesian evidence difference, $|\Delta \ln \mathcal{Z}_{\Lambda\mathrm{CDM},\mathrm{Model}}|$, for the $\omega_0\omega_a$CDM model across various dataset combinations. The colored bars represent combinations of DESI~DR2 with BBN, CC, CMB, and Ly$\alpha$ data, together with different Type~Ia supernova samples (Pantheon$^+$, DES-SN5Y, DES-SN5Y with $z>0.01$, and Union3). The shaded regions correspond to the revised Jeffreys’ scale, indicating \textit{weak}, \textit{moderate}, \textit{strong}, and \textit{decisive} Bayesian evidence.}\label{fig_17}
\end{figure*}

Fig.~\ref{fig_17} shows the corresponding values of the Bayesian evidence for the $\omega_0\omega_a$CDM model corresponding to each dataset. 
One can observe that when combining DESI~DR2 + CMB with DES-SN5Y, the $\omega_0\omega_a$CDM model shows values $|\Delta \ln \mathcal{Z}|$ between 3 and 4.5, indicative of strong Bayesian evidence according to the revised Jeffreys’ scale. The combination DESI~DR2 + CMB + DES-SN5Y with free $\sum m_{\nu}$ shows the strongest overall support, with $|\Delta \ln \mathcal{Z}| = 4.49$, corresponding to decisive evidence. When the low-$z$ SNe~Ia are excluded from the DES-SN5Y sample, the Bayesian evidence decreases to $|\Delta \ln \mathcal{Z}| < 1$, indicating an inconclusive statistical preference.

Together, the $\Delta \chi^2$ and Bayesian analyzes indicate that, at the statistical level, these findings also support our earlier results, showing that the apparent preference for dynamical dark energy is primarily driven by the inclusion of the low-$z$ supernova sample.

\section{Discussion and Conclusions}\label{sec_5}
In this paper, we provide a detailed review of whether there is or is not a preference for dynamical dark energy in light of the recent DESI DR2 measurements. To do so, we consider the $\omega_0\omega_a$CDM model and also vary the parameters $\sum m_{\nu}\,[\mathrm{eV}]$ and $N_{\mathrm{eff}}$. We then use an MCMC analysis and use the several datasets as discussed in Sec.~\ref{sec_2}, to constrain the parameters of the $\omega_0\omega_a$CDM model.

Our results show that, although DESI~DR2 significantly improves the precision of the BAO, neither $\Lambda$CDM nor $\omega_0\omega_a$CDM reduces the sound horizon in the direction required to ease the Hubble tension. Across all Pantheon$^{+}$ based combinations, $r_d$ remains in the range $147$-$149$~Mpc, far from the $\sim 7\%$ reduction needed, and becomes even more inconsistent once CMB data are included. The inferred values of $H_0$ remain likewise near $h \simeq 0.672$-0.676, preserving a tension $2\sigma$ with SH0ES and increasing to nearly $4\sigma$ when CMB constraints are added. As dark energy is negligible at recombination, dynamical dark energy cannot modify early-time physics or reduce the sound horizon. Moreover, DESI~DR2 favors $\omega(z) > -1$, implying $\rho_{\mathrm{DE}}(z)/\rho_{\mathrm{DE},0} > 1$, which drives $H_0$ towards lower values rather than higher ones. This behavior is generic and reflects the negative correlation between dark energy evolution and the present expansion rate.

Furthermore, our results show that DESI DR2 pushes the posterior toward the quadrant $\omega_0 > -1$ and $\omega_a < 0$, indicating a mild-to-moderate preference for dynamical dark energy. The significance of this preference varies across the different dataset combinations, remaining below the $3\sigma$ level in most cases, although combinations involving DESI DR2, CMB, and DES–SN5Y or Union3 reach values up to $\sim 3.8\sigma$. Allowing $\sum m_{\nu}$ and $N_{\mathrm{eff}}$ to vary does not alter this preference: the favored region remains in the dynamical dark energy quadrant, and neither extension provides evidence for significant departures from the standard neutrino sector. In particular, constraints on $\sum m_{\nu}$ remain consistent with $\lesssim 0.1$\,eV at the 95\% confidence level, with nonzero values detected only at the $\sim 1.5\sigma$ level.

While DESI~DR2 based combinations do exhibit a mild-to-moderate preference for dynamical dark energy, the statistical significance remains well below the threshold required to rule out $\Lambda$CDM. This is also supported by the systematic diagnostics, which show that the dynamical dark energy driven by the DES-SN5Y sample is in fact driven by the inclusion of low-$z$ SNe~Ia in DES-SN5Y. This indicates that the preference for dynamical dark energy over $\Lambda$CDM in DES-SN5Y is biased by low-$z$ SNe~Ia affected by calibration systematics rather than arising from the cosmological data as a whole.

The evolution of the dark energy equation of state $\omega(z)$ and the corresponding fractional dark energy density $f_{\mathrm{DE}}(z)$ shows clear signatures of dynamical behaviour across all dataset combinations. The evolution of $\omega(z)$ lies below the $\Lambda$CDM value ($\omega = -1$) at higher redshifts, indicating a phantom regime, and gradually rises above $-1$ toward low redshift, entering a quintessence-like phase. The dashed horizontal line marks the boundary $\omega(z) = -1$, and the point where the curve crosses this line corresponds to the phantom crossing, and this kind of behaviour is characterized by Quintom-B type dark energy. Also, the range $-1 < \omega < 1$ can be characterized by a single scalar-field model.

The evolution of $f_{\mathrm{DE}}(z)$ shows that the effective dark energy density increases toward low redshift, with mild departures from a constant value at intermediate redshifts that mirror the phantom-like features seen in $\omega(z)$. At high redshift, $f_{\mathrm{DE}}(z)$ naturally approaches zero, reflecting the fact that dark energy is essentially negligible in the early Universe.

Statistically speaking, both the $\Delta \chi^2$ and Bayesian evidence analyses confirm that the preference for dynamical dark energy arises almost entirely from the inclusion of the low-$z$ DES-SN5Y SNe Ia. When these low-$z$ SNe~Ia are included, the $\omega_0\omega_a$CDM model appears strongly favored, with notable improvements in the fit and values of $|\Delta \ln \mathcal{Z}|$ in the strong to decisive range. However, once the low-$z$ subsample is removed, this improvement vanishes and the Bayesian evidence drops to an inconclusive level ($|\Delta \ln \mathcal{Z}| < 1$).

The DESI collaboration offers valuable insights into the nature of dark energy. The preference for dynamical dark energy across several dataset combinations is intriguing, yet it remains far from sufficient to challenge the standard $\Lambda$CDM model. At present, the statistical evidence does not reach the level required to claim a genuine departure from a cosmological constant ($\Lambda$), particularly once the potential systematics associated with low-$z$ SNe Ia samples are taken into consideration. In this sense, DESI collaboration provides important hints, but not definitive answers, reinforcing the need for more precise and independent data before any firm conclusions can be drawn about the dynamics of dark energy.

Looking ahead, the decisive progress will come from the next generation of Stage~IV surveys. Upcoming measurements from DESI’s future data releases, the Rubin Observatory, Euclid, the Nancy Grace Roman Space Telescope, the Simons Observatory, and PFS will dramatically sharpen our constraints on the expansion history, the growth of structure, and the high redshift Universe. These surveys will test whether the hints of dynamical dark energy persist or fade with improved data. Only after this wealth of forthcoming observations becomes available will we be able to determine whether $\Lambda$CDM truly breaks down, or whether the current deviations simply reflect statistical fluctuations or residual systematics.
\begin{acknowledgements}
SC acknowledges the Istituto Nazionale di Fisica Nucleare (INFN) Sez. di Napoli,  Iniziative Specifiche QGSKY and MoonLight-2  and the Istituto Nazionale di Alta Matematica (INdAM), gruppo GNFM, for the support. This paper is based upon work from COST Action CA21136  Addressing observational tensions in cosmology with systematics and fundamental physics (CosmoVerse), supported by COST (European Cooperation in Science and Technology). We also acknowledge Dr. Isidro G\'omez-Vargas and Dr. Vipin Kumar Sharma for their fruitful discussions, and Prof. Shao-Jiang Wang for his guidance on the plots used in Fig. 16.
\end{acknowledgements}

\appendix

\section{\textbf{DESI Dark Energy Characteristics}}\label{appendix}

\subsection{Quintom-B Type Dark Energy}\label{appendix_a1}
\begin{figure*}[htb]
\begin{subfigure}{.48\textwidth}
\includegraphics[width=\linewidth]{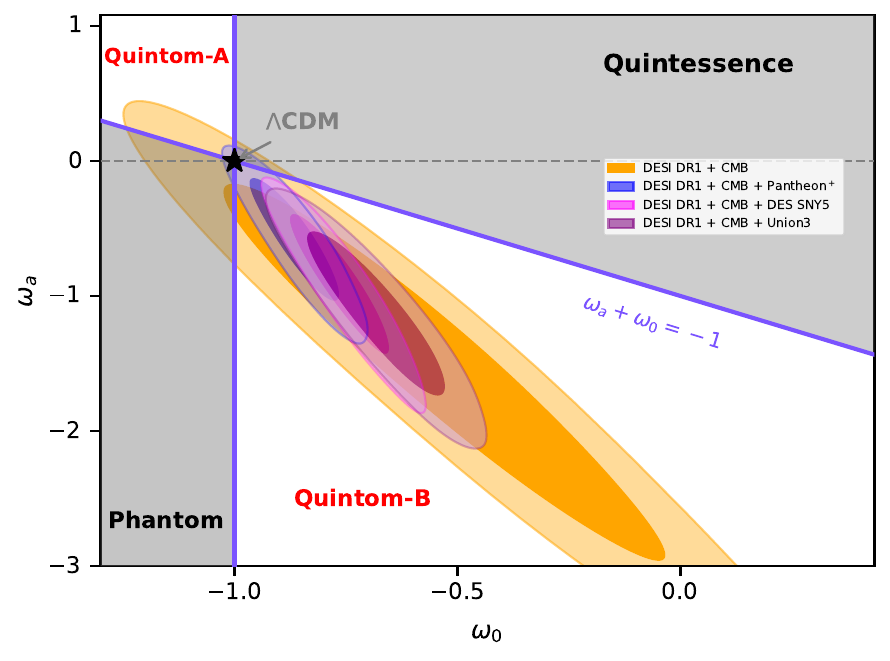}
\end{subfigure}
\hfil
\begin{subfigure}{.47\textwidth}
\includegraphics[width=\linewidth]{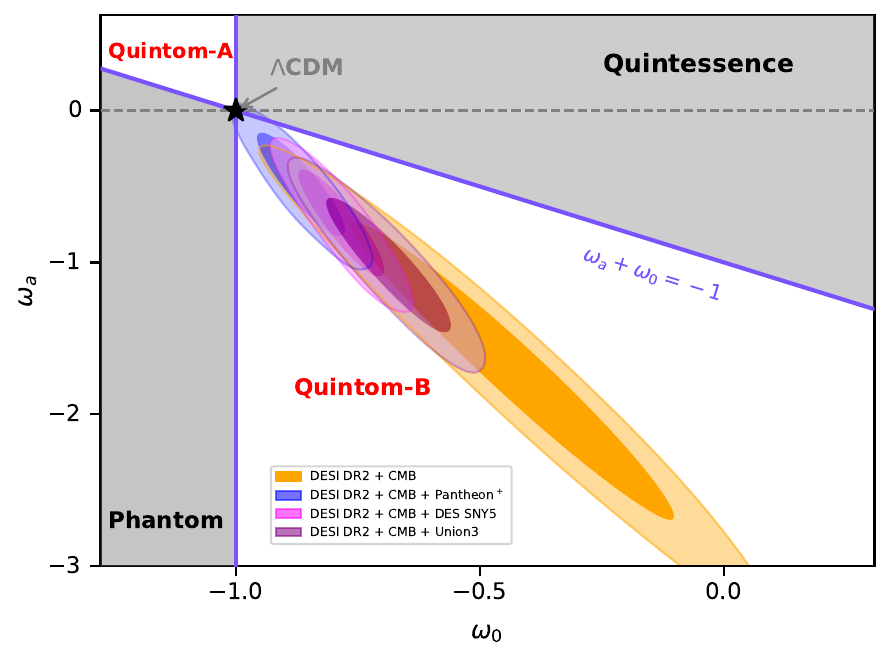}
\end{subfigure}
\caption{This figure shows the 68\% and 95\% confidence contours in the $\omega_0$--$\omega_a$ plane for the $\omega_0\omega_a$CDM model using DESI DR1 (left) and DESI DR2 (right) combined with CMB and several Type~Ia supernova 
datasets (Pantheon$^+$, DES--SN5Y, and Union3). The plane is divided into four regions associated with distinct dark-energy models: Quintessence ($\omega>-1$), Phantom ($\omega<-1$), Quintom A (crossing the 
$\omega=-1$ boundary from above), and Quintom B (crossing from below) The black star marks the $\Lambda$CDM limit at $(\omega_0, \omega_a)=(-1,0)$.}\label{fig_18}
\end{figure*}

In this Appendix, we discuss the possible characteristics of the DESI dark-energy
 constraints. In Fig.~\ref{fig_18}, we show the $\omega_0$--$\omega_a$ 
quadrant using the combination of DESI~DR2 with CMB data and several Type~Ia 
supernova samples (Pantheon$^+$, DES--SN5Y, DES--SN5Y, and Union3). We divide the dark-energy parameter space into four models:
\begin{itemize}
    \item \emph{Quintessence}: $\omega>-1$ in all epochs, i.e.\ $\omega_0>-1$ and 
    $\omega_0+\omega_a>-1$ \cite{ratra1988cosmological}.
    
    \item \emph{Phantom}: $\omega<-1$ at all epochs, i.e.\ $\omega_0<-1$ and 
    $\omega_0+\omega_a<-1$ \cite{caldwell2002phantom}.
    
    \item \emph{Quintom-A}: $\omega>-1$ in the past, but $\omega<-1$ today, corresponding 
    to $\omega_0<-1$ and $\omega_0+\omega_a>-1$ \cite{cai2025quintom,yang2025gaussian}.
    
    \item \emph{Quintom-B}: $\omega<-1$ in the past, but $\omega>-1$ today, corresponding 
    to $\omega_0>-1$ and $\omega_0+\omega_a<-1$ \cite{cai2025quintom,yang2025gaussian}.
\end{itemize}

As can be seen in Fig.~\ref{fig_18}, the contours of each data set combination fall in the region with $\omega_a<0$ and $\omega_0>-1$, indicating the preference for the dynamical dark energy scenario characterized by \emph{Quintom-B}

\subsection{DESI DR1 vs DESI DR2 Dark Energy}\label{appendix_a2}
\begin{figure*}[htb]
\begin{subfigure}{.48\textwidth}
\includegraphics[width=\linewidth]{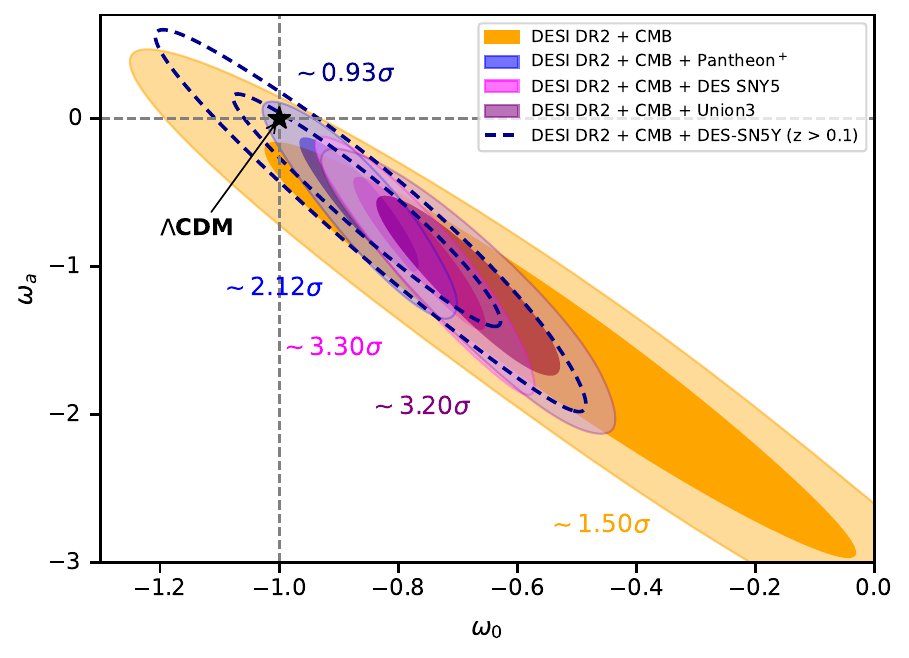}
\end{subfigure}
\hfil
\begin{subfigure}{.47\textwidth}
\includegraphics[width=\linewidth]{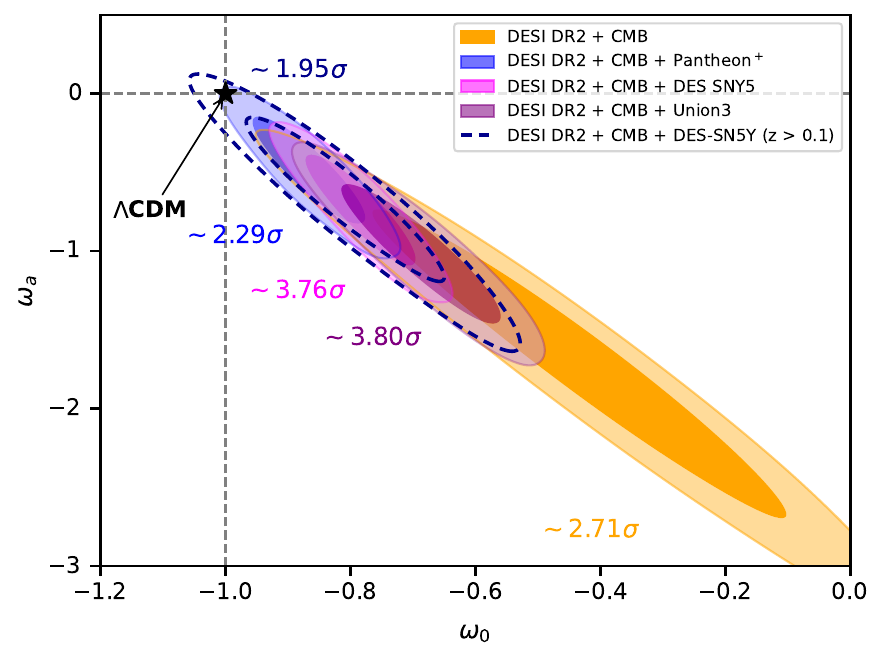}
\end{subfigure}
\caption{The figure shows the $\omega_0$–$\omega_a$ quadrant of the $\omega_0\omega_a$CDM model to compare the DESI DR1 and DESI DR2 preference for dynamical dark energy using CMB data and several Type~Ia supernova samples (Pantheon$^+$, DES–SN5Y, DES–SN5Y $(z>0.1)$, and Union3).}\label{fig_19}
\end{figure*}

Fig.~\ref{fig_19} shows the comparison between DESI Data Release 1 (left panel) and Data Release 2 (right panel) in the parameter space $\omega_0$-$\omega_a$, each combined with CMB data and several Type~Ia supernova samples (Pantheon$^+$, DES–SN5Y, DES–SN5Y with $z>0.1$ and Union3). The contours represent the confidence regions of 68\% and 95\% and highlight how the improvements in the DESI~DR2 measurements compare with those of DESI~DR1.

In both releases, the contours move away from the $\Lambda$CDM point $(\omega_0,\omega_a)=(-1,0)$ and fall within the region defined by $\omega_0 > -1$ and $\omega_a < 0$. This part of the parameter space corresponds to a \textit{Quintom-B}–type evolution (as discussed in Appendix~\ref{appendix_a1}), where dark energy behaves like a phantom component in the past but transitions to a quintessence-like behavior at the present epoch.

The statistical preference for the dynamical dark energy from DESI DR1 to DESI DR2 increases from $0.9\sigma$–$3.3\sigma$ to $1.9\sigma$–$3.8\sigma$, depending on the choice of the supernova sample used. A key aspect of this comparison is the role of low-$z$ SNe Ia in the preference for dynamical dark energy, as discussed in Sec.~\ref{sec_4d}. When the full DES–SN5Y sample is included, the contours show a deviation from the cosmological constant point, indicating a preference for dynamical dark energy of up to $\sim 3\sigma+$ in DESI DR1 and $\sim 3.8\sigma$ in DESI DR2. When the low-$z$ SNe~Ia sample is excluded from the DES–SN5Y dataset, DESI DR1 shows a deviation below $1\sigma$, and DESI DR2 also shows a deviation below $1\sigma$ in the preference for dynamical dark energy.

We also tested with the Pantheon$^+$ sample, which contains a larger number of low-$z$ well-calibrated SNe~Ia and shows a preference for dynamical dark energy of about $2.2\sigma$ with DESI DR1 and $2.3\sigma$ with DESI DR2. This indicates that it is still too early to conclude that the $\Lambda$CDM model is favored or that dynamical dark energy models are preferred by current datasets.

The above behavior shows that the departure from $\Lambda$CDM in both surveys is biased by the low-$z$ SNe~Ia samples. Moreover, DESI DR2, which is better than DESI DR1, shows a greater preference for dynamical dark energy compared to DR1, but still it is too early to say that the $\Lambda$CDM model is disfavored, and one should wait for the Stage IV surveys, especially the final DESI DR2 releases, before making any strong conclusions

\bibliographystyle{elsarticle-num}
\bibliography{mybib}

\end{document}